%% file: Robust_Cpt_Rev2.tex
\documentclass[a4paper,12pt]{article}
\input{header.tex}

\usepackage{colortbl,bm}
\usepackage[ruled,vlined]{algorithm2e}
\usepackage{enumerate}
\usepackage{subfig}
\usepackage{stmaryrd}
\usepackage{natbib}

\newcommand{\vtau}{\bm{\tau}}
\usepackage{authblk}
\usepackage{lscape}
\usepackage{amssymb}
\usepackage[titletoc,toc,title]{appendix}
\usepackage{bibunits}

\newcommand{\Oc}{\mathcal{O}}
\newcommand{\FC}{Q\hspace{+.1ex}}
\newcommand{\Fc}{\tilde{c}\hspace{+.1ex}}

  \let\hl\relax
\date{}
\title{Changepoint Detection in the Presence of Outliers}
\hl{
\author[1,$\dag$]{Paul Fearnhead}
\author[2,3]{Guillem Rigaill}
\affil[1]{Department of Mathematics and Statistics, Lancaster University}
\affil[2]{Institute of Plant Sciences Paris-Saclay, UMR 9213/UMR1403, CNRS, INRA, Universit{\'e} Paris-Sud, Universit{\'e} d'Evry,
Universit{\'e} Paris-Diderot, Sorbonne Paris-Cit{\'e}}
\affil[3]{Laboratoire de Math\'{e}matiques at Mod\'{e}lisation d'Evry (LaMME), Universit{\'e} d'Evry Val d'Essonne, UMR CNRS 8071, ENSIIE, USC INRA }
\affil[$\dag$]{Correspondence: p.fearnhead@lancaster.ac.uk}}

\begin{document}
\maketitle 
\begin{center}
 {\bf Abstract}
\end{center}
Many traditional methods for identifying changepoints can struggle in the presence of outliers, or when the noise is heavy-tailed. Often they will infer additional changepoints in order to fit the outliers. To overcome this problem, data often needs to be pre-processed to remove outliers, though this is difficult for applications where the data needs to be analysed online. We present an approach to changepoint detection that is robust to the presence of outliers. The idea is to adapt existing penalised cost approaches for detecting changes so that they use loss functions that are less sensitive to outliers. We argue that loss functions that are bounded, such as the classical biweight loss, are particularly suitable -- as we show that only bounded loss functions are robust to arbitrarily extreme outliers. We present an efficient dynamic programming algorithm that can find the optimal segmentation under our penalised cost criteria. Importantly, this algorithm can be used in settings where the data needs to be analysed online. We show that we can consistently estimate the number of changepoints, and accurately estimate their locations, using the biweight loss function. We demonstrate the usefulness of our approach for applications such as analysing well-log data, detecting copy number variation, and detecting tampering of wireless devices.


\begin{bibunit}[apalike]

 \section{Introduction}
\label{sec:introduction}

Changepoint detection has been identified as one of the major challenges for modern, big data applications \cite[]{Frontiers2013}. The problem arises when analysing data that can be ordered, for example time-series or genomics
data where observations are ordered by time or position on a chromosome respectively. Changepoint detection refers to locating points in time or position where some aspect of the data of interest, such as location, scale or distribution, 
changes. There has been a recent explosion in methods for detecting changes \cite[e.g.][and references therein]{Frick:2014,Fryzlewicz:2014,cao2015changepoint,Haynes:2016,ma2016pairwise}  
in recent years, in part motivated by the range of applications for which changepoint detection is important. Exemplar areas
of application include bioinformatics \cite[]{Olshen:2004,Futschik:2014}, ion channels \cite[]{Hotz:2013}, climate records \cite[]{Reeves:2007}, oceonagraphic data \cite[]{Killick:2010,Killick:2012} 
and finance \cite[]{Kim:2012}.

\begin{figure}
 \centering \includegraphics[scale=0.4]{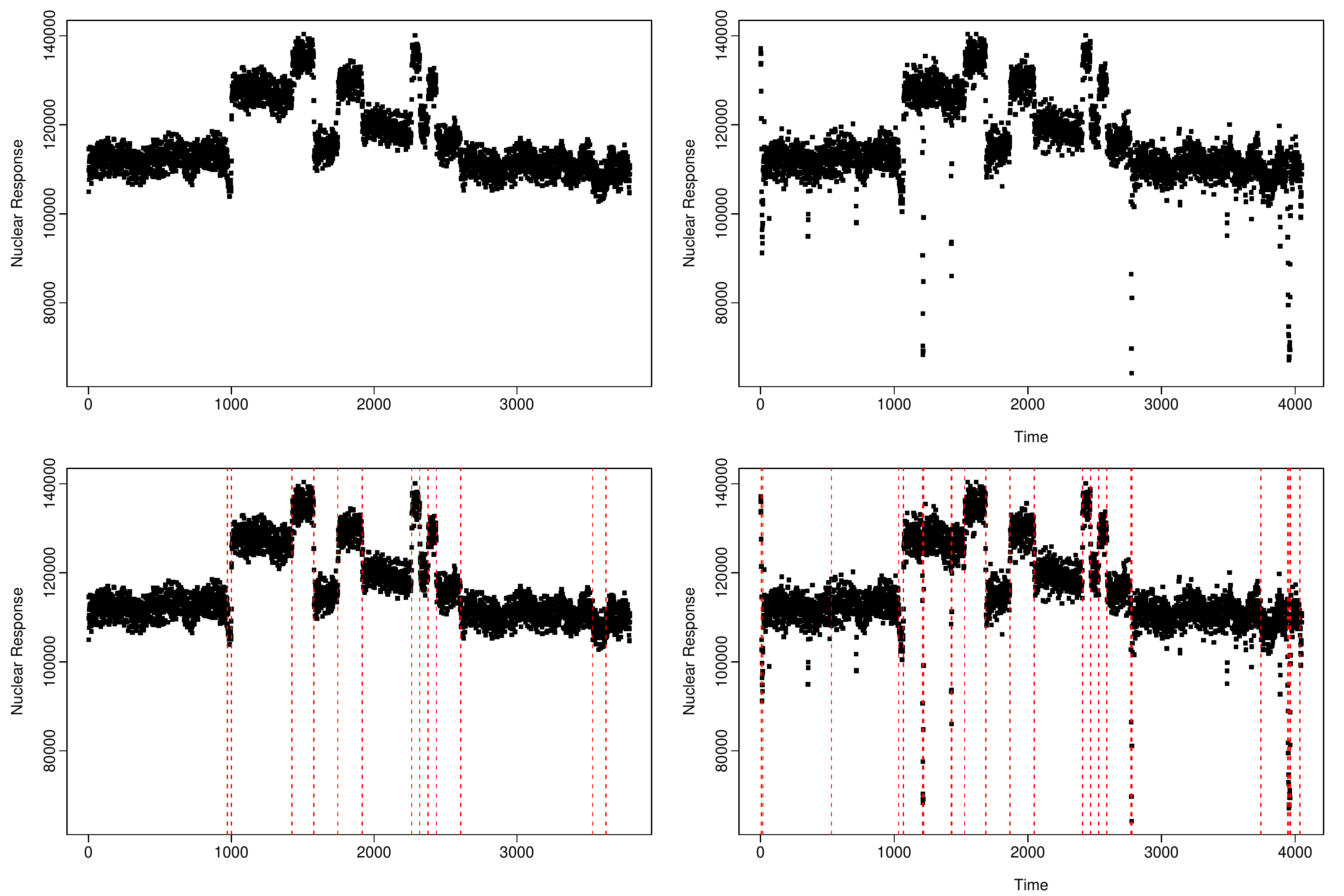}
 \caption{\label{Fig:Oil1} Well-log data: data with outliers removed (left column) and with outliers included (right column). Bottom row shows segmentations of the data under a least squares loss. 
 }
\end{figure}

What has received less attention is the problem of distinguishing between changepoints and outliers. 
%Outliers will tend to be rare observations that, for some reason, are very different from the main body of observations. Often they will occur singularly, though, depending on how they are caused, they may come in short clusters. 
To give an example of the issue outliers can cause when attempting to detect changepoint, consider the problem of detecting changes in well-log data. An example of such data, taken
originally from \cite{Fitzgerald:1996}, is shown in Figure \ref{Fig:Oil1}. This data was collected from a probe being lowered into a bore-hole. As it is lowered the probe takes measurements of the rock
that it is passing through. As the probe moves from one type of rock strata to another, there is an abrupt change in the measurements. It is these changes in rock strata that we wish to detect. The real motivation for
collecting this data was to detect these changes in real-time.  This would enable changes in rock strata that are being drilled through to be quickly detected, so that appropriate changes to the settings of the drill 
can be made. 

The data in the top-left plot in Figure \ref{Fig:Oil1} has been analysed by many different change detection methods \cite[e.g.][]{Fitzgerald:1996,Fearnhead:2006SC,Adams:2007,Wyse:2011,Ruggieri:2016}. However, this plot actually shows data that has been pre-processed to remove outliers. 
The real data that was collected by the probe is shown in the top-right plot of Figure \ref{Fig:Oil1}. There are a number of short periods of time where the probe mis-functions, and very low measurements are 
recorded. These are examples of what we are calling outliers. The real challenge with detecting the changes is to distinguish between actual changes and these outliers. Most existing methods for changepoint detection are 
unable to do so; hence the reason that most analysis of this data has used the ``cleaned'' data set in the top-left plot. For example in the bottom row of Figure \ref{Fig:Oil1} we show the results of estimating the changepoints based on minimising a square-error-loss criteria with a penalty for each
detected changepoint. Whilst this method performs well when analysing the cleaned dataset, it is unable to distinguish between changes and outliers when analysing the real data. 

This lack of robustness for detecting a change in mean in the presence of outliers for many changepoint methods stems from explicit or implicit modelling assumptions of Gaussian noise. For example, methods
based on a likelihood-ratio test for detecting a change \cite[]{Worsley:1979}, or that use a penalised likelihood approach to detect multiple changes \cite[]{Killick:2012}, or that do a Bayesian analysis \cite[]{Yao:1984}, 
may be based on a Gaussian likelihood and thus explicitly assume Gaussian noise.
Alternative methods, such as using an $L_2$ (square error) loss or a cusum based approach \cite[]{Page:1954}, may not make such an assumption explicitly. However the resulting method are closely related to
those based on a Gaussian likelihood \cite[see e.g.][]{Hinkley:1971},
and thus are implicitly making similar assumptions. Whilst these methods show some robustness to heavier-tailed noise \cite[]{Lavielle:2000}, in practice they can seriously over-estimate the number of changes in the presence of outliers.

Our approach is based on ideas from robust statistics, namely replacing an $L_2$ loss with an alternative loss function that is less sensitive to outliers. We then use such a loss function within a
penalised cost approach to estimating multiple changepoints. The use of alternative loss functions as a way to make changepoint detection robust to ouliers has been considered before 
\cite[e.g.][]{Huvskova:1989,Huvskova:1991,Huvskova:2005,Huvskova:2013}. That work
derives cusum-like tests for a single changepoint. Such a test for a single changepoint can then be used with binary segmentation to find multiple changes. As we discuss more fully in Section \ref{Sec:Alt}, this
approach suffers from the draw-back that the test statistic is based upon how well the data can be modelled as not having a change, and does not directly compare this with
how well we can fit the data with one or more changepoints. Thus it could spuriously infer a change if we have a cluster of outliers at consecutive time-points, even if the value of those outliers are not consistent
with them coming from the same distribution. 
%Secondly, for detecting multiple changepoints, this method suffers from the difficulties that binary segmentation can have, for example a lack of power to detect a short segment if the distribution before and after that segment are similar.

One challenge with the penalised cost approach that we suggest is minimising this cost, which we need to do to infer the changepoints. We show how recent efficient dynamic programming algorithms 
\cite[]{Maidstone:2016,rigaill2015pruned} can
be adapted to solve this minimisation problem. Our algorithm can use any loss function provided we are interested in the change of univariate parameter, such as the location parameter for univariate data, and the
loss function is piecewise quadratic. Importantly these algorithms are sequential in nature, and thus can be directly applied in situations which need an online analysis of the data.
%We derive bounds on the computational complexity of this algorithm, and present empirical results that suggest the expected computational cost of the algorithm is linear in the number of data points. 

Whilst our approach can be used with a range of loss functions, we particularly recommend using a loss function that is bounded. We present a theoretical result that shows that we need a bounded loss function
if we wish our method to be robust to any single outlier. The simplest such loss function is the biweight loss \cite[]{Huber:2011} which is the pointwise minimum of an $L_2$ loss and a constant. 
We show that, under mild conditions, we can consistently estimate the number of changepoints, and accurately estimate their locations, if we use a penalised cost approach with the biweight loss.

To illustrate the usefulness of our approach, with the biweight loss, in practice, we present its use for three distinct applications. The first is for the online analysis of the well-log data of Figure \ref{Fig:Oil1}.
Secondly we show that it out-performs existing methods for  detecting copy number variation. This includes performing better than methods that pre-process the data in an attempt to remove outliers. By comparison
our approach is easier to implement as it does not require any pre-processing steps. Finally we consider the problem of detecting tampering of wireless security devices. Results here show our method can reliably
distinguish between actual tampering events and changes in the data caused by short-term environmental factors. 

Proofs of results are given in the Appendices in the supplementary material. 
\hl{Code implementing the new methods in this paper is available from \url{https://github.com/guillemr/robust-fpop}.} %\textcolor{red}{ADD CODE DETAILS}.

%SOMETHING ON THE PROBLEM WE SOLVE/ALGORITHMIC INNOVATION
%\textcolor{blue}{GR: I guess we should state that we propose the first exact algorithm to solve
%this type of problem. I am wondering whether we should say that our approach is numerically stable.}

%OUTLINE OF PAPER

%LINK TO ROBUST STATS/M-ESTIMATION

%BUT FOCUS ON THE biweight COST -- EXPLAIN WHY THIS IS NEEDED FOR FULLY ROBUST.

%ALGORITHMIC INNOVATION. 

\section{Model Definition}
\label{sec:Model}
Assume we have data ordered by some covariate, such as time or position along a chromosome.
Denote the data by $\mathbf{y}=(y_1,\hdots,y_n)$. We will use the notation that, for $s\leq t$, the set of observations from time $s$ to time $t$ is $\mathbf{y}_{s:t}=(y_{s},...,y_t)$. 
If we assume that there are $k$ changepoints in the data, this will correspond to the data being split into $k+1$ distinct segments. 
We let the location of the $j$th changepoint be $\tau_j$ for $j=1,\hdots,k$, and set $\tau_0=0$ and $\tau_{k+1}=n$. The $j$th segment will consist of data points $y_{\tau_{j-1}+1},\ldots,y_{\tau_j}$. We let 
$\bm{\tau}=(\tau_0,\ldots,\tau_{k+1})$ be the set of changepoints.

The statistical problem we are considering is how to infer both the number of changepoints and their locations. We assume the changepoints correspond to abrupt changes in the location, that is mean, 
median or other quantile, of the data. We will focus on a minimum penalised cost approach to the problem. This approach encompasses  penalised likelihood approaches to changepoint 
detection amongst others. %See REF FOR DISCUSSION OF THIS.. SEEMS THIS IS A CASE OF M-ESTIMATION?

To define our penalised cost, we first introduce a loss function for a single observation, $y$, and a segment-specific location parameter $\theta$. We denote this as $\gamma(y;\theta)$. For a penalised likelihood
approach this loss would be equal to minus the log-likelihood. The class of losses we will consider are discussed below. 

We can now define the cost associated with a segment of data, $y_{s:t}$. This is 
\[
 \mathcal{C}(y_{s:t})=\min_\theta \sum_{i=s}^t \gamma(y_i;\theta),
\]
the minimum, over the segment-specific parameter $\theta$, of the sum of the losses associated with each observation in the segment. The penalised cost for a segmentation is then
\begin{equation} \label{eq:1}
 Q(y_{1:n};\tau_{1:k})=\sum_{i=0}^k \left\{\mathcal{C}(y_{\tau_i+1:\tau_{i+1}})+\beta\right\},
\end{equation}
where $\beta>0$ is a chosen constant that penalises the introduction of changepoints. We estimate the number and position of the changepoints by the value of $k$ and $\tau_{1:k}$ that minimise this penalised
cost. The value of $\beta$ has a substantial impact on the number of changepoints that are estimated \cite[see][for examples of this]{Haynes:2015}, 
with larger values of $\beta$ leading to fewer estimated changepoints.  

%Under strong assumptions on the true data generating model, there are choices of $\beta$ that can lead to consistent estimation of the number of changepoints
%SOME FURTHER DISCUSSION/REFS.

For inferring changes in the mean of the data, it is common to use the squared-error loss function \cite[e.g.][]{Yao:1989,Lavielle:2000}
\[
 \gamma(y;\theta)=(y-\theta)^2.
\]
In this case, the penalised cost approach corresponds to a penalised likelihood approach where the data within a segment are IID Gaussian with common variance.
Minimising a penalised cost of this form is closely related to binary segmentation procedures based on cusum statistics \cite[e.g.][]{Vostrikova:1981,Bai:1997,Fryzlewicz:2014}, 
as discussed in \cite{Killick:2012}. %If the data within a segment
%is generated as IID Gaussian with variance $\sigma^2$, then minimising the penalised cost is known to consistently estimate the number and location of the changepoints, assuming in-fill asymptotcs
Use of the square-error loss function results in an approach that is very sensitive to outliers. For example, this loss function was the one used in the analysis of the well-log data in Figure \ref{Fig:Oil1}, where we
saw that it struggles to distinguish outliers from actual changes of interest.

\subsection{Penalised Costs based on M-estimation}

\begin{figure}
 \centering \includegraphics[scale=0.5]{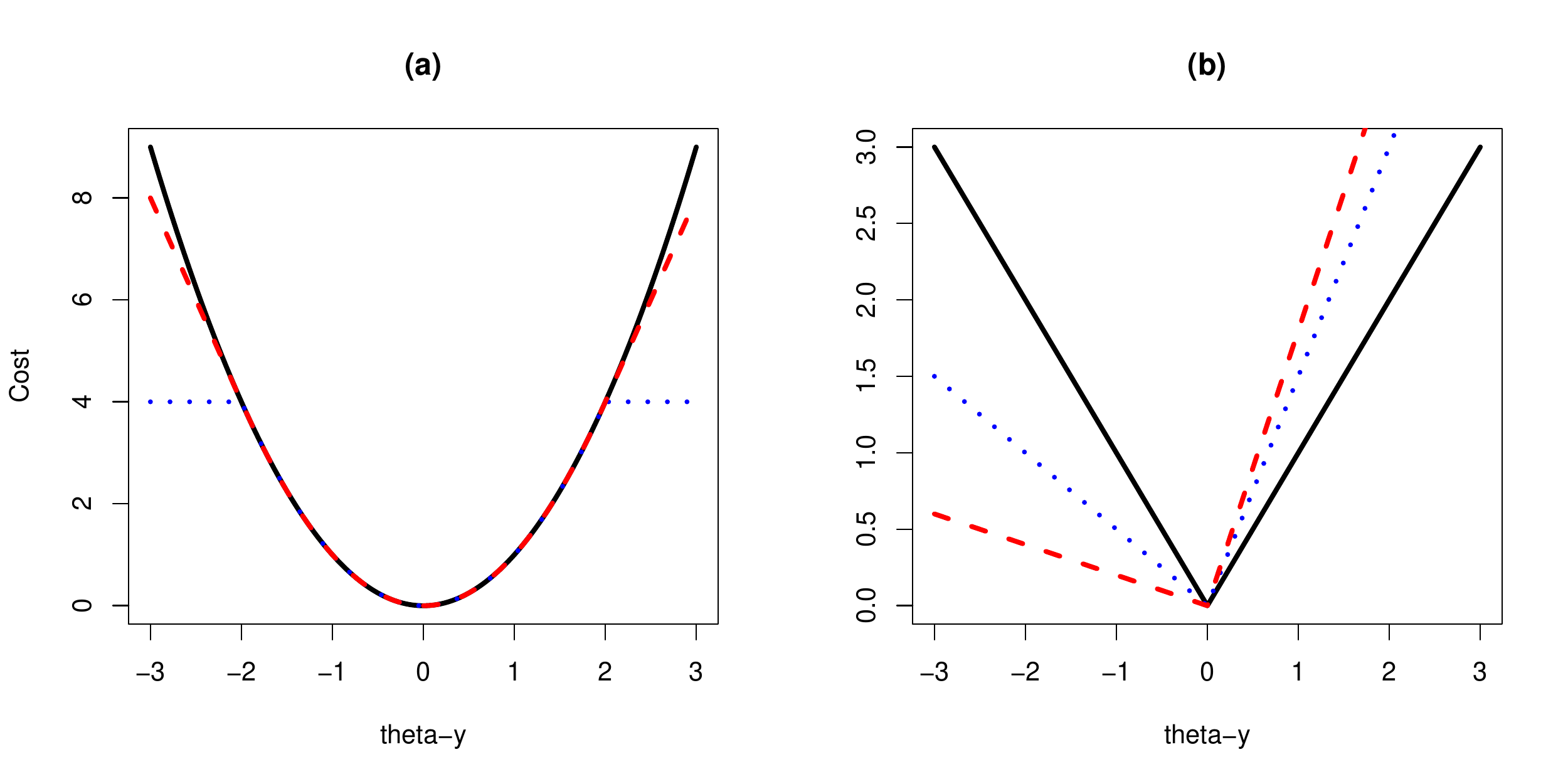}
 \caption{\label{Fig:1} Example of different losses. (a) The square error loss (full-line), and the related Huber loss (red dashed) and biweight loss (blue dotted). 
 (b) The absolute error loss (full-line), and its generalisation for detecting change in quantiles for $u=0.1$ (red dashed) and $u=0.25$ (blue dotted).}
\end{figure}

To develop a changepoint approach that can reliably detect changepoints in
the presence of outliers we need a loss function that increases at a slower rate in $|y-\theta|$.  
Standard examples \cite[]{Huber:2011} are absolute error, $\gamma(y;\theta)=|y-\theta|$, Huber loss
\[
 \gamma(y;\theta)=\left\{ \begin{array}{cl} (y-\theta)^2 & \mbox{if } |y-\theta|<K, \\ 2K|y-\theta|-K^2 & \mbox{otherwise,} \end{array} \right.
\]
and the biweight loss,
\begin{equation} \label{eq:biweight}
 \gamma(y;\theta)=\left\{ \begin{array}{cl} (y-\theta)^2 & \mbox{if } |y-\theta|<K, \\ K^2 & \mbox{otherwise,} \end{array} \right. 
\end{equation}
or if interest lies in changes in the $u$th quantile for $0<u<1$,
\[
 \gamma(y;\theta)=\left\{ \begin{array}{cl} 2u(y-\theta) & \mbox{if } y>\theta, \\ 2(1-u)(\theta-y) & \mbox{otherwise.} \end{array} \right. 
\]
These are summarised in Figure \ref{Fig:1}. 

%PLOT NEEDS CORRECTING FOR WINSORISED.

We will develop an algorithm for finding the best segmentation under a penalised cost criteria that can deal with any of these choices for the loss. In practice we particularly advocate the use
of the biweight loss.
For a penalised cost approach to detecting changepoints to be robust to extreme outliers we will need the loss function to be bounded. For unbounded loss functions, such as the absolute error loss or Huber loss, a penalised cost approach will place an outlier in a segment on its own if that outlier is sufficiently extreme. This is shown by the following result.
\begin{theorem} \label{thm:1}
Assume that the loss function satisfies  $\gamma (y; \theta) = g(|y-\theta|)$ where $g(0)=0$ and $g(\cdot)$ is an unbounded, increasing function.  
Choose any $t \in \{1,\ldots,n\}$ and fix the set of other observations, $y_s$ for $s\neq t$. Then there exists values of $y_t$ such that the segmentation that minimises the penalised cost (\ref{eq:1}) will have changepoints at $t-1$ and $t$. 
\end{theorem}

If we choose a loss function, such as the biweight loss, that is bounded, then this will impose a minimum segment length on the segmentations that we infer using the penalised cost function. 
Providing this minimum segment length is greater than 1, our inference procedure will be robust to the presence of extreme outliers -- unless these outliers cluster at similar values, and for 
a number of consecutive time-points greater than our minimum segment length.
\begin{theorem} \label{thm:2}
If the loss function satisfies $0\leq \gamma(y;\theta) \leq K$, and we infer changepoints by minimising the penalised cost function (\ref{eq:1}) with penalty $\beta$ for adding a changepoint, then all inferred segments will be of length greater than $\beta/K$.
\end{theorem}

The other conclusion to draw from this result is that, for any choice of $K$ and $\beta$, we would want the minimum segment length to be smaller than any segment we expect, or that we wish to detect, in the data.
Any real segments shorter than the minimum segment length are unlikely to be detected, with the observations in such short segments being identified as outliers instead. Furthermore our procedure can lose power 
to detect real segments that are only slightly longer than the minimum segment length (see empirical results for scenario 4 in Section \ref{sec:accuracy}).

\subsection{Consistency under Biweight Loss}

As mentioned above, and as suggested by Theorem \ref{thm:1}, a particular focus will be on the use of the biweight loss (\ref{eq:biweight}). Here we give conditions under which we can consistently estimate the number and location of changepoints  when using this loss. 

We will consider the standard in-fill asymptotics, as we let the number of data points, $n$, increase. To be able to consistently estimate the number of changepoints we will need the penalty for adding a changepoint, $\beta$ in (\ref{eq:1}), to increase with $n$. We will thus denote the choice of penalty for a given number of data points to be $\beta_n$.

 We assume a fixed number of changepoints, $k_0$, and fixed constants $0<u_1<\cdots<u_{k_0}<1$ so that for a data set of size $n$ we have the $i$th changepoint at $\tau_i=\lfloor nu_i \rfloor$, for 
$i=1,\ldots,k_0$. As above we let $\tau_0=0$ and $\tau_{k_0+1}=n$. We further assume fixed segment-specific location parameters, $\mu_0,\ldots,\mu_{k_0}$, with the obvious constraint that 
$\mu_i \neq \mu_{i-1}$ for $i=1,\ldots,k_0$. Finally we let $Z_1,Z_2,\ldots$ be IID noise random variables, so that for $t=1,\ldots,n$ the observations are realisations of
\[
Y_t=\mu_i+Z_t,
\]
where $i$ is such that $\tau_{i}<t\leq\tau_{i+1}$.

Our results require two mild conditions on the distribution of the noise random variables. Firstly introduce the mean of the loss function, $M(\theta)=\mbox{E}\{\gamma(Z_i;\theta)\}$.  We assume $M(\theta)$ takes it minimum value at $\theta=0$. We can make this assumption without loss of generality, as if $M(\theta)$ has its minimum at $\theta^*$ we can just re-parameterise our model with new noise random variables set to $Z_i-\theta^*$ and with location parameters re-defined to be $\mu_i+\theta^*$.

Our first condition is that there exists constants $c_1>0$ and $c_2>0$ such that
\begin{equation} \label{ass:1}
M(\theta)=\mbox{E}\left[ \min\left\{ (Z_i-\theta)^2,K^2 \right\}\right] \geq M(0)+\min\left\{c_1\theta^2,c_2 \right\}.
\end{equation}
This is a weak assumption, and will hold if $M(\theta)$ has a positive second derivative for all $\theta$ in a neighbourhood around $0$ and that $M(\theta)-M(0)\geq c_2>0$  
for all $\theta$ outside this region. The latter requirement is a common assumption made to ensure identifiability of estimates of a location parameter when using a given loss function. 

Our second condition is slightly stronger. Let $p=\Pr(|Z_i|> K)$ and $\sigma^2=\mbox{E}(Z_i^2 \mid |Z_i|\leq K)$, then we need
\begin{equation} \label{ass:2}
K^2(1-2p)-(1-p)\sigma^2>0.
\end{equation}
This condition can be achieved by taking $K$ large enough. If the noise has finite variance then, using Chebyshev's inequality, it is easy to show that any choice with $K>\sqrt{3}\mbox{E}(Z_i^2)$ will ensure this 
condition holds. However we do not need the noise to have a variance. For example it is sufficient to choose $K>\sqrt{3}\mbox{E}(\min\{Z_i^2,K^2\})$, or, if $Z_i$ has a unimodal density function with mode at $0$, 
then $\sigma^2\leq K^2/3$  and it suffices to choose $K$ so that $p=\Pr(|Z_i|> K)<2/5$. By comparison, we would recommend taking $K$ sufficiently large that $|Z_i|>K$ is relatively rare, and thus $p\approx 0$. 
In line with Theorem \ref{thm:1}, this condition does not depend on the distribution of the noise conditional on $|Z_i|>K$.

\begin{theorem} \label{thm:Consistency}
Assume (\ref{ass:1}) and (\ref{ass:2}) and the model described above. For a given $n$ let $\hat{k}_n$ be the estimate of the
number of changepoints, and $\hat{\tau}_1,\ldots,\hat{\tau}_{\hat{k}_n}$  their estimated locations, obtained by minimising the penalised cost (\ref{eq:1}) using the biweight loss function and a penalty $\beta_n$. 

There exists constants $C_1>0$ and $C_2>0$ 
such that as $n\rightarrow \infty$, if $\beta_n>C_1\log(n)$ and $\beta_n=o(n)$ then
\[
\Pr\left[\hat{k}_n=k_0 \mbox{ and } \max_{i=1,\ldots,k_0} \left\{ \min_{j=1,\ldots,\hat{k}_n} \left| \tau_i-\hat{\tau}_j \right| \right\} \leq C_2 \log(n) 
\right]\rightarrow 1.
\]
\end{theorem}

%The proof of this theorem is given in Appendix \ref{App:Consistency} in the supplementary material.
The theorem shows that for an appropriate choice of $\beta_n$ we can obtain a consistent estimate of the true number of parameters, and that the error in estimating any of the changepoint locations will
be less than $C_2 \log(n)$ with probability 1. The latter order of error is in line with asymptotic results for the accuracy of changepoint estimates using wild binary segmentation with the cusum test 
\cite[]{Fryzlewicz:2014}. We require much weaker conditions on the distribution of the noise, but our result assumes stronger conditions on the number of changes, the segment lengths and the size of change of mean at each changepoint than, for example, results in \cite{Fryzlewicz:2014} and \cite{baranowski2016narrowest}. 
The result supports the use of a penalty, $\beta_n$, that is proportional to $\log(n)$, a choice that is common for other penalised cost procedures, 
but it does not specify the constant of proportionality.

%MOVE THIS TO SIMULATION SCENARIO

%If $K$ is large, then the biweight loss is similar to the standard square error loss. Thus it a natural default penalty choice is one that as $K\rightarrow\infty$ is the same as a default penalty for the square error loss. One such choice is $\beta_n=2\mbox{E}(\gamma(Z_i))\log(n)$, which tends to the standard BIC penalty for square error loss as $K$ increases \cite[see][for results on the consistency of estimating the number of changepoints using the BIC penalty]{yao1988estimating}. This is the choice we will use within the simulation study below; though we discuss the choice of penalty further in the Discussion section. 

\subsection{Alternative Robust Changepoint Methods} \label{Sec:Alt}

There have been other proposed $M$-procedures for robust detection of changepoints \cite[]{Huvskova:1989,Huvskova:1991,Huvskova:2005,Huvskova:2013}. 
These differ from our approach in that they are based on sequentially applying tests for single changepoints. One approach is to use a Wald-type test. 
For a convex loss function $\gamma(y;\theta)$ which depends only on $y-\theta$, define $\gamma(y;\theta)=\rho(y-\theta)$ and define $\phi$ to be the first derivative of $\rho$. 
Then we can estimate a common $\theta$ for data $y_{1:n}$ by minimising
\[
\sum_{i=1}^n \rho(y_i-\theta),
\]
with respect to $\theta$. In many cases this is equivalent to solving
\[
\sum_{i=1}^n \phi(y_i-\theta)=0.
\]
If $\hat{\theta}$ denotes the estimate we obtain, 
we can define residuals as $\phi(y_i-\hat{\theta})$, and their partial sums, or cusums, by
\[
S_m=\sum_{i=1}^m \phi(y_i-\hat{\theta}).
\]
A Wald-type test is then based on a test-statistic of the form
\[
T_n=\max_{1\leq m \leq n-1} \frac{n}{m(n-m)}S_m^2,
\]
where the term $n/(m(n-m))$ is introduced so that the variability of the term on the right-hand side will be similar for each value of $m$.
Large values of $T_n$ are taken as evidence for a change. The position of a changepoint is then inferred at the position $m$ which maximises the right-hand side. 
To detect multiple changepoints, this Wald-type test needs is currently  used within a binary segmentation procedure; though it can also be used with improved versions
of binary segmentation, such as 
wild binary segmentation \cite[]{Fryzlewicz:2014}.

There are two main differences between the Wald-type test approach and our penalised cost approach. The first is that the Wald-type test statistic is appropriate only for convex loss functions.
So, for example, the biweight loss is not appropriate for use with this approach. To see this note that the derivative of the biweight loss satisfies $\phi(x)=0$ for $|x|>K$. 
Thus large abrupt changes in the data will lead to $M$-residuals which are 0, and hence provide no evidence for a change in the test statistic. 

Secondly any loss function which increases linearly in  $|y-\theta|$ for sufficiently large $|y-\theta|$ will result in  $\phi(y_i-\theta)$ being constant for large $|y_i-\theta|$. 
Thus, large residuals will have a bounded contribution to the test statistic. To see this consider the Wald-type test with the Huber loss. To calculate this test statistic we first
calculate our estimate of the location parameter for the data, $\hat{\theta}$, assuming the data is from a single segment. The $i$th residual is then $K$ if $y_i>\hat{\theta}+K$, $-K$ if $y_i<\hat{\theta}-K$, 
and $y_i-\hat{\theta}$ otherwise. The cusum statistic is just the sum of these residuals. This is equivalent to winsorizing the data, where we shrink extreme positive or negative values to be $K$ above or below our 
estimate of the location parameter, and then using a cusum test for detecting a changepoint. The actual value of the data points that are above $\hat{\theta}+K$ or below $\hat{\theta}-K$ will not 
affect the cusum values, and hence not affect the value of the Wald-type test statistic.

The use of  Huber loss within a Wald-type test will thus have a similar robustness to extreme outliers that bounded loss functions have for the penalised cost approach. 
%For detecting a single change, a Wald-type test with Huber loss and a penalised cost approach with biweight loss will behave similarly if they use the same, large, $K$. 
The main difference is that the Wald-type test statistic does not consider whether the data after a putative changepoint is consistent with data from a single segment. 
Thus a cluster of outliers of the same sign that occur concurrently but which are very different in value, such as we observe for the well-log data, 
will produce a similar value for the test-statistic as a set of concurrent observations that are very different to the other data points but are also very similar to one another. 
By comparison, the penalised cost based approach would, correctly, say the latter provided substantially more evidence for the presence of a change. %An example of this is give in FIGURE TO ADD.

%The penalised cost approach generalises more naturally to detecting multiple changepoints. The use of these Wald-type tests for detecting a single changepoint is currently 
%used within a binary segmentation procedure in order to infer multiple changepoints. Though it could also be used within improved versions of binary segmentation, such as 
%wild binary segmentation \cite[]{Fryzlewicz:2014}.
%However binary segmentation can sometimes perform poorly as it tries to add changepoints one at a time, 
%and cannot change earlier changepoints that have been detected. See \cite{Fryzlewicz:2014} for discussion of the problems with binary segmentation, together with a possible improvement.

\section{Minimising the Penalised Cost}

An issue with detecting changepoints using any of these loss functions, is how can we efficiently minimise the resulting penalised cost over all segmentations? 
We present an efficient dynamic programming algorithm for performing this minimisation exactly. This algorithm is an extension of the pruned DP algorithm of \cite{rigaill2015pruned} and the FPOP
algorithm of \cite{Maidstone:2016} \cite[see also][]{Johnson:2013} to the robust loss functions. We will call the resulting algorithm R-FPOP.

\subsection{A Dynamic Programming Recursion}

We develop a recursion for finding the minimum cost (\ref{eq:1}) of segmenting data $y_{1:t}$ for $t=1,\ldots,n$. In the following we let $\vtau$ denote a vector of changepoints. Furthermore we let $\mathcal{S}_t$ denote the set of possible changepoints for the $y_{1:t}$, so
\[
\mathcal{S}_t=\left\{ 
\vtau=\tau_{1:k}: 0<\tau_1<\cdots<\tau_{k}<t
\right\}.
\]
Note that $\mathcal{S}_t$ has $2^{t-1}$ elements. Define
\[
Q_t=\min_{\vtau\in\mathcal{S}_t} Q(y_{1:t};\tau_{1:k})=\min_{\vtau\in\mathcal{S}_t} \sum_{i=0}^k \left\{\mathcal{C}(y_{\tau_i+1:\tau_{i+1}})+\beta\right\},
\]
where here and later we use the convention that $k$ is the number of changepoints in $\vtau$, and that $\tau_0=0$ and $\tau_{k+1}=t$.
First we introduce the minimum penalised cost of segmenting $y_{1:t}$ conditional on the most recent segment having parameter $\theta$,
\[
Q_t(\theta)=\min_{\vtau\in\mathcal{S}_t} \left[ 
\sum_{i=0}^{k-1} \left\{\mathcal{C}(y_{\tau_i+1:\tau_{i+1}})+\beta\right\}
+\sum_{j=\tau_k+1}^t \gamma(y_j;\theta) +\beta
\right],
\]
where we take the first summation on the right-hand side to be 0 if $k=0$.
Trivially we have $Q_t=\min_\theta Q_t(\theta)$ and $Q_1(\theta)=\gamma(y_1;\theta)+\beta$.

The idea is to recursively calculate $Q_t(\theta)$ for increasing values of $t$. To do this, we note that each element in $\mathcal{S}_t$ is either an element in $\mathcal{S}_{t-1}$ or an element in $\mathcal{S}_{t-1}$ with the addition of a changepoint at $t-1$. So
\begin{eqnarray}
Q_t(\theta)&=&\min_{\vtau\in\mathcal{S}_{t-1}} \left[\min\left\{ 
\sum_{i=0}^{k-1} \left(\mathcal{C}(y_{\tau_i+1:\tau_{i+1}})+\beta\right)
+\sum_{j=\tau_k+1}^t \gamma(y_j;\theta) +\beta, \nonumber \right. \right.\\ & & \left.\left.
\sum_{i=0}^{k} \left(\mathcal{C}(y_{\tau_i+1:\tau_{i+1}})+\beta\right)+\gamma(y_t;\theta)+\beta
\right\}
\right] \nonumber \\
&=& \min\left\{ 
\min_{\vtau\in\mathcal{S}_{t-1}} \left[
\sum_{i=0}^{k-1} \left(\mathcal{C}(y_{\tau_i+1:\tau_{i+1}})+\beta\right)
+\sum_{j=\tau_k+1}^{t-1} \gamma(y_j;\theta) +\beta\right], \nonumber \right. \\ & & \left.
\min_{\vtau\in\mathcal{S}_{t-1}} \left[
\sum_{i=0}^{k} \left(\mathcal{C}(y_{\tau_i+1:\tau_{i+1}})+\beta\right)+\beta
\right]
\right\}+\gamma(y_t;\theta) \nonumber
\\
&=&
\min\left\{
Q_{t-1}(\theta),Q_{t-1}+\beta 
\right\}+\gamma(y_t;\theta). \label{eq:DP}
\end{eqnarray}
The first equality comes from splitting the minimisation into the minimisation over the changepoints for $y_{1:t-1}$ and then whether there is or is not a changepoint at $t-1$. 
The second equality comes from interchanging the order of the minimisations, and taking out the common $\gamma(y_t;\theta)$ term. The final equality comes from the definitions of $Q_{t-1}(\theta)$ and $Q_{t-1}$. 
The right-hand side just depends on $Q_{t-1}(\theta)$, as $Q_{t-1}=\min_{\theta} Q_{t-1}(\theta)$. 

\subsection{Solving the Recursion}

We now show how we can efficiently solve the dynamic programming recursion from the previous section for loss functions like those introduced in Section \ref{sec:Model}. We make the assumption that the loss for any observation, $\gamma(y_t;\theta)$,
viewed as function of $\theta$, can be written as a piecewise quadratic in $\theta$. Note that by quadratic we include the special cases of linear or constant functions of $\theta$, and this definition covers all the loss functions introduced in Section \ref{sec:Model}.

%MAY NEED TO DEFINE NUMBER OF SEGMENTS IN GAMMA -- USE $L$.

As the set of piecewise quadratics is closed under both addition and minimsation, it follows that $C_t(\theta)$ can be written as a piecewise quadratic for all $t$. We summarise $C_t(\theta)$ by $N_t$ intervals
$(a^{(t)}_i,b^{(t)}_i]$, and associated quadratics $q^{(t)}_i(\theta)$. We assume that the intervals are ordered, so $a^{(t)}_1=-\infty$, $a^{(t)}_i=b^{(t)}_{i-1}$ for $i=2,\ldots,N_t$ and $b^{(t)}_{N_t}=\infty$. %(Note we have slightly abused notation, as the final interval will be open at the upper-end point.)
To make this summary of $C_t(\theta)$ unique we further assume that $q_i^{(t)}(\theta)\neq q_{i-1}^{(t)}(\theta)$ for $i=2,\ldots,N_t$. If this were not the case we could merge the neighbouring intervals.
%This is true of all the cost functions introduced in the previous section. NOTATION?

We can split (\ref{eq:DP}) into two steps. The first is
\begin{equation} \label{eq:min1}
Q_{t}^*(\theta)=\min\left\{
Q_{t-1}(\theta),Q_{t-1}+\beta 
\right\},
\end{equation}
and the second is
\[
Q_t(\theta)=Q_{t}^*(\theta)+\gamma(y_t;\theta).
\]

For the first step we first calculate $Q_{t-1}$ by first minimising the $N_{t-1}$ quadratics defining $Q_{t-1}(\theta)$ on their respective intervals, and then calculating the minimum of these minima. We then
solve the minimisation problem (\ref{eq:min1}) on each of the $N_{t-1}$ intervals. For interval $i$, the solution will either be $q_i^{(t)}(\theta)$, $Q_{t-1}+\beta$ or we will need to split the interval into two or 
three smaller intervals, on which the solution will change between $q_i^{(t)}(\theta)$ and $Q_{t-1}+\beta$. Thus we will end with a set of $N_{t-1}$, or more, ordered intervals and corresponding quadratics that define
$Q_{t}^*(\theta)$. We then prune these intervals by checking whether any neighbouring intervals both take the value  $Q_{t-1}+\beta$, and merging these if they do.  This will lead to a new set of $N_t^*$, say, 
ordered intervals, and associated quadratics, $q_{t,i}^{*}(\theta)$ say.%, for $i=1,\ldots,N^*t$, associated with each one.

For each of the $N_t^*$ intervals from the output of the minimisation problem we then add $\gamma(y_t;\theta)$ to the corresponding $q^*_{t,i}(\theta)$. This may involve splitting the $i$th interval into two
or more smaller intervals if one or more of the points of change of the function $\gamma(y_t;\theta)$ are contained in it. This will lead to the $N_t$ intervals and corresponding quadratics that define $Q_t(\theta)$.

The above describes how we recursively calculate $Q_t(\theta)$. In practice we also want to then extract the optimal segmentation under our criteria. This is straightforward to do. For each of the intervals 
corresponding to different pieces of $Q_t(\theta)$ we can associate a value of the most recent changepoint prior to $t$. When we evaluate $Q_t$, we need to find which interval contains this value, and then the
optimal value for the most recent changepoint prior to $t$ is the value associated with that interval. We can store these optimal values for all $t$, and after processing all data we can recursively track back
through these values to extract the optimal segmentation. So we would first find the value of the most recent changepoint prior to $n$, $\tau$ say, then find the value of the most recent changepoint prior to $\tau$. 
We repeat this until the most recent changepoint is at 0, corresponding to no earlier changepoints.

\begin{figure}
 \centering \includegraphics[scale=0.5]{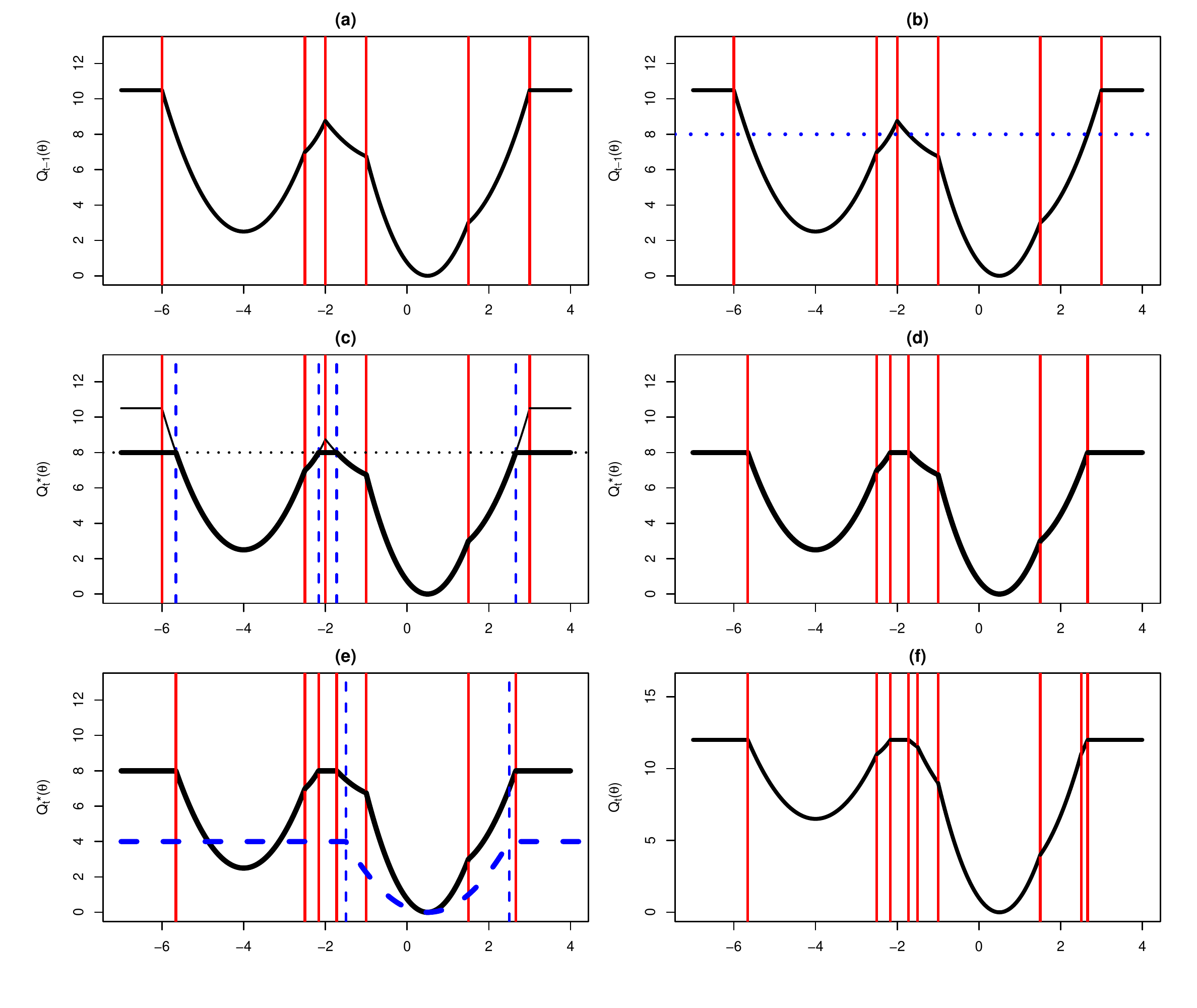}
 \caption{\label{Fig:R-FPOP_ex} Example of one iteration of R-FPOP: (a) $Q_{t-1}(\theta)$ (black solid line), and set of intervals stored (split by vertical red lines) at start of iteration. (b) Find the pointwise
 minimum of $Q_{t-1}(\theta)$ and $Q_{t-1}+\beta$ (blue dashed line). (c) This is done by solving the minimisation on each interval, which splits some intervals into two or three. New splits are shown by blue dashed vertical lines. 
 (d) Merge neighbouring intervals if they both take the value $Q_{t-1}+\beta$. (e) Now add the loss for the new observation (blue dashed curve). (f) This further splits intervals at the points where the form
 of $\gamma(y_t;\theta)$ changes, the blue vertical lines in plot (e). Shown is the final representation of $Q_t(\theta)$. At all stages only piecewise quadratic functions need to be stored.}
\end{figure}

Pseudo-code for R-FPOP is given in Appendix \ref{App:R-FPOP}. An example of the steps involved in one iteration is given in Figure \ref{Fig:R-FPOP_ex}.

\section{Computational Cost of R-FPOP} \label{sec:CompCost}

We now present results which bound the computational cost and storage requirements of R-FPOP. As above we will assume that $\gamma(y;\theta)$ can be written as a piecewise quadratic with $L$ pieces. The bounds that
we get differ depending on whether, for a given $y$, $\gamma(y;\theta)$ is convex in $\theta$. We first consider the convex case, which includes all the examples in Section \ref{sec:Model} except the biweight loss.

\begin{theorem} \label{thm:3}
If $\gamma$ is convex in $\theta$ and defined in $L$ pieces R-FPOP stores at most $2t-1 + t(L-1)$ 
quadratics and intervals at step $t$.
%the worst case complexity of Fpop is $\Oc(n^2)$ in time and $\Oc(n)$ in space.
\end{theorem}

\begin{corollary}\label{corollary:lgamma}
If $\gamma$ is convex in $\theta$ and defined in $L$ pieces, the space complexity of R-FPOP is $\Oc(n)$,
and the time complexity of R-FPOP is $\Oc(n^2)$.
\end{corollary}

For the biweight loss, which is not convex, we get worse bounds on the complexity of R-FPOP.
\begin{theorem} \label{thm:4}
For the biweight loss R-FPOP stores $\Oc(t^2)$ intervals at step $t$.
\end{theorem}

\begin{corollary} \label{cor:2}
For the biweight loss R-FPOP has worst-case space complexity that is $\Oc(n^2)$, and time complexity that is $\Oc(n^3)$.
\end{corollary}

These results give worst-case bounds on the time and storage complexity of R-FPOP. Below we investigate empirically the time and
storage cost %of R-FPOP in Section \ref{sec:Results}, where we 
and observe an average computational cost that is linear in $n$ 
when the number of changepoints is large and less than quadratic when there is no changepoint. %and storage requirements that are constant. %CHECK THIS IS A FAIR SUMMARY! MAYBE THIS SHOULD APPEAR LATER?

\section{Results} \label{sec:Results}

\subsection{Simulation Study: Computational Cost}

This paper is mostly concerned with the statistical performances of our robust estimators. 
Thus an in-depth analysis of the runtime of our approach is outside the scope of this paper. 
In this section we just aim at showing that our approach is easily applicable to 
large profiles ($n=10^3$ to $n=10^6$) in the sense that its runtime is comparable to other
commonly used approach like FPOP \cite[]{Maidstone:2016}, PELT \cite[]{Killick:2012}, WBS \cite[]{Fryzlewicz:2014} or smuceR \cite[]{Frick:2014}.

\begin{figure}
 \centering 
\begin{tabular}{ccc}
No change, $K=0$ & Many changes, $K= \frac{n}{100}$ \\
%% Gr: here same graph as before i erase the legend with xournal and replace out by biweight (change in ggplot2)
\includegraphics[width=6.5cm, trim = 0cm 0cm 3.4cm 0cm, clip]{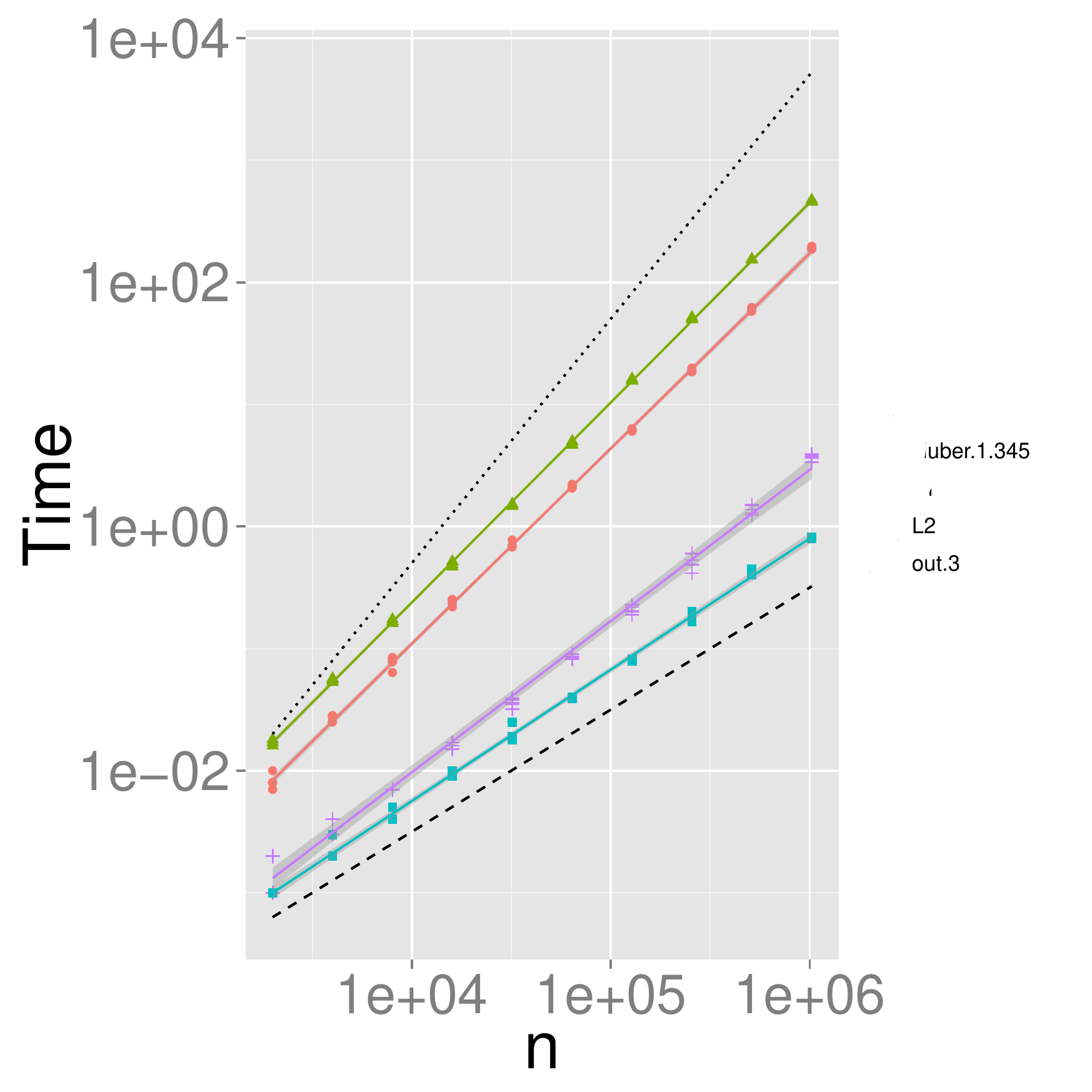} & 
\includegraphics[width=6.5cm, trim = 0cm 0cm 3.4cm 0cm, clip]{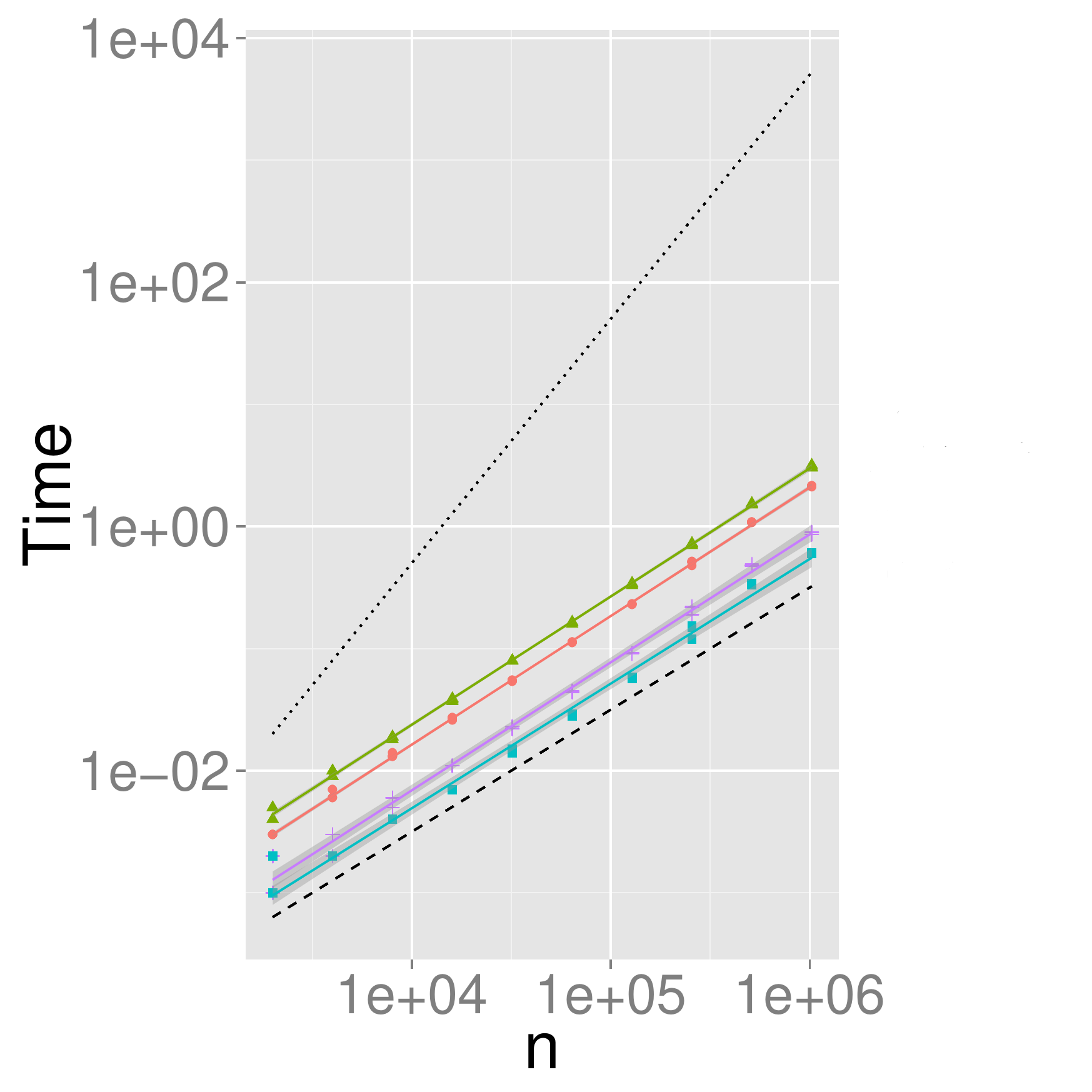} &
\includegraphics[width=3.1cm, trim = 13.7cm 3cm 0cm 7cm, clip]{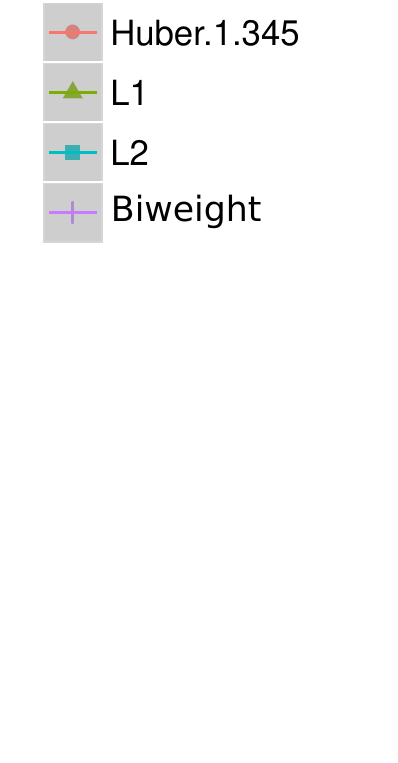}
\end{tabular}
 \caption{Runtime in seconds of R-FPOP for different loss functions. We simulated profiles with $n$ going from $2000$ to $1024000$, with or without changes and using IID Gaussian noise. 
The axes use a log-scale, and we have added lines of slope 1 (dashed) and 2 (dotted).}\label{figure:runtime}
\end{figure}

We used a standard laptop with an Intel Core i7-3687U CPU with 2.10GHz x 4 Core and 7.7 Gb of Ram.
For the biweight loss, for a profile of length $n=10^6$ and in the absence of any true change the runtime 
is around 4 seconds (slightly larger than FPOP see Figure \ref{figure:runtime} left $L_2$). 
As a matter of comparison on the same computer the runtime of competitor methods WBS, PELT  and smuceR  for a profile of length $n=10^5$
are respectively around 7 seconds, 40 seconds and 175 seconds. 
For an increasing number of changes runtimes are smaller (see Figure \ref{figure:runtime} right). 
Runtimes for the $L_1$ and Huber loss are quite a bit larger:
in the absence of changes and for $n=10^6$ the $L_1$ runtime is around 500 seconds and the Huber runtime is around 200 seconds (see Figure \ref{figure:runtime} left).

Most importantly, we see that with many changepoints, the average CPU cost of all penalised cost approaches increases only linearly with the number of data points (parallel to the dashed black line in Figure \ref{figure:runtime} right). 
With no changepoints, the average CPU cost increases faster in particular for the $L_1$ and Huber losses however it is less than quadratic (slopes smaller than the dotted black line 
in Figure \ref{figure:runtime} left). The CPU cost of the biweight loss is very close to the CPU cost of the $L_2$ loss.
%Most importantly, we see that both with no changepoints and many changepoints, the average CPU cost of all penalised cost approaches increases only linearly with the number of data points.

\subsection{Simulation Study: Accuracy} \label{sec:accuracy}

\begin{figure}
 \centering \includegraphics[scale=0.45]{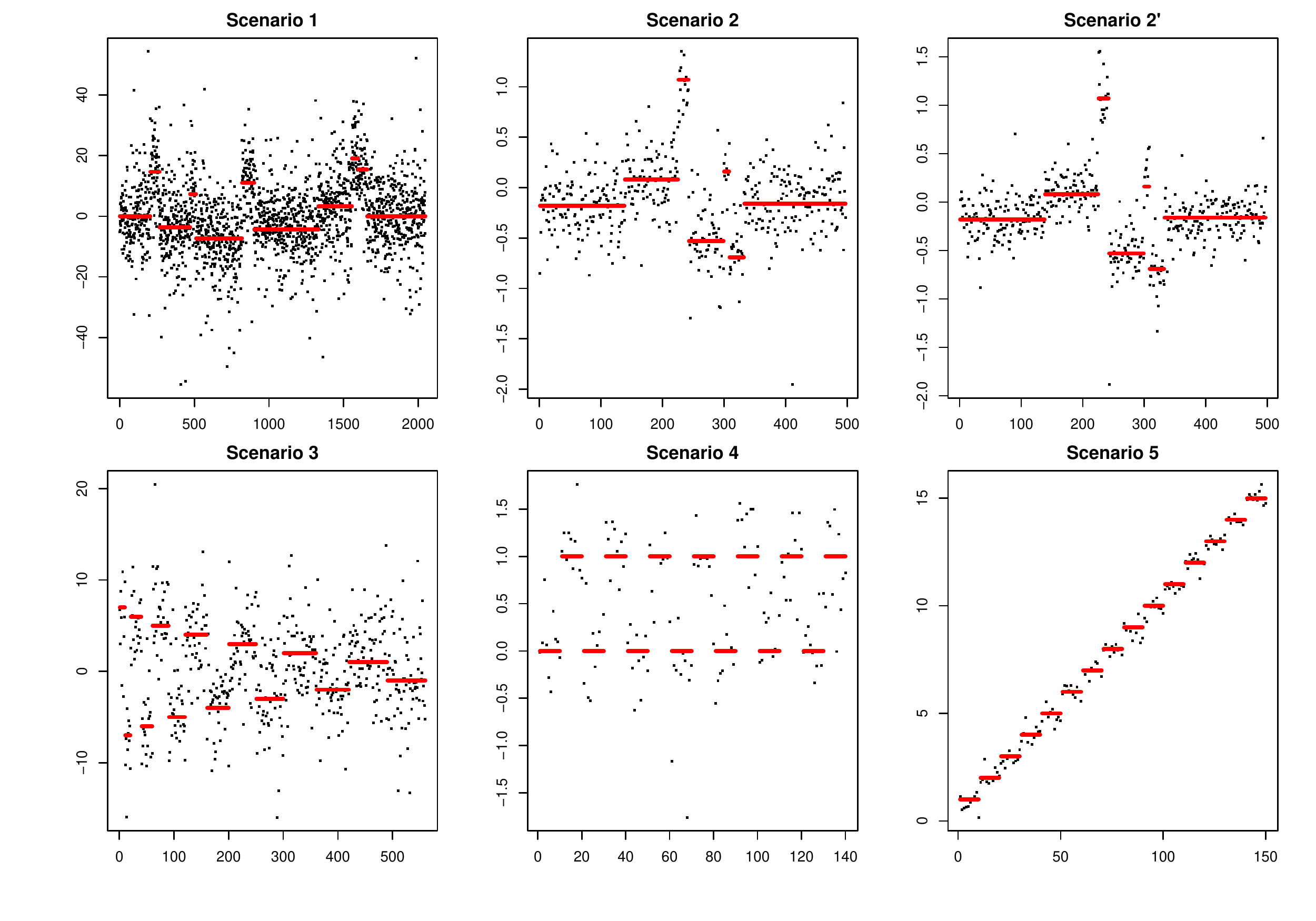}
 \caption{\label{Fig:scenarios} The signal, and example data, for each of the scenarios considered for the simulation study. Data was generated with the noise having a $t$-distribution with 5 degrees of freedom.}
\end{figure}

We assessed the performance of our robust estimators using the simulation benchmark proposed in the WBS paper 
\citep{Fryzlewicz:2014}. In that paper 5 scenarios are considered. These vary in length from $n=150$ to $n=2048$ and contain a variety 
of short and long segments, and a variety of sizes of the change in location from one segment to the next. 
We considered an additional scenario 
from \cite{Frick:2014} corresponding to scenario 2 of WBS with a standard deviation of 0.2 rather than 0.3. 
In our simulation study we are interested to see how the presence of outliers or heavy tailed noise affect different changepoint methods, and so
we will test each method assuming  t-distributed noise.
The underlying signals and example data for the three scenarios are shown in Figure \ref{Fig:scenarios}.

 For all approaches we need to choose the value $K$ in the loss function and the
penalty/threshold for adding a changepoint. These will depend on the standard deviation of the noise. Our approach is to estimate this standard deviation using the median absolute deviation of the differenced 
time-series, as in \cite{Fryzlewicz:2014}, which we denote as $\hat{\sigma}$. 
We compared our various robust estimators (Huber and biweight loss) to  binary segmentation using the robust cusum test \cite[]{Huvskova:1989}, described in Section \ref{Sec:Alt} (Cusum).
For the biweight loss we chose $K=3\hat{\sigma}$, so that extreme residuals according to a Gaussian model are treated as outliers. For the Huber loss we chose $K=1.345\hat{\sigma}$, 
a standard choice for trading statistical efficiency of estimation with robustness. 
We further set the penalty/threshold to
be $\beta=2\hat{\sigma}^2\log(n)\mbox{E}(\phi(Z)^2)$, where $\phi$ is the gradient of the loss function and $Z$ is a standard Gaussian random variable. This is based on the Schwarz information
criteria, adapted to account for the variability of loss function that is used \cite[see, e.g., theoretical results in][for further justification of this]{Huvskova:2012}, and for the biweight loss this
is inline with Theorem \ref{thm:Consistency}, which suggested the use of a penalty that is proportional to $\log(n)$.
We also compared to just using the standard square-error loss: implemented using FPOP \citep{Maidstone:2016}; and to  
the WBS \citep{Fryzlewicz:2014} approach that uses a standard cusum test statistic for detecting changepoints.
Again we used $\beta=2\hat{\sigma}^2\log(n)\mbox{E}(\phi(Z)^2)$, which in this case simplifies to the standard BIC penalty $\beta=2\hat{\sigma}^2\log(n)$,
and is the value that gave the best results for these methods across the 6 scenarios when there is normal noise 
\cite[see][]{Maidstone:2016}.

We  consider analysing data where the noise was from a t-distribution. We vary the degrees of freedom
from 3 to 100 to see of how varying how heavy-tailed the noise is affects the performance of different methods.

\begin{figure}
\center{
\begin{tabular}{cccc}
Scenario 1 & Scenario 2 & \\
\includegraphics[width=5.1cm, trim = 0cm 0cm 4.8cm 0cm, clip]{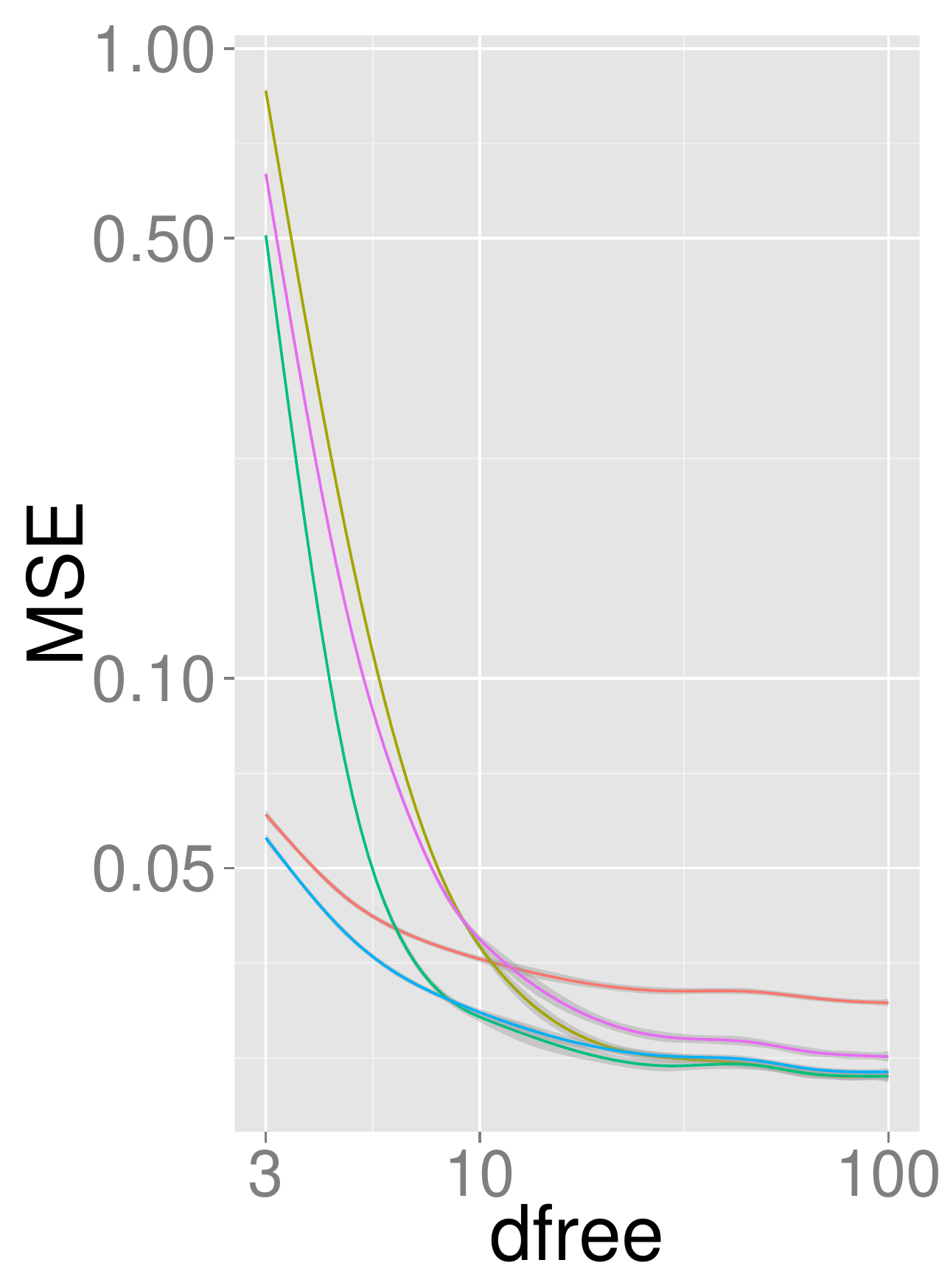} &
\includegraphics[width=5.1cm, trim = 0cm 0cm 4.8cm 0cm, clip]{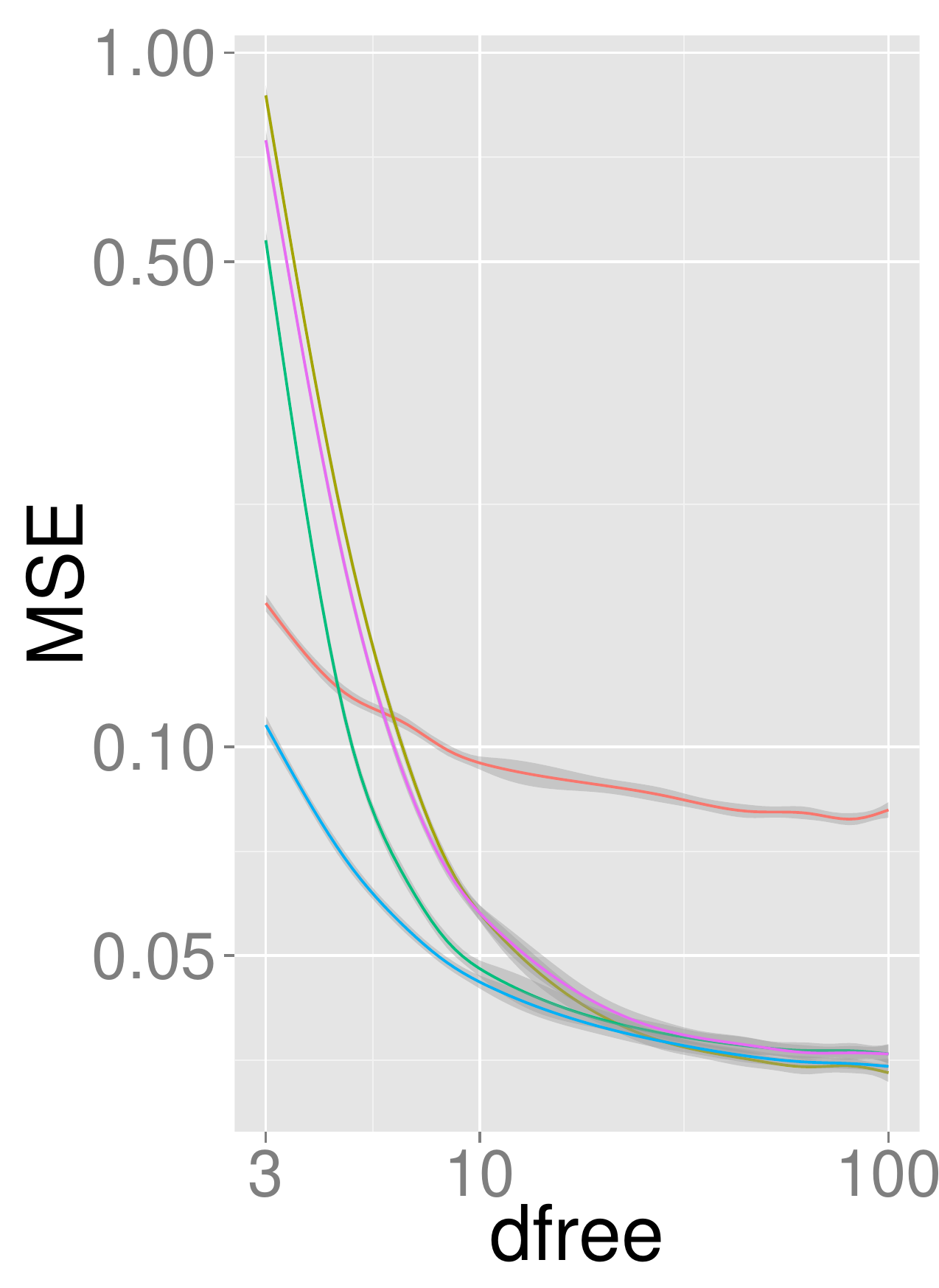} & 
\includegraphics[width=3cm, trim = 13cm 5cm 0cm 6.7cm, clip]{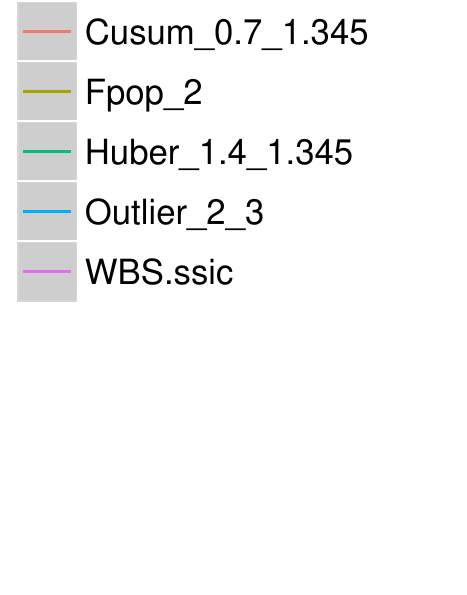} \\
Scenario 2' & Scenario 3 & \\
\includegraphics[width=5.1cm, trim = 0cm 0cm 4.8cm 0cm, clip]{M_select_Simu_3_MSE_1.pdf} &
\includegraphics[width=5.1cm, trim = 0cm 0cm 4.8cm 0cm, clip]{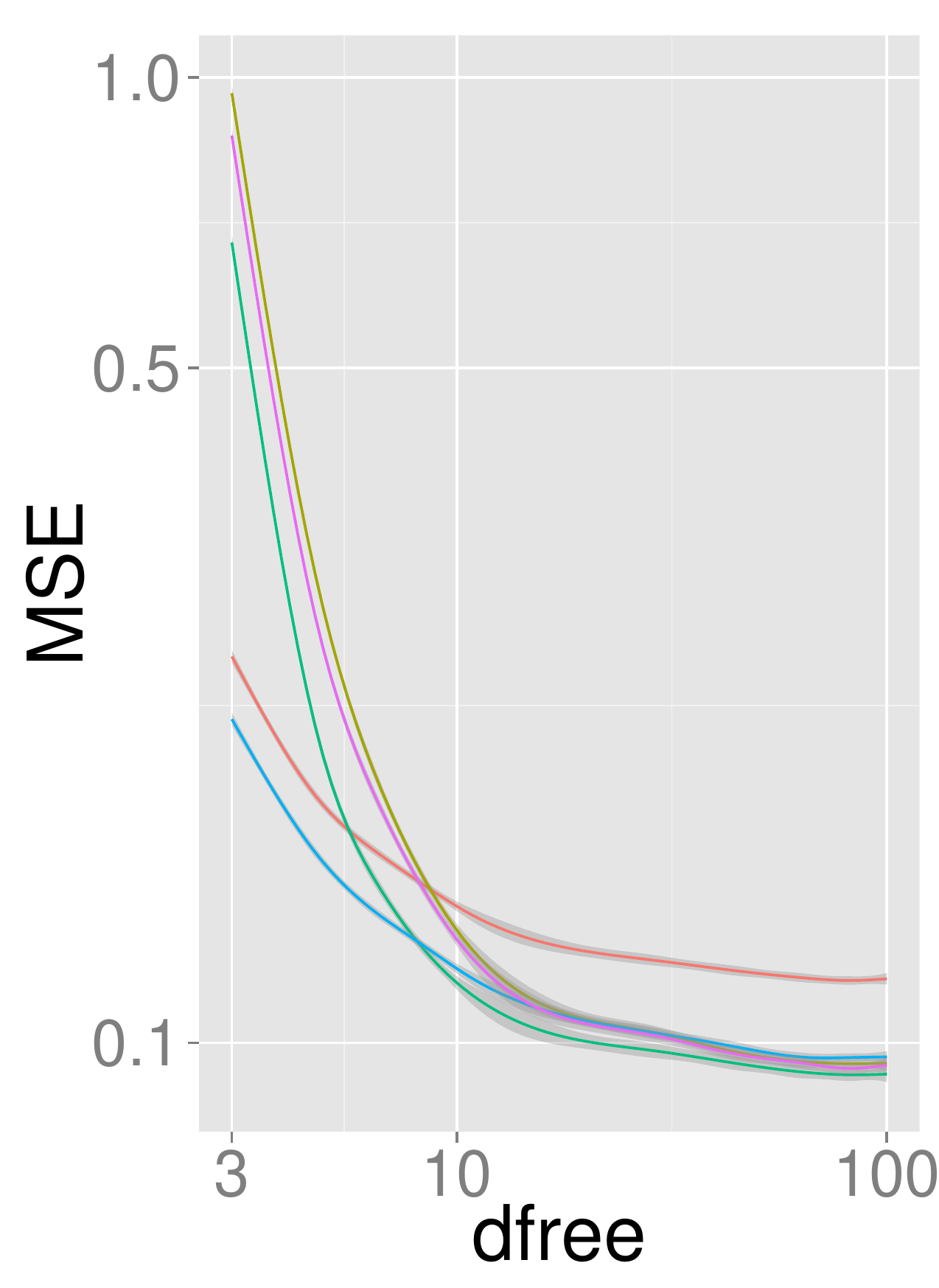} &
\\
Scenario 4 & Scenario 5 & \\
\includegraphics[width=5.1cm, trim = 0cm 0cm 4.8cm 0cm, clip]{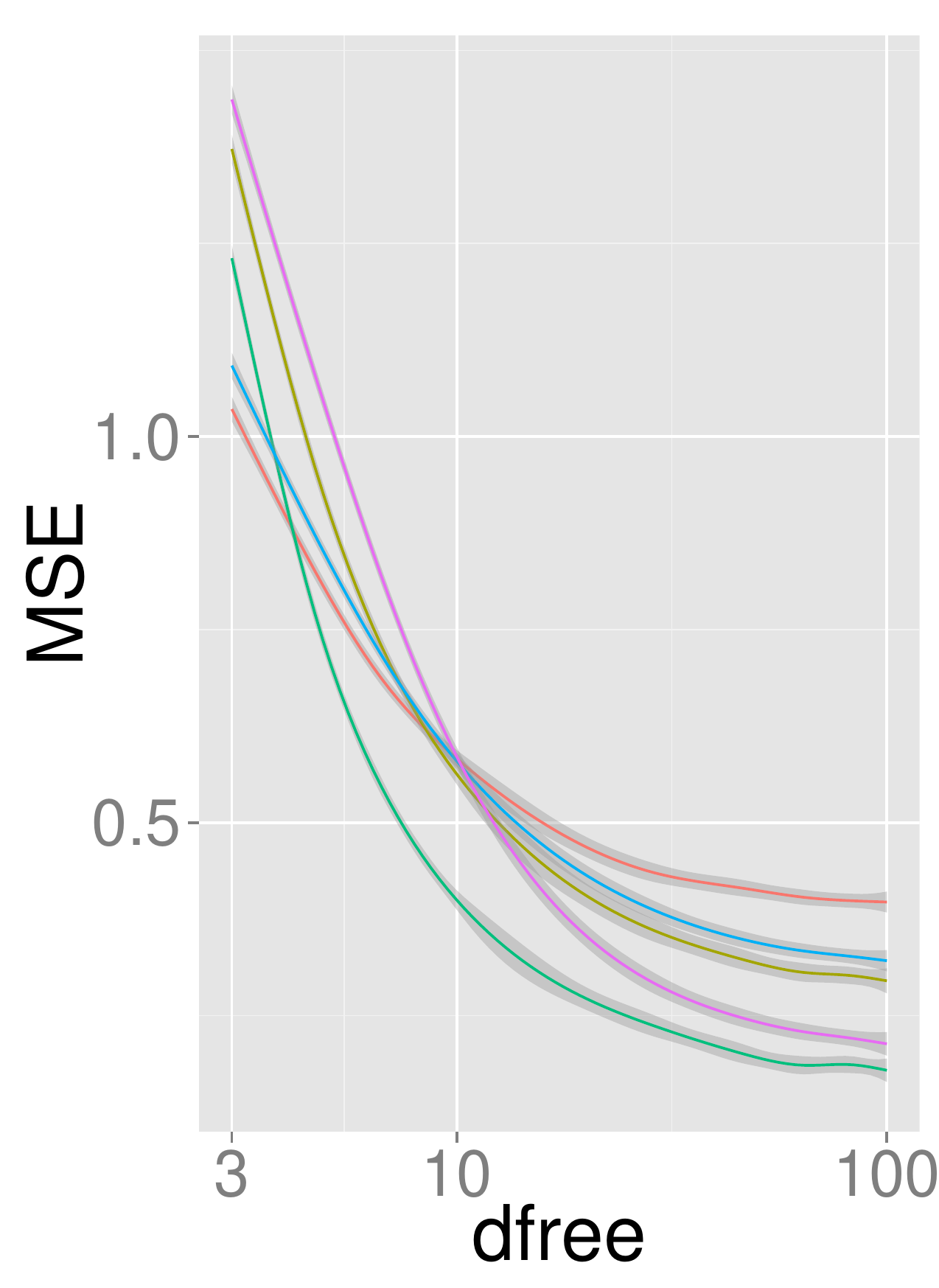} &
\includegraphics[width=5.1cm, trim = 0cm 0cm 4.8cm 0cm, clip]{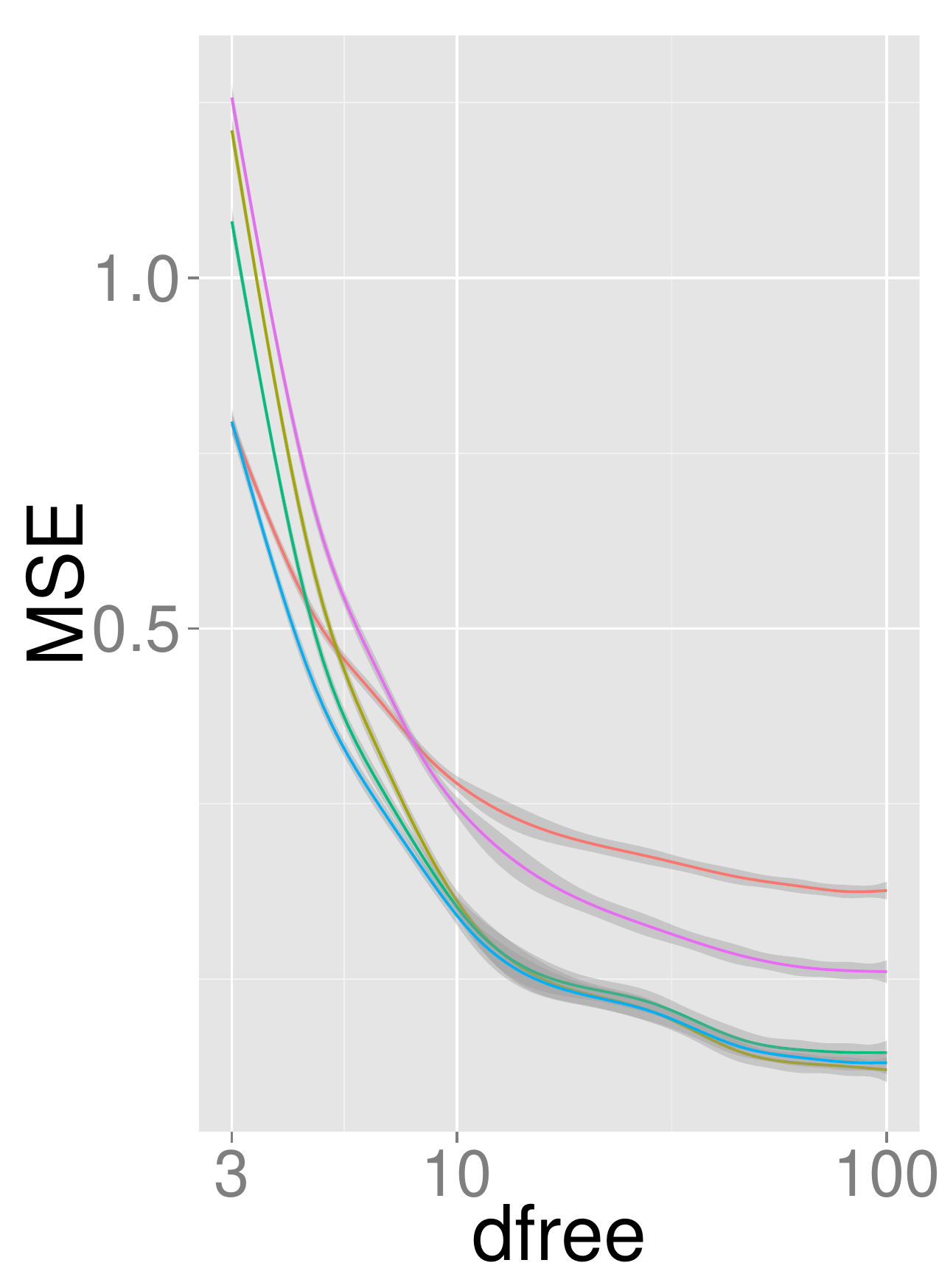} &
\\
\end{tabular}}
\caption{Smoothed log MSE of all tested approaches on the 6 scenarios using a student-noise with the degrees of freedom ranging from 3 to 100.
%We made 1600 simulations per degree of freedom. 
\label{Fig:R1}}
\end{figure}

\begin{figure}
\center{
\begin{tabular}{cccc}
Scenario 1 & Scenario 2 & \\
\includegraphics[width=5.1cm, trim = 0cm 0cm 4.8cm 0cm, clip]{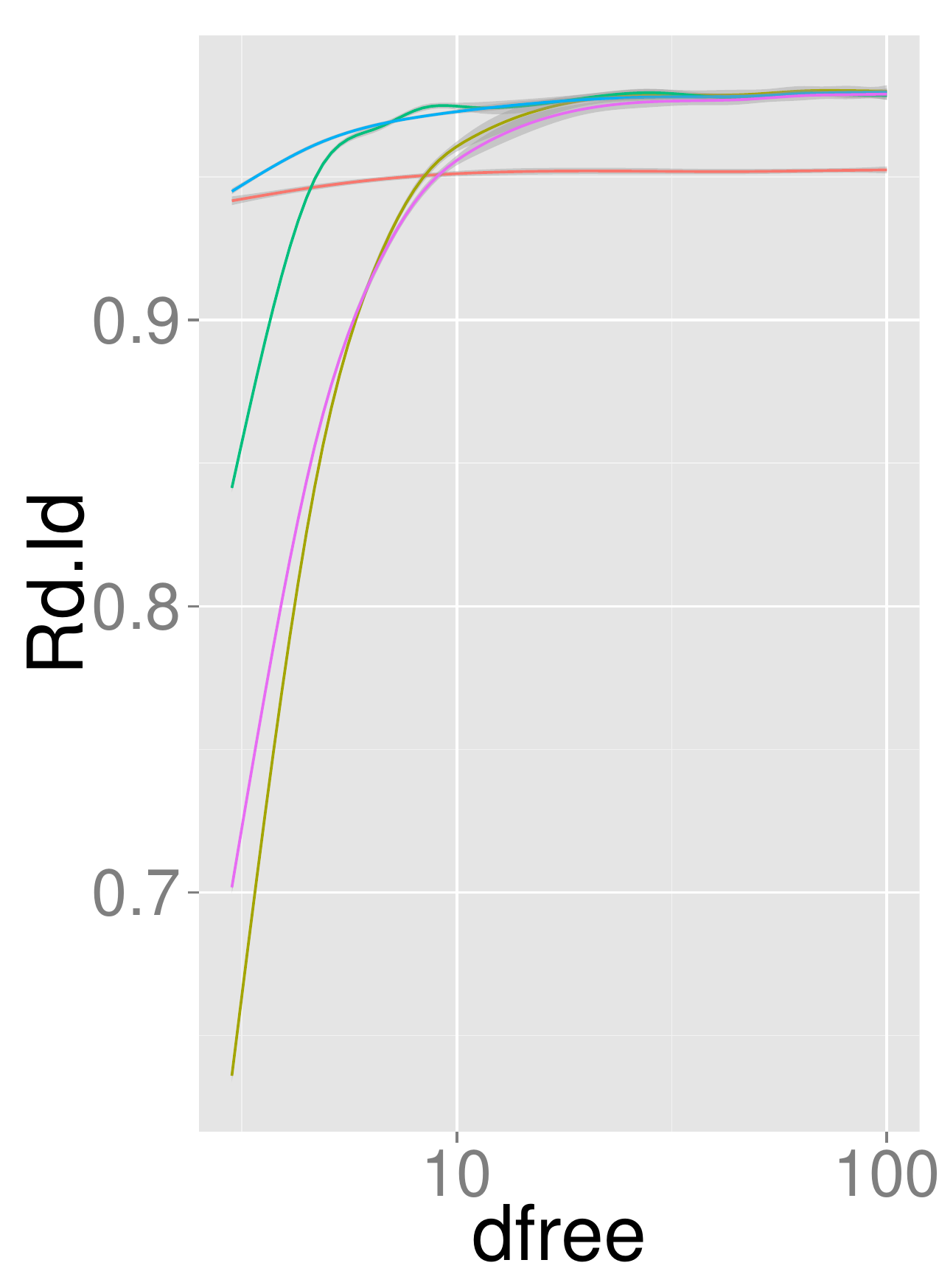} &
\includegraphics[width=5.1cm, trim = 0cm 0cm 4.8cm 0cm, clip]{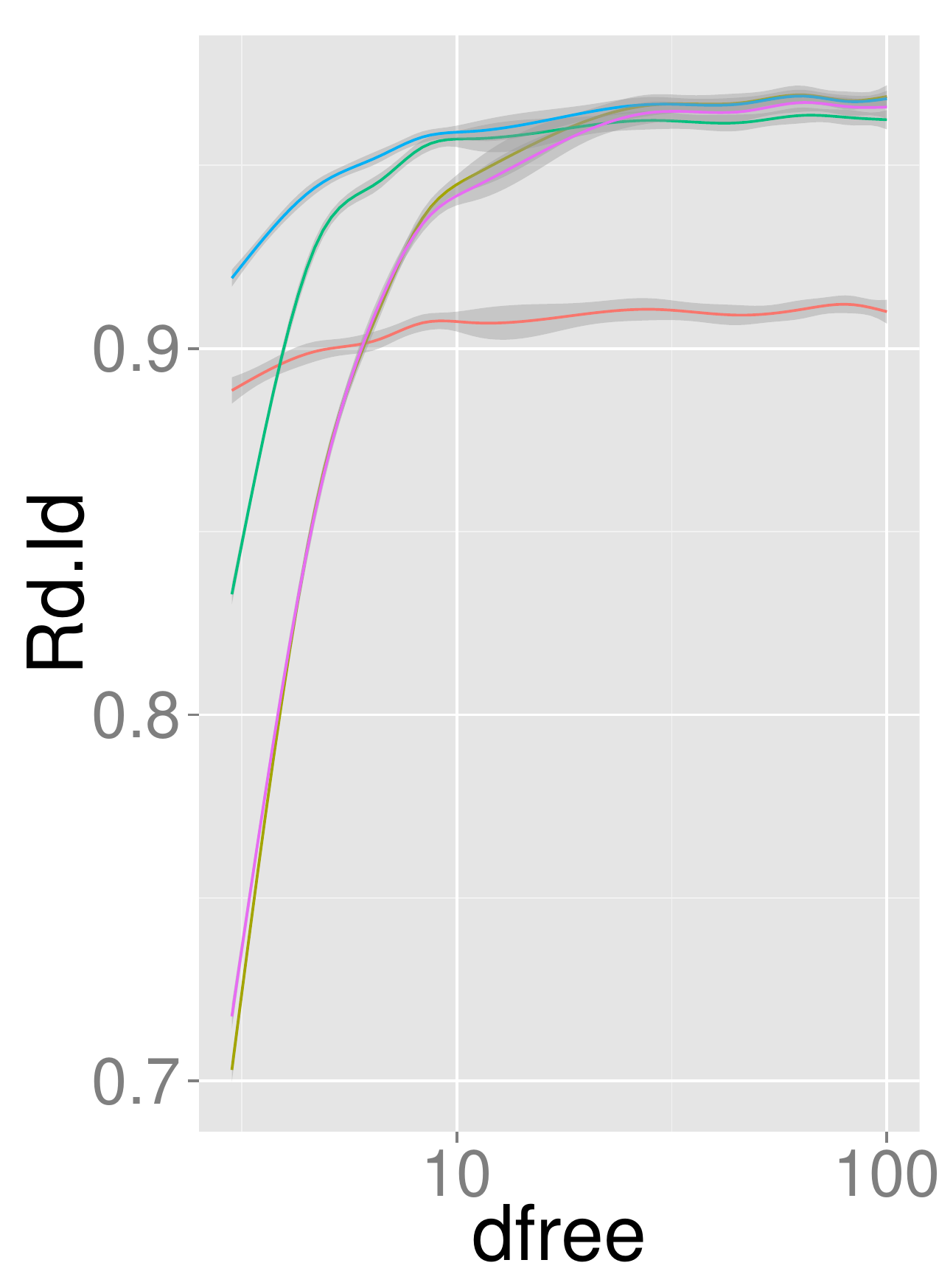} & 
\includegraphics[width=3cm, trim = 13cm 5cm 0cm 6.7cm, clip]{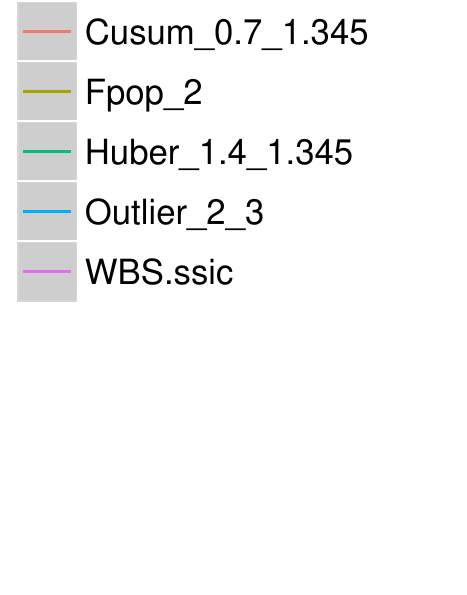} \\
Scenario 2' & Scenario 3 & \\
\includegraphics[width=5.1cm, trim = 0cm 0cm 4.8cm 0cm, clip]{R_select_Simu_3_Rd_Id_1.pdf} &
\includegraphics[width=5.1cm, trim = 0cm 0cm 4.8cm 0cm, clip]{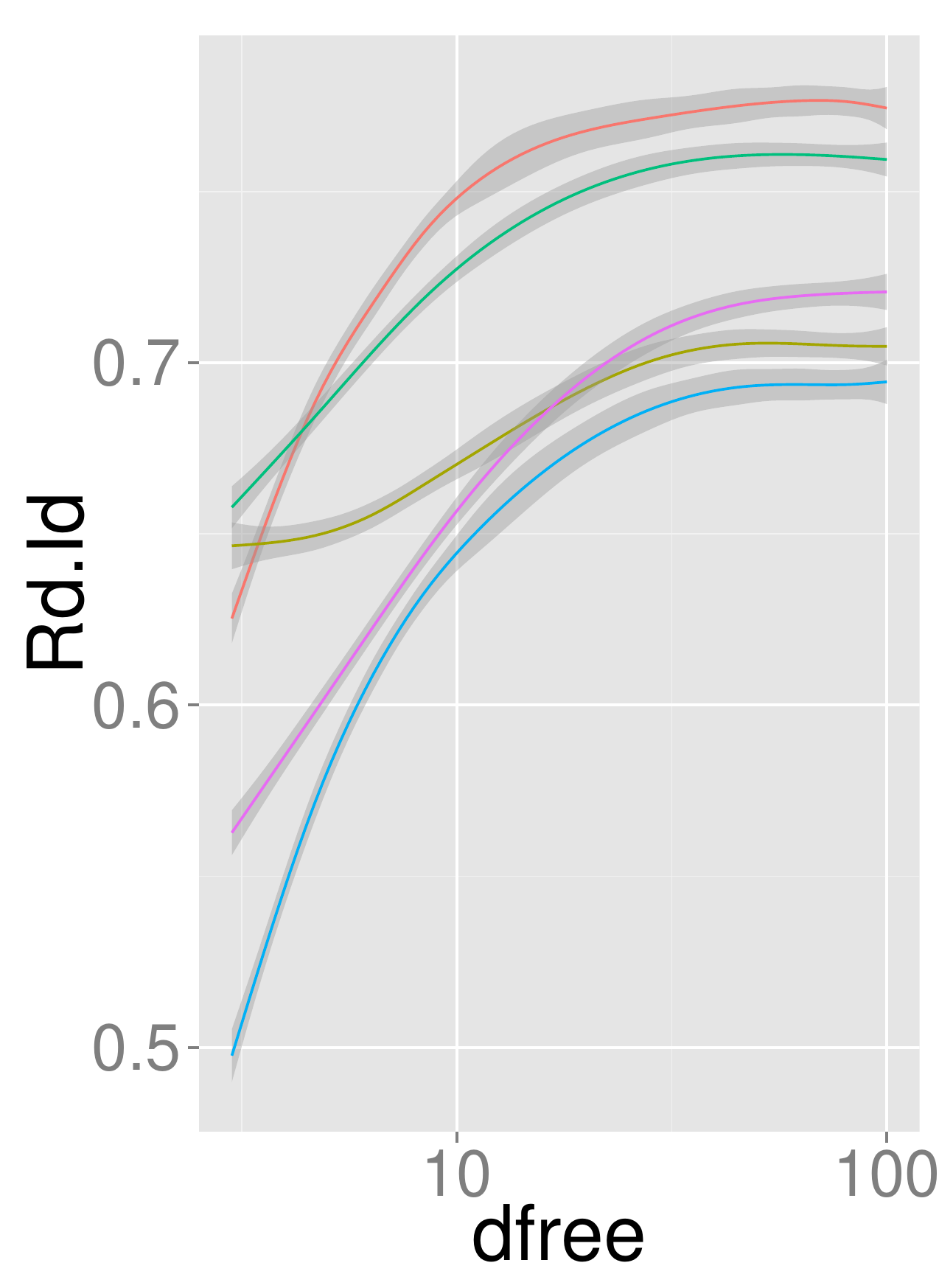} &
\\
Scenario 4 & Scenario 5 & \\
\includegraphics[width=5.1cm, trim = 0cm 0cm 4.8cm 0cm, clip]{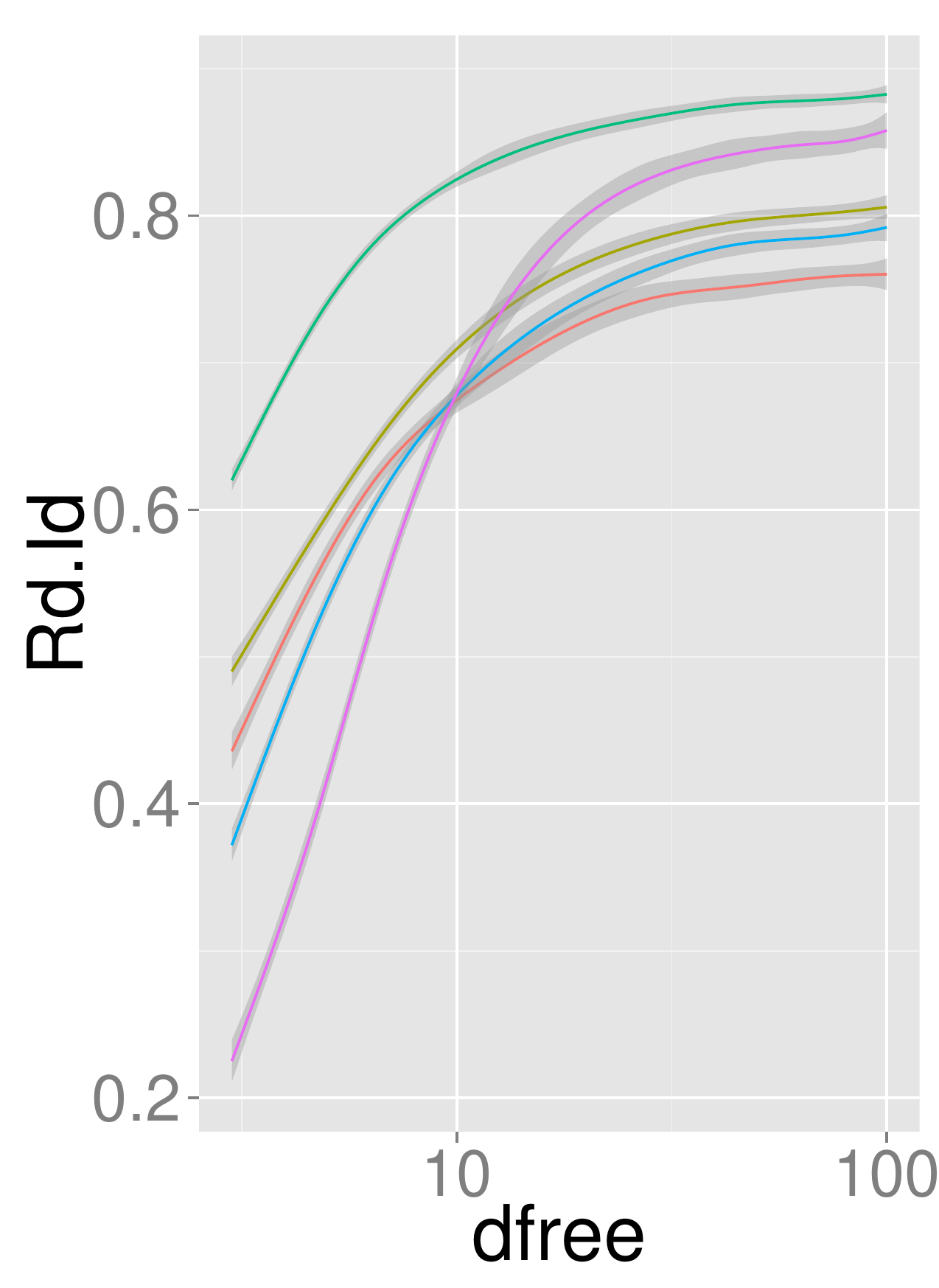} &
\includegraphics[width=5.1cm, trim = 0cm 0cm 4.8cm 0cm, clip]{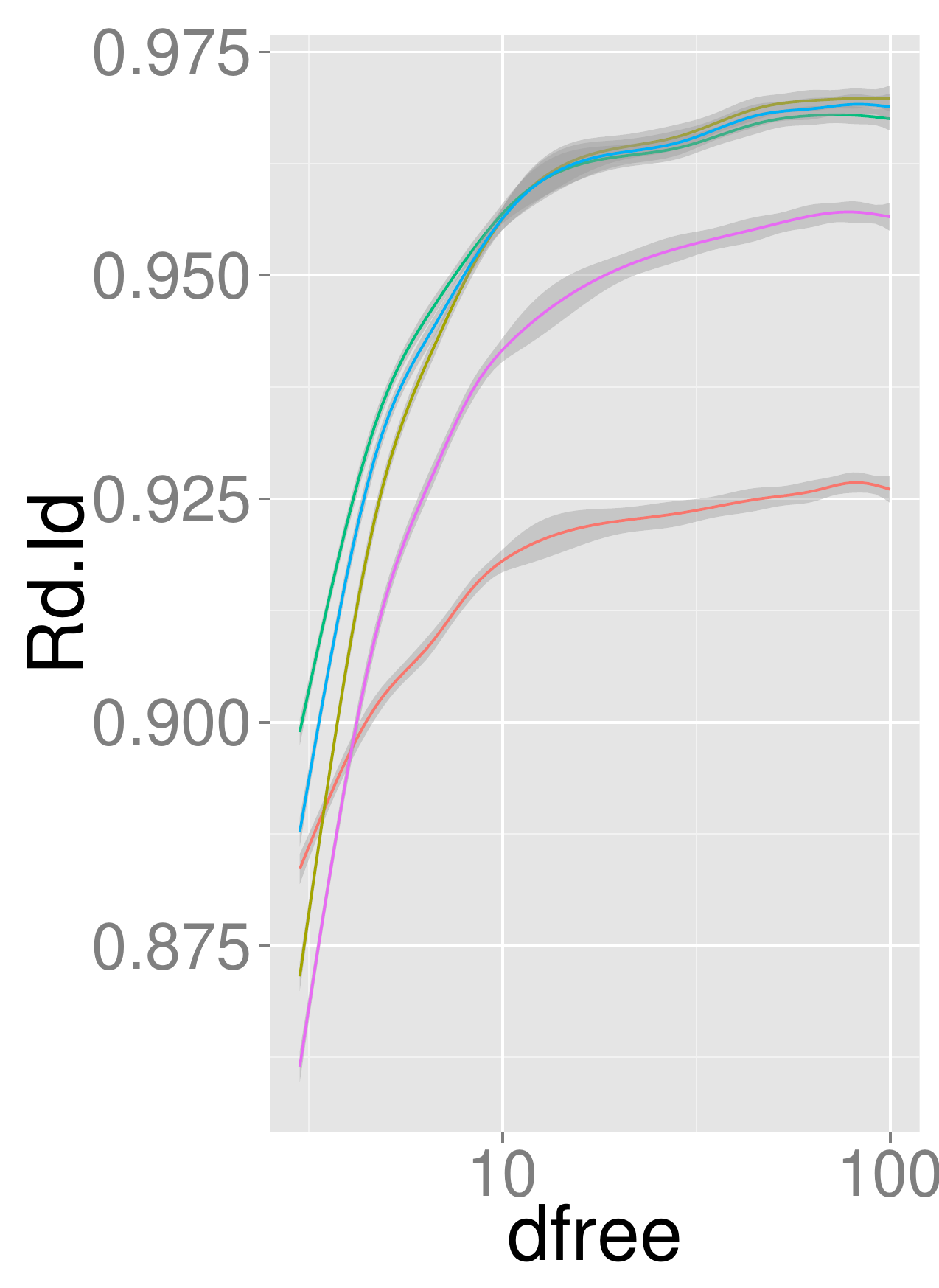} &
\\
\end{tabular}}
\caption{Smoothed normalized Rand-index of all tested approaches on the 6 scenarios  using  student-noise with the degrees of freedom varying from 3 to 100.
%We made 1600 simulations per degree of freedom.
\label{Fig:R2}}
\end{figure}

In Figures \ref{Fig:R1} and \ref{Fig:R2}  we show the results of all approaches as a function of the degrees of freedom. We compare methods based on how they estimate
the underlying piecewise constant mean function, measured in terms of mean square error; and how well they estimate the segmentation, measured using the normalized rand-index. 
%Our estimate of the mean function is based on estimating the mean within each inferred segment as the sample mean of data in that segment. 
The normalized rand-index measures the overlap between
the true segmentation and the inferred segmentation, with larger values indicating a better estimation of the segmentation. 

In terms of mean square error, for almost all scenarios we consider the biweight loss performs best when the degrees of freedom is small. 
It also appears to lose little in terms of accuracy when the degrees of freedom is large, and
the noise is close to Gaussian. The robust cusum approach also performs well when the degrees of freedom are small, but in most cases it shows a marked drop in accuracy relative to the alternative methods 
when the noise is close to Gaussian. The one scenario where the biweight loss performs poorly when the noise is close to Gaussian is Scenario 4. In this case we have short segments, only slightly larger than 
the minimum segment length for the biweight loss, with the segment mean being the
same for all odd segments. We can get a reasonable fit under the biweight loss by, for example, ignoring the changepoints and treating all observations in the even
segments as outliers. The problem of distinguishing between this case and the presence of actual changepoints causes the poor performance.

The results in terms of the quality of the segmentation, as measured using the rand-index, are more mixed. The biweight loss is clearly best in scenarios 1 and 2, but performs
poorly for scenario 3. Here the use of the Huber loss appears to give best results across the different scenarios. Again we see that the use of the L2 loss, using either FPOP or WBS, performs poorly when the degrees
of freedom are small.

%\begin{figure}
%\center{
%\begin{tabular}{cccc}
%Scenario 1 & Scenario 2 & \\
%\includegraphics[width=5.1cm, trim = 0cm 0cm 4.8cm 0cm, clip]{output_img_student/forpaper/R_select_Simu_1_Rd_Id_2.pdf} &
%\includegraphics[width=5.1cm, trim = 0cm 0cm 4.8cm 0cm, clip]{output_img_student/forpaper/R_select_Simu_2_Rd_Id_2.pdf} & 
%\includegraphics[width=3cm, trim = 13cm 5cm 0cm 6.7cm, clip]{output_img_student/forpaper/R_select_Simu_3_Rd_Id_2.pdf} \\
%Scenario 2' & Scenario 3 & \\
%\includegraphics[width=5.1cm, trim = 0cm 0cm 4.8cm 0cm, clip]{output_img_student/forpaper/R_select_Simu_3_Rd_Id_2.pdf} &
%\includegraphics[width=5.1cm, trim = 0cm 0cm 4.8cm 0cm, clip]{output_img_student/forpaper/R_select_Simu_4_Rd_Id_2.pdf} &
%\\
%Scenario 4 & Scenario 5 & \\
%\includegraphics[width=5.1cm, trim = 0cm 0cm 4.8cm 0cm, clip]{output_img_student/forpaper/R_select_Simu_5_Rd_Id_2.pdf} &
%\includegraphics[width=5.1cm, trim = 0cm 0cm 4.8cm 0cm, clip]{output_img_student/forpaper/R_select_Simu_6_Rd_Id_2.pdf} &
%\\
%\end{tabular}}
%\caption{Smoothed normalized Rand-index of all tested approaches on the 6 scenarios (longer profiles : 
%length of each segment multiplied by 2 and sd multiplied by $\sqrt{4}$) using a student-noise with a degree of freedom varying from 3 to 100. We made 1600 simulations per degree of freedom.}
%\end{figure}

\subsection{Online analysis of well-log data}

We return to the  well-log data of Figure \ref{Fig:Oil1}. For this data, due to the presence of substantial outliers, we choose to use the biweight loss function. 
%, as to be robust to such outliers we need a bounded loss function (see Theorem \ref{thm:1}). 
We set the threshold, $K$ in (\ref{eq:biweight}), to be twice an estimate of the standard deviation of the observation noise. We set $\beta$
to be 70 times the estimated variance of the noise. This is larger than that of the BIC penalty, but this is needed due to the presence of auto-correlation in the observation noise \cite[]{Lavielle:2000}, and is the same penalty used
for the analysis presented in Figure \ref{Fig:Oil1}. 
%Choice of this penalty is important in practice, and for this type of application we would recommend testing a range of penalties 
%\cite[using the CROPS algorithm of][]{Haynes:2015} on some training data to learn an appropriate penalty \cite[]{Rigaill:2013}. 

\begin{figure}
 \centering \includegraphics[scale=0.45]{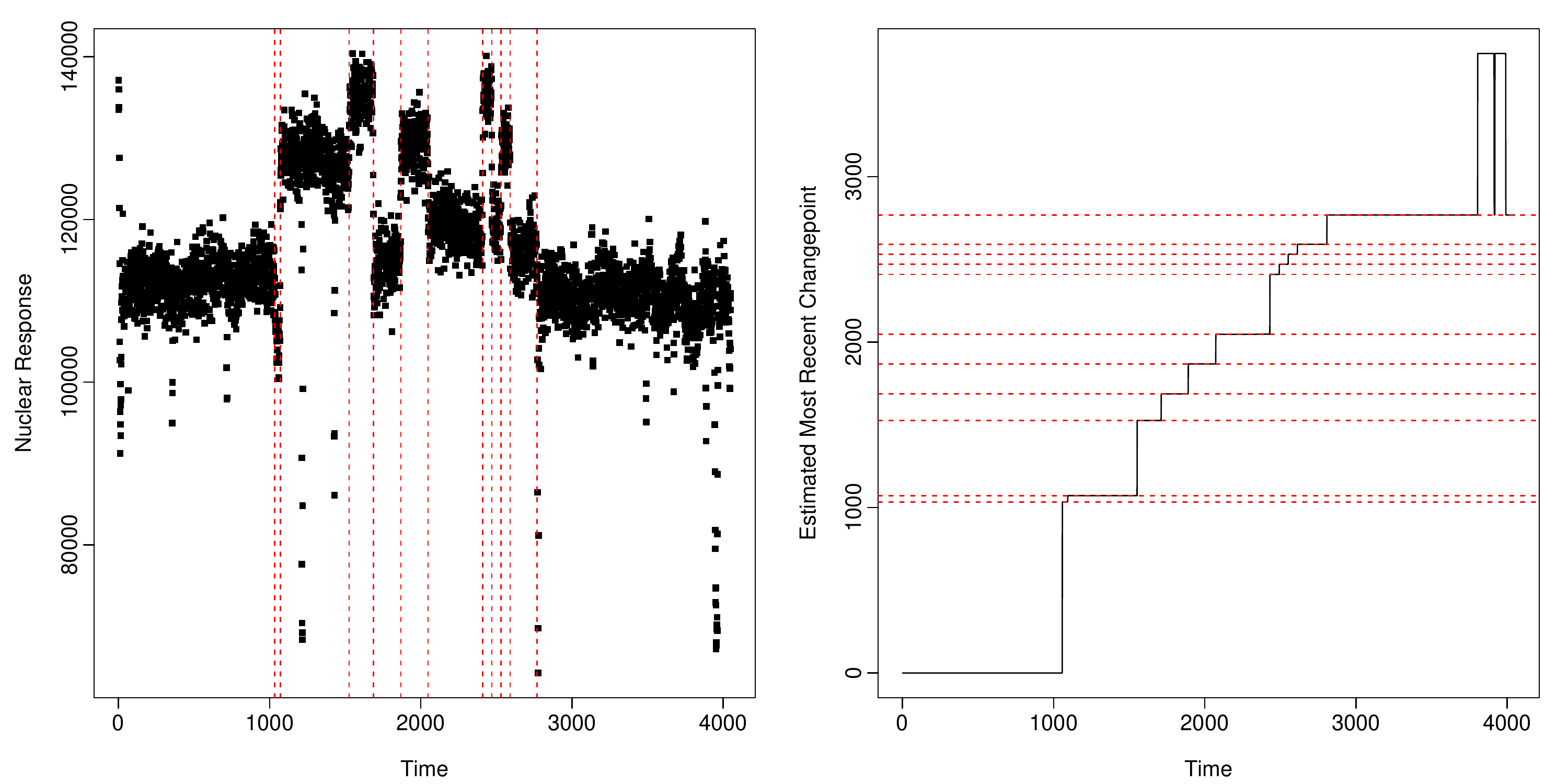}
 \caption{\label{Fig:Oil2} Estimated changepoints from batch analysis of the well-log data (left-hand plot) under biweight loss. Estimate of location of most-recent changepoint from online analysis (right-hand plot).
 The black line shows the estimate of the most-recent changepoint against the number of data points analysed. The red dashed horizontal lines show the locations of the changepoints detected from the batch analysis.
 }
\end{figure}

Figure \ref{Fig:Oil2} shows the estimated changepoints we obtain from a batch analysis of the data. As we can see, using the biweight penalty makes the changepoint detection robust to the presence of the outliers. All 
obvious changes are detected, and we do not detect a change at any point where the outliers cluster. 

As mentioned in the introduction, the motivation for analysing this data requires an online analysis. We present output from such an online analysis in the right-hand plot of Figure \ref{Fig:Oil2}. Here we plot the 
estimate of the most recent changepoint prior to $t$, given data $y_{1:t}$, as a function of $t$. To help interpret the result we also show the locations of the changepoints inferred from the batch analysis. We see that
we are able to quickly detect changes when they happen, and we have only one region where there is some fluctuation in where we estimate the most recent changepoint. Whilst by eye the plot may suggest we immediately 
detect the changes, there is actually some lag. This is inevitable when using the biweight loss, due to the presence of a minimum segment length that can be inferred (see Theorem \ref{thm:2}). The lag in detecting
the changepoint is between 21 and 27 observations for all except the final changepoint. The final inferred changepoint is less pronounced, and is not detected until after a lag of 40 observations. This lag
can be reduced by increasing $K$, but at the expense of less robustness to outliers. The region of fluctuation over the estimate of the most recent changepoint corresponds to uncertainty about whether there
are changepoints in the last inferred segment (corresponding to the final two changepoints inferred in the bottom-left plot of Figure \ref{Fig:Oil1}). One disadvantage of detection methods that involve minimising
a penalised cost, and of other methods that produce a single estimate of the changepoint locations, is that they do not quantify the uncertainty in the estimate.

\subsection{Estimating Copy Number Variation}

\begin{figure}
\center{
\begin{tabular}{cc}
Tumor Fraction = 1 & Tumor Fraction = 0.5  \\
\includegraphics[width=6.5cm, trim = 0cm 0cm 0cm 0cm, clip]{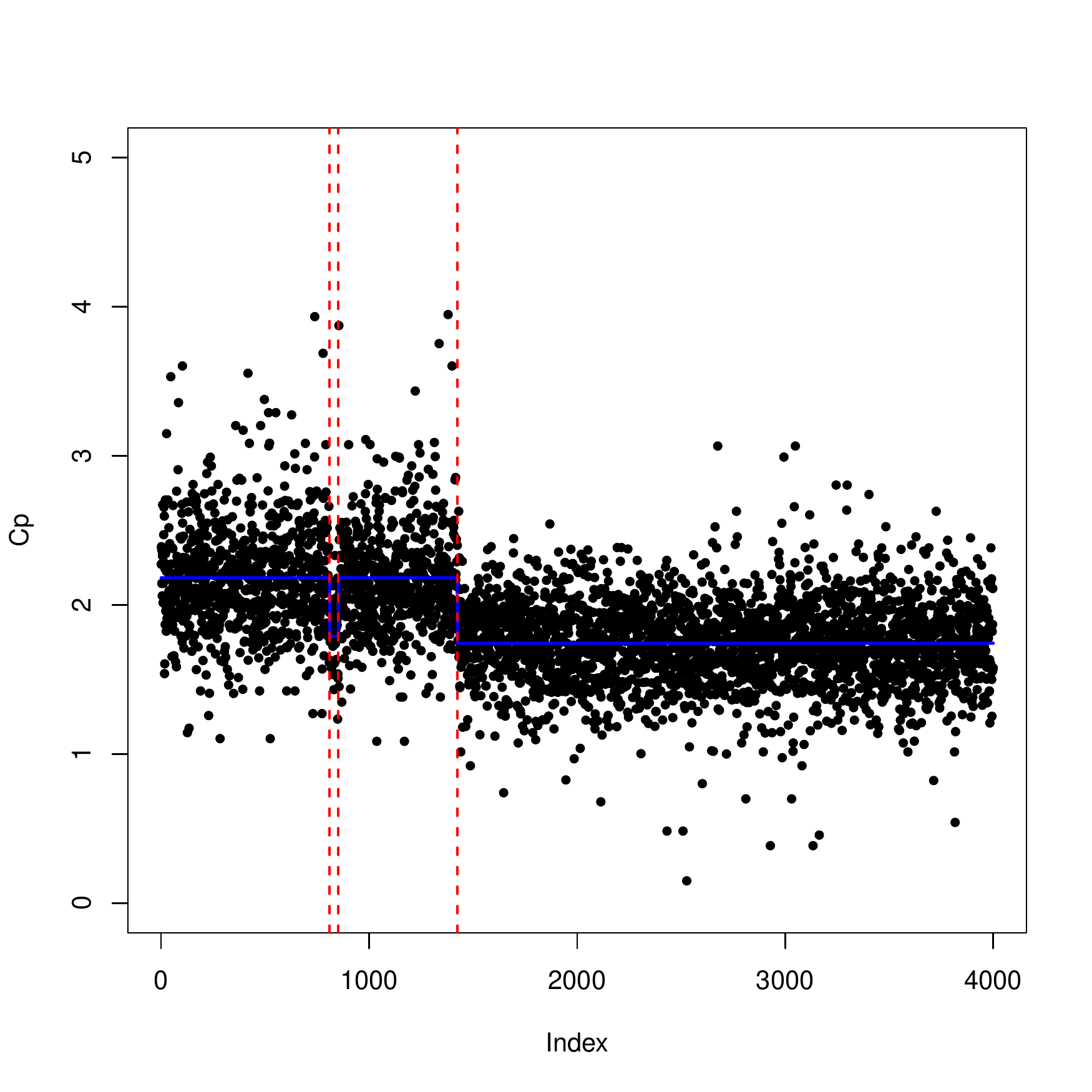} &
\includegraphics[width=6.5cm, trim = 0cm 0cm 0cm 0cm, clip]{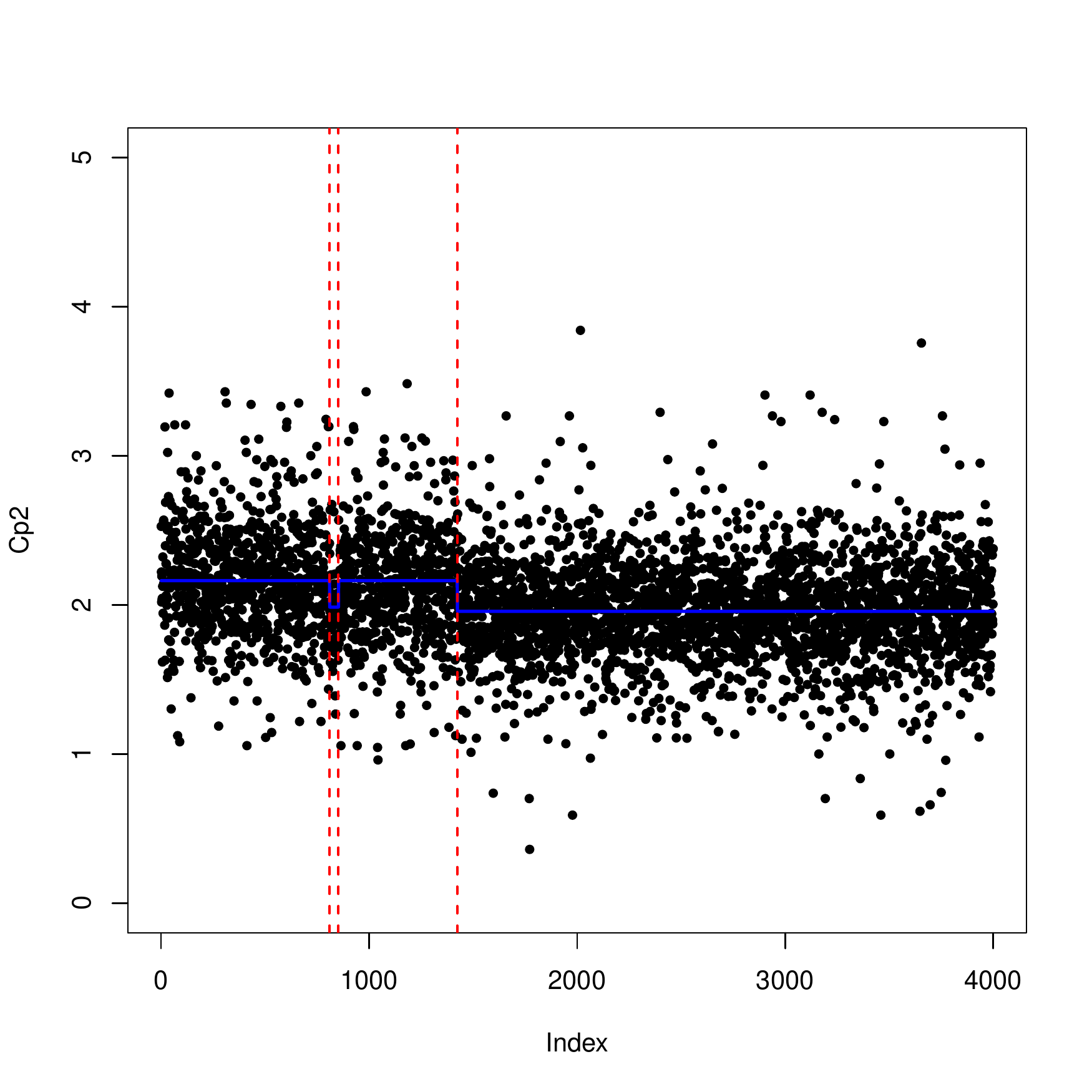} \\
\end{tabular}
}
\caption{Two DNA copy number profiles obtained using the jointseg package with a tumor fraction of 1 (left) and
0.5 (right). The true change-points are represented
with red dotted lines. It can be seen
that a number of data-points are quite far from from the blue line. The size of each jump is larger when the tumor fraction is larger.\label{fig:simu}}
\end{figure}

Healthy human cells have two copies of DNA. 
In tumor cells, parts of chromosomes of various sizes
(from kilobases to a chromosome arm) may be
deleted or amplified several times, and this can lead to the copy number of the DNA from such regions being different from 2. 
Copy numbers (CN) can be measured
using microarray or sequencing experiments. They are piecewise constant
along the genome, and interest lies in detecting whether, and where, the copy number changes.
For many samples we would have a mixture of healthy and
tumor cells, and the signal to noise ratio for changes in copy number will go down with the tumor fraction.
The detection of changes in copy number is further complicated by the presence of outliers. 
We illustrate
this in Figure \ref{fig:simu} using output from the jointseg package \citep{pierre2014performance} which enables simulation of
realistic CN profiles by resampling real datasets for which the truth is known.

A standard way to analyse such data is to use the smooth.CNA function of the well known DNAcopy package \cite[]{Bengtsson:2016}. This function shrinks outliers towards the value of its neighbors. Once this
is done one can run a preferred segmentation approach. As we will see below, this heuristic preprocessing  procedure greatly improves changepoint detection. We want to compare such a two-stage approach to
a simpler analysis where we analyse data using our penalised cost approach with the biweight loss.

To assess the performance of our approach on DNA copy number data we used the jointseg package. %\citep{pierre2014performance}. 
%Using this package we simulated realistic DNA copy number
% profiles by resampling real DNA copy number profiles. 
We simulated profiles of length $n=4000$ with 
10 change-points with segments of at least 40 data-points. The package propose two real datasets,
GSE11976 and GSE29172, to resample from. For both we considered four levels of difficulty corresponding to different 
 tumor fractions:  0.34, 0.50, 0.79 and 1 for GSE11976; and 0.3, 0.5, 0.7 and  1 for GSE29172.

We consider four approaches: FPOP (L2), FPOP after using smooth.CNA to remove outliers (Rout L2), robust binary segmentation (Cusum)
and our biweight loss with a threshold value of 3.
All approaches are implemented for a range of penalty values. For every simulated profile and each run of a method we computed the number of true positive (TP) and false positive (FP) change-points.
For all true change-point we counted one TP if there is at least one change-point identified within a window of 15 data-points. 
We then computed the number of FPs as the number of predicted changes minus the number of TPs. We then average, over 200 simulated profiles, the number of TPs and FPs per approach, penalty value and difficulty to recover ROC curves.

Overall our robust biweight loss outperforms the L2 loss following outlier removal and the Cusum
approach. % in 5 out of the 8 scenario we considered. 
For low tumor fractions (0.3 and 0.5 GSE29172 and 0.34 GSE11976) the biweight loss is possibly slightly better than the Cusum approach. 
For a tumor fraction of 1 the biweight loss is slightly better
than the L2 following outlier removal. In other cases it is clearly better.
%In two cases (GSE11976 with TF= 0.3 and GSE29172 with TF=0.5) it is as effecient as the L2 loss following outlier removal
%and clearly better than the Cusum approach. For the GSE29172 dataset with a tumor fraction of 0.3 it worse than the L2 following outlier removal and better than the Cusum.
Results are shown for the two datasets and a tumor fraction of 0.7 and 0.79 in Figure \ref{fig:DNA_copy_TF0p5}.
Results for other tumor fractions are provided in figures in Appendix \ref{App:all results}.

\begin{figure}
\center{
\begin{tabular}{ccc}
GSE11976 (TF=0.79) & GSE29172 (TF=0.7) &\\
\includegraphics[width=6.1cm, trim = 0cm 0cm 3.5cm 0cm, clip]{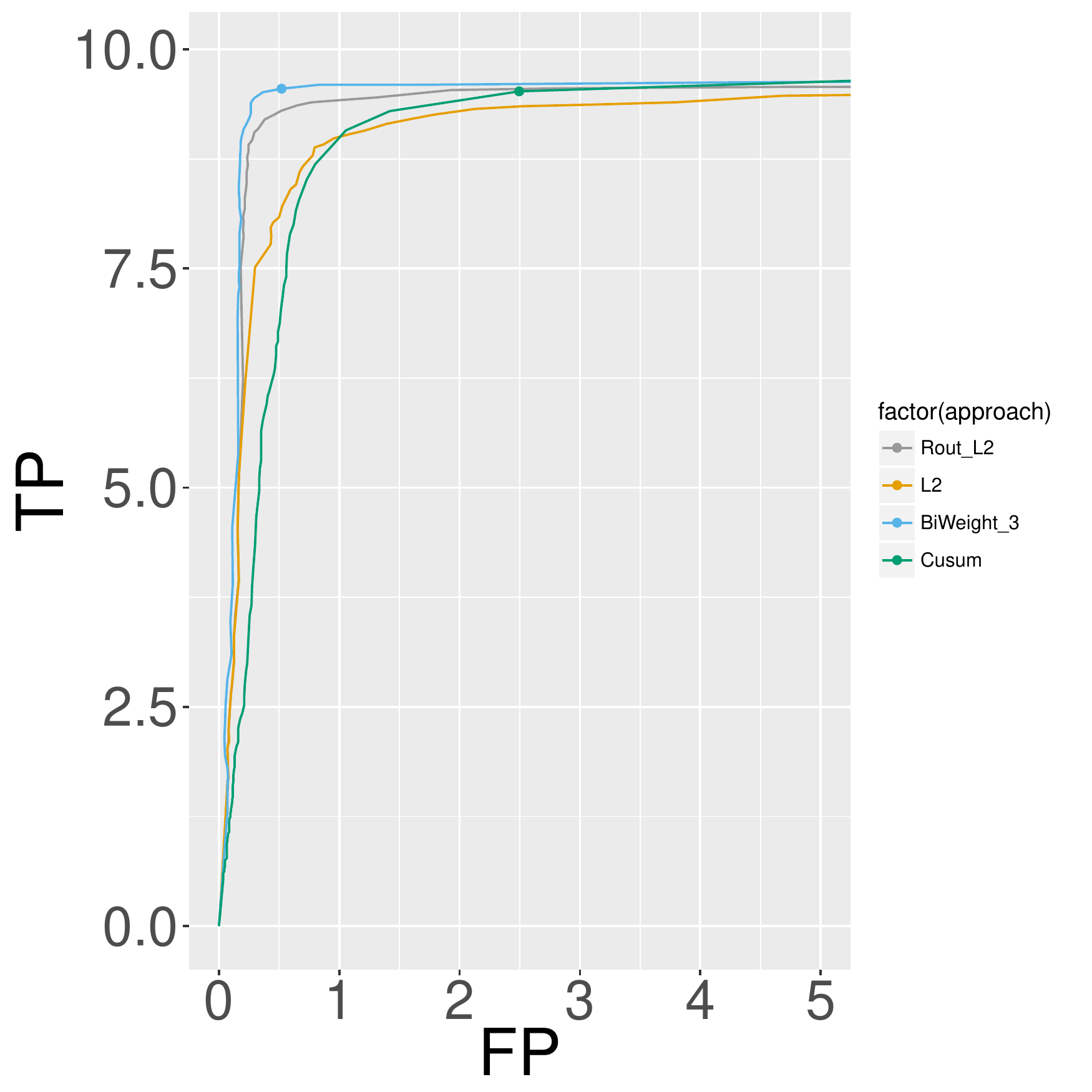} &
\includegraphics[width=6.1cm, trim = 0cm 0cm 3.5cm 0cm, clip]{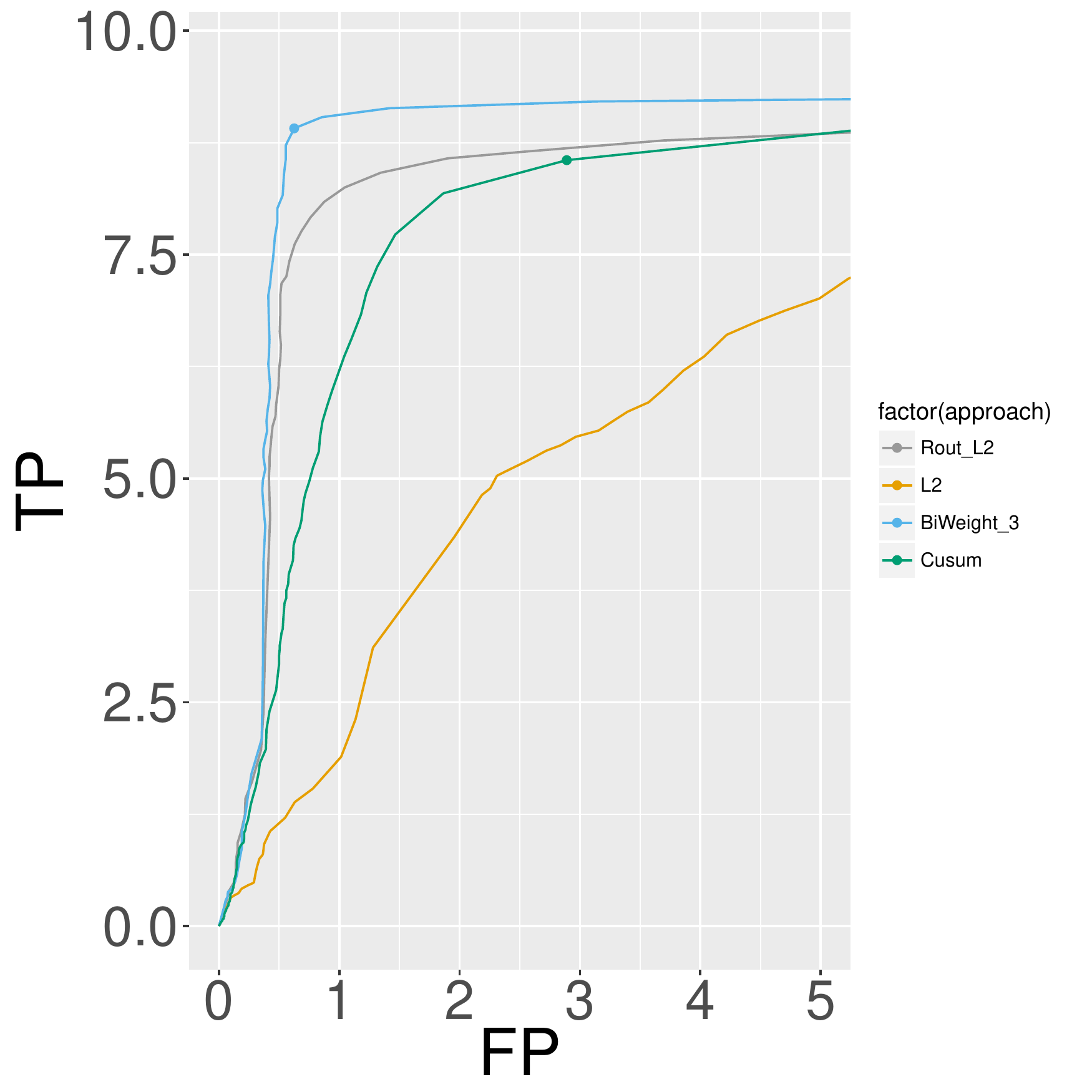} &
\includegraphics[width=3cm, trim = 14cm 6cm 0cm 7cm, clip]{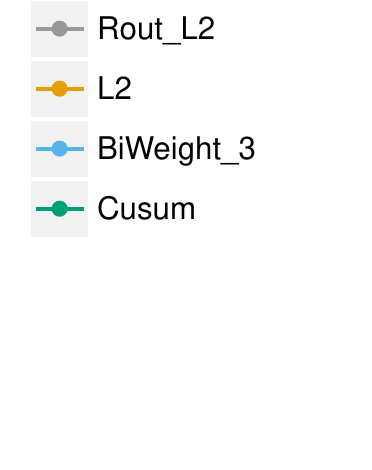} 
\end{tabular}}
\caption{Average ROC curve on the GSE11976 and GSE29172 datasets for a tumor fraction of respectively 0.79 and 0.7, 
for the Cusum, L2, L2 with outlier removal (Rout L2) and our robust biweight loss (Biweight 3).}\label{fig:DNA_copy_TF0p5}
\end{figure}

\subsection{Wireless Tampering}

We now consider an application which looks at security of the Internet of Things (IoT). Many IoT devices use WiFi to communicate. Often, for example with surveillance systems, these need a high level of security. 
Thus it is important to be able to
detect if a device has been tampered with. WiFi signals include a ``preamble'' which is used by the receiver to determine channel state. One approach that can be used to detect tampering is to monitor channel state
variation \cite[]{Bagci:2016}. 
Abrupt changes in it could indicate some tampering event. However changes can also be caused by less sinister events, such as movement of people within the communication environment. 
Thus the challenge is to detect a change caused by tampering as opposed to any ``outliers'' caused by such temporary environmental factors.

\begin{figure}
 \centering \includegraphics[scale=0.4]{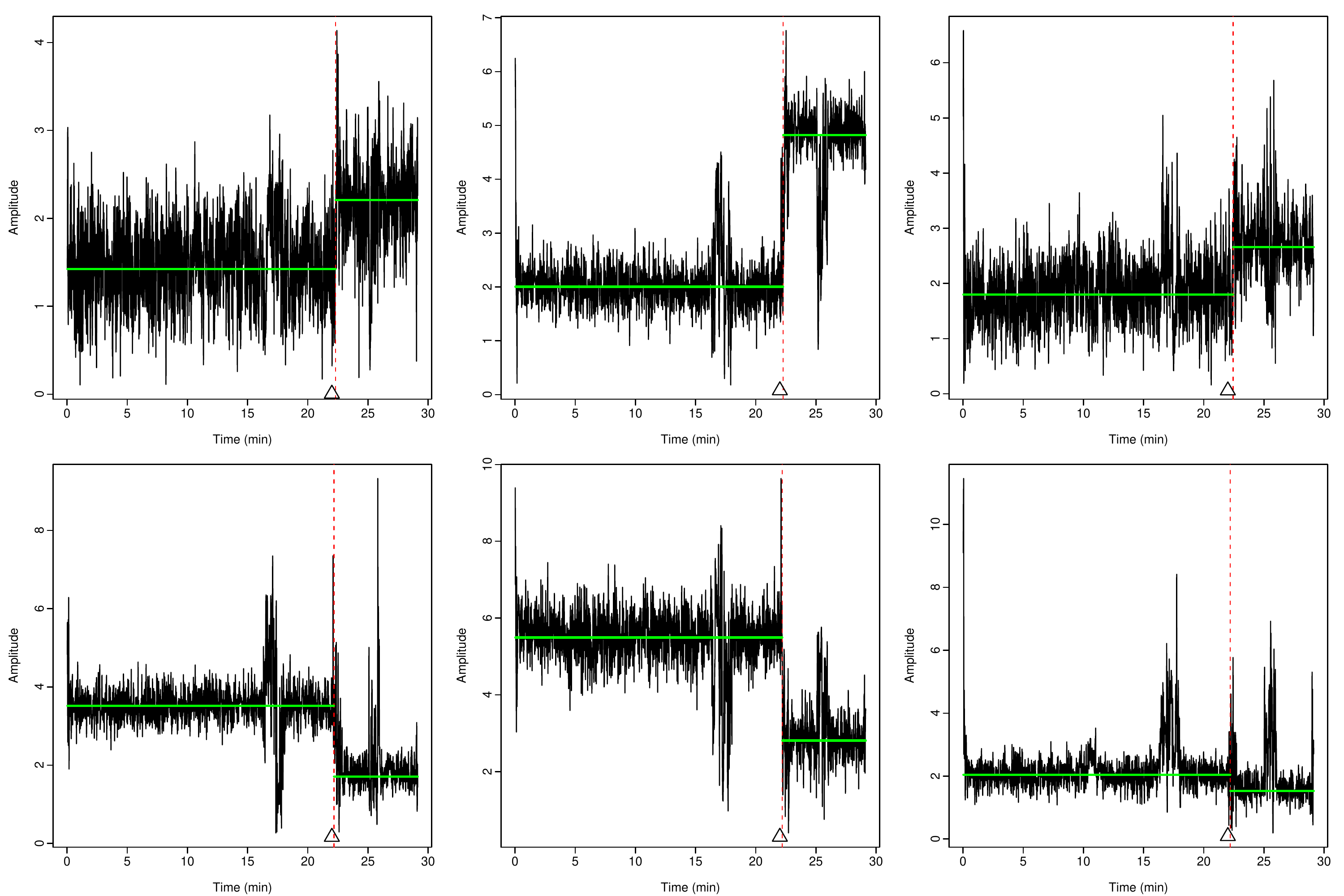}
 \caption{\label{Fig:Wireless} Examples from the analysis of the wireless tampering data. We show six examples of the data, with different structure before and after a change, and with different patterns of
 outliers caused by temporary environmental factors. In each case there is a single changepoint, after 22 minutes (denoted by the triangle). The inferred changepoint (vertical dashed line) and inferred mean function
 (green full horizonal line) from our method with the biweight loss function are shown in each case.
 }
\end{figure}

Figure \ref{Fig:Wireless} shows some time-series of channel state information (CSI) that has been extracted from the preamable from  a signal sent by a single IoT device. 
This data is taken from \cite{Bagci:2015}, where a
controlled experiment was performed, with an actual tampering event occurring after 22 minutes. Before this tampering event, there was movement of people around the device, which has a short-term effect on the
time-series data. 

In practice the channel state information from an IoT device is multi-dimensional, and we show time-series for 6 out of 90 dimensions. 
Whilst ideally we would jointly analyse the data from all 90 time-series that we get from the device, we will just consider analysing each time-series individually. Our interest is to see how viable it is
to use our approach, with the biweight loss, to accurately distinguish between tampering event and any effects due to temporary environmental factors. The six time-series we show each show different patterns, both
in terms of the change caused by tampering, and the effect of people walking near the device. As such they give a thorough testing of any approach.
We implemented %used 
the biweight loss with the SIC penalty for a change, and with $K$ chosen so that the minimum segment length (see Theorem \ref{thm:2}) corresponds to a period of 20 seconds. Results are shown in Figure
\ref{Fig:Wireless}, where we see that we accurately only detect the change that corresponds to the tampering event in all cases.

\section{Discussion}

We have presented an algorithm for detecting changepoints by minimising a penalised cost which measures fit to the data by a loss function that is piecewise quadratic. In particular we have shown that by using
bounded loss functions we can develop algorithms that are robust to the presence of arbitrarily large outliers. We particular recommend the use of the biweight loss function, and have shown that using such a loss
function can lead to consistent esimation of the number of changepoints and accurate estimation of their location under weak conditions on the noise distribution.

If we use the biweight loss we have to choose an appropriate value for $K$. To some extent this is a modelling decision, but a reasonable default is to choose this to be around 2 to 3 times an estimate of the 
standard deviation of the noise. This will mean the loss performs similarly to the square error loss, as most observations will be within $K$ of the segment location parameter, but with the added benefit of 
robustness to extreme outliers. 

We have shown that using the biweight loss with a penalty for adding a changepoint that is $C_1 \log(n)$ for some suitable constant $C_1$ can lead to consistent estimation of the number of changepoints. If $K$ is chosen 
as suggested, it is natural to choose $C_1$ to be similar to choices that are known to work well with the square-error loss, as we did within Section \ref{sec:accuracy}. Such choices are not guaranteed to produce large
enough constants to ensure consistency. If this is a concern, it is possible to use the idea behind that strengthened Schwarz information criteria of \cite{Fryzlewicz:2014}, and choose a penalty $C_1 (\log n)^{1+\epsilon}$ for some small
$\epsilon>0$.

Care must be taken if there are violations
of the IID assumption for the noise. In such cases it is known that consistent estimation of the number of changepoints is still possible if we appropriately inflate the penalty \cite[]{Lavielle:2000}, and
we would suggest using a similar inflation when using the biweight loss. Choosing how much to inflate is difficult in practice, and thus it makes sense to try a range of penalties \cite[which can be done efficiently, 
e.g. using the CROPS algorithm of][]{Haynes:2015}. For applications which involve analysing multiple similar data sets, we would recommend using a small set of training data to help choose an appropriate constant
\cite[see e.g.][]{Rigaill:2013}.

Finally, the joint choice of $K$ and $\beta$ can be informed by the minimum segment length that can be inferred for such a choice; see Theorem \ref{thm:2}. To have robustness to extreme outliers we need this 
minimum segment length to be greater than 1. Equally it should be chosen to be smaller than the shortest segment we wish to identify. This choice is linked to the question of how many similar observations would we
require before we would classify them as coming from a new segment as opposed to being correlated outliers.

\hl{
{\bf Acknowledgements} We thank Utz Roedig and Ethem Bagci for supplying and discussions around the wireless tampering data, and to Lawrence Bardwell for help with analysing this data. 
This work was supported by EPSRC grant EP/N031938/1 (StatScale) and an ATIGE grant from G\'{e}nopole.
}
%\bibliography{refs}
\putbib[refs]
\end{bibunit}
\pagebreak

\begin{center}
 {\bf Supplementary Material for Changepoint Detection in the Presence of Outliers}
\end{center}

\begin{appendices}
\setcounter{page}{1}

\begin{bibunit}[apalike]

\section{Proof of Theorem \ref{thm:Consistency} } \label{App:Consistency}

We first introduce notation, and present some basic properties that will be used within the proof. Let $\gamma(y;\theta)$ be the biweight loss, so $\gamma(y;\theta)=\min\{ (y-\theta)^2,K^2\}$, and we will keep $K$ fixed throughout. To simplify notation we will write $\gamma(y;0)=\gamma(y)$, and further note that $\gamma(y;\theta)=\gamma(y-\theta;0)=\gamma(y-\theta)$. We also note that $\gamma(y)$ is bounded, $0\leq \gamma(y) \leq K^2$, and satisfies a Lipschitz property
\[
|\gamma(y_1)-\gamma(y_2)|\leq 2K|y_1-y_2|.
\]

Let $Z_1,Z_2,\ldots$ be independent, identically distributed (IID) random variables whose distribution is that of the residual process within our model. The following random functions will play an important role
\[
X_i(\theta)= \gamma(Z_i-\theta)-\gamma(Z_i).
\]
By assumption (\ref{ass:1}), we have
\[
\mbox{E}\{X_i(\theta)\}=M(\theta)-M(0)=\geq \min\{c_1\theta^2,c_2\},
\]
for constants $c_1>0$ and $c_2>0$. Define $v(\theta)=\mbox{E}\{X_i(\theta)^2\}$.
By the bounded and Lipschitz properties of $\gamma(y)$ we have
\[
v(\theta)=\mbox{E}\left\{X_i(\theta)^2\right\} \leq \min\{4K^2\theta^2,K^4\}.
\] 
We will be interested in sums of $X_i(\theta)$, and thus define
\[
S_l(\theta)=\sum_{i=1}^l X_i(\theta).
\]
%It follows immediately from the IID assumption on the $X_i$s that we can upper bound the variance of $S_l(\theta)$ by $l\{4K^2\theta^2,K^4\}$.
To simplify notation in the following we will use $C_1$, $C_2$, $\alpha$, $\alpha'$ etc. to denote constants, and allow these constants to differ for different results and for different parts of the proofs.

Our consistency results states that asymptotically we estimate the correct number of changepoints and that each changepoint location is estimated within some degree of accuracy. 
The proof of consistency requires two preliminary results -- the proofs of which appear at the end of this section. The first is used to bound how much the un-penalised cost can be reduced by adding extra changepoints, 
and is used to show that asymptotically we do not over-estimate the number of changepoints.
\begin{lemma} \label{lem:1}
Under assumptions (\ref{ass:1}) and (\ref{ass:2}),
there exists strictly positive constants $\alpha$, $C_1$ and $\delta$ such for any $l\leq n$,
\[
\Pr\left\{ \min_\theta S_l(\theta) < -\alpha \log(n)\right\} \leq C_1n^{-2-\delta}.
\]
\end{lemma}

The second lemma will give us a probabilistic bound on the increase in the cost we get if we miss a true changepoint. This will be used to show both that asymptotically we cannot under-estimate the number of changepoints, and that any estimated segmentation must have an estimated changepoint ``close to" a real changepoint.
\begin{lemma} \label{lem:2}
Let $S_l'(\theta)$ be an independent copy of $S_l(\theta)$. Then under assumptions (\ref{ass:1}) and (\ref{ass:2}), for any $\Delta>0$, there exists positive constants, $C_1$, $C_2$ and $\alpha$ such that
\[
\Pr\left( \min_{\theta}\left\{ S_l(\theta)+S_l'(\theta-\Delta) \right\}  \leq  l \alpha \right) \leq  C_1\exp\{-C_2 l\}.
\]
\end{lemma}

Remember that our asymptotic results are for esimators of the changepoints obtained by minimising a penalised cost
\[
Q(y_{1:n};\hat{\tau}_{1:k})=\sum_{i=0}^k \left\{\mathcal{C}(y_{\hat{\tau}_i+1:\hat{\tau}_{i+1}})+\beta_n\right\},
\]
where we allow the penalty for adding a changepoint, $\beta_n$, to depend on $n$.  To simplify the notation we will write this
cost function as $Q(\hat{\tau}_{1:k})$ from now on. We will further introduce the notation
\[
Q_0(\hat{\tau}_{1:k})=\sum_{i=0}^k \mathcal{C}(y_{\hat{\tau}_i+1:\hat{\tau}_{i+1}}),
\]
to denote the un-penalised cost of a segmentation. We will extend both these functions to be defined when the argument is a set of un-ordered changepoints. Thus for an unordered vector $\hat{\tau}_{1:k}$, if we let $\hat{\tau}_{(1)}<\hat{\tau}_{(2)}<\cdots<\hat{\tau}{(k)}$ be the ordered changepoint locations then, for example, 
\[
Q_0(\hat{\tau}_{1:k})=\sum_{i=0}^k \mathcal{C}(y_{\hat{\tau}_{(i)}+1:\hat{\tau}_{(i+1)}}),
\]
where as before $\tau_{(0)}=0$ and $\tau_{(k+1)}=n$. Finally we will use, for example, $Q_0(\hat{\tau}_{1:k},\hat{\tau}'_{1:k'})$ to be the unpenalised cost for a segmentation with changepoints given by the union of $\hat{\tau}_{1:k}$ and $\hat{\tau}'_{1:k'}$.

{\bf Proof of Theorem \ref{thm:Consistency} }

The proof has three parts. Each showing that the probability of ``bad" segmentations, for a different definition of ``bad", goes to 0 as $n$ increases. In each case the idea is to show that the penalised cost for such ``bad" segmentations must be greater than
some other segmentation. The segmentation we compare with will either be the true segmentation of the data, or a slight adaptation of the ``bad" segmentation which adds one or more true changepoints.  Furthermore we can upper bound the un-penalised cost of, say, the correct segmentation as
\begin{equation} \label{eq:Q0}
 Q_0(\tau_{1:k_0})\leq \sum_{i=1}^n \gamma(Z_i),
\end{equation}
where $Z_1,\ldots,Z_n$ are the  noise random variables used to generate the data. This upper bound comes from the fact it is the cost if we fix the segment location 
parameter for each segment to be the true value for that segment, as opposed to minimising the cost over all possible location parameter values. A similar bound will 
also be used to bound the contribution to the cost function from individual segments that are subsets of a true segment. We also repeatedly used the fact that if we add
changepoints then the un-penalised cost can never increase.

{\bf Part 1}

The first part is to show that the estimated number of changepoints, $\hat{k}_n$, satisfies
\[
\Pr(\hat{k}>k_0)\rightarrow 0
\]
as $n\rightarrow\infty$, provided $\beta_n>C_1 \log(n)$ for a suitable $C_1$. To do this we use Lemma \ref{lem:1}.

Let ${E}_n$ be the event that 
\[
\min_{1\leq s < t \leq n} \min_{\theta} \sum_{i=s}^t \left\{\gamma(Z_i-\theta)-\gamma(Z_i)\right\} > - \alpha \log(n),
\]
where $\alpha$ is defined as in Lemma \ref{lem:1}. Then by Lemma \ref{lem:1} it follows that
 $\Pr(E_n)\rightarrow 1$ as $n\rightarrow \infty$. 
 
Now assume that $E_n$ occurs and
 consider a segmentation $\hat{\tau}_{1:\hat{k}}$ with $\hat{k}>k$. We have
\begin{eqnarray*}
Q_0(\hat{\tau}_{1:\hat{k}})  &\geq& Q_0(\hat{\tau}_{1:\hat{k}},\tau_{1:k_0}) \\
&=& \sum_{j=0}^{k^*} \mathcal{C}(Y_{\tau_{(j)}^*+1:\tau_{(j+1)}^*})\\
&=& \sum_{j=0}^{k^*} \min_\theta \sum_{i=\tau_{(j)}^*+1}^{\tau_{(j+1)}^*} \gamma(Z_i-\theta),
\end{eqnarray*}
where $k^*\leq k_0+\hat{k}$ is the number of distinct changepoints in the union of $\hat{\tau}_{1:\hat{k}}$ and $\tau_{1:k_0}$, and $\tau_{(1)}^*,\ldots,\tau_{(k^*)}^*$ are the ordered changepoint locations. 

As $E_n$ occurs, and using (\ref{eq:Q0}),
\begin{eqnarray*}
Q_0(\hat{\tau}_{1:\hat{k}}) - Q_0(\tau_{1:k_0}) &\geq& Q_0(\hat{\tau}_{1:\hat{k}})  - \sum_{i=1}^n \gamma(Z_i)\\
&=& \sum_{j=0}^{k^*} \min_\theta \sum_{i=\tau_{(j)}^*+1}^{\tau_{(j+1)}^*} \left\{
\gamma(Z_i-\theta)-\gamma(Z_i) \right\}\\
&>& -(k^*+1) \alpha \log(n).
\end{eqnarray*}
Thus for the penalised costs,
\[
Q(\hat{\tau}_{1:\hat{k}})-Q(\tau_{1:k_0}) > -(k^*+1) \alpha \log(n) + (\hat{k}-k_0)\beta_n.
\]
Now as $k^*\leq k_0+\hat{k}$ we have that the right-hand side will be positive if $\beta_n > 2(k_0+1)\alpha\log(n)$. 
For such a $\beta_n$ we have that the true segmentation will have a lower penalised cost than any segmentation with more than $k_0$ changepoints if $E_n$ occurs. As $\Pr(E_n)\rightarrow 1$ we have that $\Pr(\hat{k}_n>k_0)\rightarrow 0$ as required.

{\bf Part 2}

The second part of the proof is to show that asymptotically we cannot underestimate the number of changepoints if $\beta_n=o(n)$.

Let $l$ be the largest integer less than half the length of the smallest segment. Let
$\tilde{E}_n$ be the event that
\[
\min_{j\in\{1,\ldots,k_0\}}
\min_{\theta}
\sum_{i=\tau_j-l+1}^{\tau_j+l}
\left\{
\gamma(Y_i-\theta)-\gamma(Z_i)
\right\} > \alpha' l,
\]
for some suitable $\alpha'$ which is specified below. Now
\begin{eqnarray*}
\min_{\theta'} \sum_{i=\tau_j-l+1}^{\tau_j+l} \gamma(Y_i-\theta') &=&
\min_{\theta'} \left\{
\sum_{i=1}^{l} \gamma(\mu_{j-1}+Z_{\tau_j-l+i}-\theta') + \gamma(\mu_{j}+Z_{\tau_j+i}-\theta')
\right\} \\
&=&\min_{\theta} \left\{
\sum_{i=1}^{l} \gamma(Z_{\tau_j-l+i}-\theta) + \gamma(Z_{\tau_j+i}-\theta+\Delta_{j})
\right\}
\end{eqnarray*}
where $\mu_{j-1}$ and $\mu_j$ are the segment location parameters before and after the $j$th changepoint and $\Delta_j=\mu_j-\mu_{j-1}$ is
the change at the $j$th changepoint. We have $|\Delta_j| >0$ for each $j$ as by assumption $\mu_j\neq \mu_{j-1}$.

If we choose $\alpha'>0$ to be a value such that the statement of Lemma \ref{lem:2} holds for all $\Delta=|\Delta_j|$, then, as $l\rightarrow\infty$ as $n\rightarrow\infty$, we have that $\Pr(\tilde{E}_n)\rightarrow 1$ as $n\rightarrow \infty$. 

Now assume event $\tilde{E}_n$ occurs. Consider a segmentation $\hat{\tau}_{1:\hat{k}}$ with $\hat{k}<k_0$. As $\hat{k}<k_0$
there exists a changepoint $\tau_j$ such that no estimated changepoint is within $l$ of $\tau_j$. We can bound the un-penalised cost of the segmentation $\hat{\tau}_{1:\hat{k}}$ by comparing with a segmentation that  also includes changes at $\tau_j-l$, $\tau_j$ and $\tau_{j}+l$:
\begin{eqnarray*}
Q_0(\hat{\tau}_{1:\hat{k}})  &\geq& Q_0(\hat{\tau}_{1:\hat{k}},\tau_j-l,\tau_j+l) \\
&\geq& Q_0(\hat{\tau}_{1:\hat{k}},\tau_j-l,\tau_j,\tau_j+l) + 
\min_{\theta}
\sum_{i=\tau_j-l+1}^{\tau_j+l}
\left\{
\gamma(Y_i-\theta)-\gamma(Z_i)
\right\} \\
&>&Q_0(\hat{\tau}_{1:\hat{k}},\tau_j-l,\tau_j,\tau_j+l) + \alpha' l.
\end{eqnarray*}
The second inequality comes from a bound on the reduction in the un-penalised cost from adding a changepoint at $\tau_j$. 
It is obtained using (\ref{eq:Q0}) for data $Y_{\tau_j-l+1:\tau_j+l}$ with a change at $\tau_j$.

Thus the penalised cost satisfies
\[
Q(\hat{\tau}_{1:\hat{k}}) -Q(\hat{\tau}_{1:\hat{k}},\tau_j-l,\tau_j+l) >\alpha l -3\beta_n, 
\]
where the $3\beta_n$ term comes from adding 3 changepoints.

Now $l$ increases linearly in $n$, thus the right-hand side is positive for large enough $n$ providing $\beta_n=o(n)$. Thus for large enough $n$, the segmentation $\hat{\tau}_{1:\hat{k}}$ cannot minimise the penalised cost if $\tilde{E}_n$ occurs. This
argument applies for all such segmentations with fewer than $k_0$ changes. Thus as $\Pr(\tilde{E}_n)\rightarrow 1$ we have that $\Pr(\hat{k}_n<k_0)\rightarrow 0$ as required.

{\bf Part 3}

Taken together, the first two parts of the proof show that $\Pr(\hat{k}_n=k_0)\rightarrow1$. The third part of the proof relates to the accuracy of the estimated changepoint locations. As $\Pr(\hat{k}_n=k_0)\rightarrow1$ we need consider only segmentations of the data with $k_0$ changepoints. 

We introduce an event $\bar{E}_n$ similar to $\tilde{E}_n$ used in the second part of the proof, but with $l=\lfloor C_2 \log(n)\rfloor$. Consider this event only for $n$ sufficiently large that $l$ is less than the smallest segment length. That is $\bar{E}_n$ is the event
\[
\min_{j\in\{1,\ldots,k_0\}}
\min_{\theta}
\sum_{i=\tau_j-l+1}^{\tau_j+l}
\left\{
\gamma(Y_i-\theta)-\gamma(Z_i)
\right\} > \alpha' l,
\]
where $\alpha'$ is a constant. By the same argument as in Part 2 of the proof, we can choose $\alpha'>0$ such that, by Lemma \ref{lem:2}, $\Pr(\bar{E}_n)\rightarrow 1$.

Now assume both $\bar{E}_n$ and ${E}_n$, which was defined in the first part of the proof, occur. Consider any segmentation with $k_0$ changepoints, $\hat{\tau}_{1:k_0}$, 
for which
\[
\max_{i=1,\ldots,k_0} \left\{ \min_{j=1,\ldots,\hat{k}_n} \left| \tau_i-\hat{\tau}_j \right| \right\} > C_2 \log(n). 
\]
Let $i$ be the index of  a changepoint for which 
\[
\min_{j=1,\ldots,\hat{k}_n} \left| \tau_i-\hat{\tau}_j \right| > C_2 \log(n),
\]
and let $\tau'_{1:k_0+1}=(\tau_{1:i-1},\tau_i-l,\tau_i+l,\tau_{i+1:k_0})$. This is the set of all actual changepoints except the $i$th one together with 
the two changes at a distance $l$ from $\tau_i$. We will compare the penalised cost of segmentation $\hat{\tau}_{1:k_0}$ with the cost of the true segmentation:
\begin{eqnarray*}
Q(\hat{\tau}_{1:k_0})-Q(\tau_{1:k_0}) &=&
Q_0(\hat{\tau}_{1:k_0})-Q_0(\tau_{1:k_0}) \\
 & \geq & Q_0(\hat{\tau}_{1:k_0},\tau'_{1:k_0+1})-\sum_{j=1}^n \gamma(Z_j),
 \end{eqnarray*}
where the last inequality comes from using (\ref{eq:Q0}). We can write the final term as a sum over the $k^*+1$ segments of the segmentation with change-points given by the union
of $\hat{\tau}_{1:k_0}$ and $\tau'_{1:k_0+1}$. This union contains all true changepoints except for $\tau_i$. Thus the $k^*$ segments that do not contain $\tau_i$ will contribute a term
\[
\min_{\theta} \left\{ \sum_{j=s}^t \gamma(Z_j-\theta)-\gamma(Z_j) \right\},
\]
for some suitable $t>s$, to this sum. As event $E_n$ holds, this term is bounded below by $-\alpha\log(n)$.
The remaining term is of the form
\[
\min_{\theta}
\sum_{j=\tau_i-l+1}^{\tau_i+l}
\left\{
\gamma(Y_j-\theta)-\gamma(Z_j)
\right\}
\]
and this is bounded below by $\alpha'l$. Thus we have
\[
Q(\hat{\tau}_{1:k_0})-Q(\tau_{1:k_0}) >\alpha'\lfloor C_2 \log(n) \rfloor - (2k_0+1)\alpha \log (n),
\]
where the first term on the right-hand side comes from the bound on the term for the segment that include $\tau_i$, the other term is from the $k^*\leq 2k_0+1$ other terms. Thus for $C_2>(2k_0+1)\alpha/\alpha'$ this will be positive for large enough $n$, and the penalised cost would prefer the true segmentation to $\hat{\tau}_{1:k_0}$. This will hold for all $\hat{\tau}_{1:k_0}$ for which the error in estimating at least one changepoint is greater than $C_2 \log(n)$. Thus if $E_n$ and $\bar{E}_n$ hold then for large enough $n$ all
segmentations with $k_0$ changepoints will have
\[
\max_{i=1,\ldots,k_0} \left\{ \min_{j=1,\ldots,\hat{k}_n} \left| \tau_i-\hat{\tau}_j \right| \right\} \leq C_2 \log(n), 
\]
as required.
\hfill $\Box$

We finish by giving the proofs of our two lemmas.

 { \bf Proof of Lemma \ref{lem:1}.}

The proof proceeds in two parts. The first involves considering $\min_\theta S_l(\theta)$ for $-2K\leq \theta \leq 2K$, and the second considers $|\theta|>2K$. In each case we need to show that there is a sufficiently large $\alpha$ such that the probability of the minimum of $S_l(\theta)$, for the respective ranges of $\theta$, being less than $-\alpha \log(n)$ is bounded by a constant times $n^{-2-\delta}$.

For the first part, we initially give a bound on the probability of small values for $S_l(\theta)$ for a fixed value of $\theta$. To do this we use Bennett's inequality \cite[see Theorem 2.9 in][]{boucheron2013concentration} for 
\[
\tilde{S}=\sum_{i=1}^l \{X_i(\theta)-M(\theta) \}=S_l(\theta)-lM(\theta).
\]
As $X_i(\theta)\geq -K^2$ and $\mbox{E}\{X_i(\theta)^2\}=v(\theta)$, Bennett's inequality gives us
\[
\Pr(\tilde{S}\leq -t)\leq \exp \left( - \frac{t^2}{2(lv(\theta)+K^2t/3)}
\right).
\]
Now the event $S_l(\theta) \leq -\alpha' \log(n)$ is equivalent to the event $\tilde{S} \leq -\alpha' \log(n)-l M(\theta)$, so
\begin{eqnarray*}
\Pr\left\{S_l(\theta) \leq -\alpha' \log(n)\right\}& \leq &\exp \left( - \frac{\{\alpha' \log(n) + l M(\theta)\}^2}{2[lv(\theta)+K^2\{\alpha'n \log(n)+l M(\theta)\} /3]}\right) \\
&=& \exp\left\{-\alpha' \log(n) \left[ 
\frac{\alpha' \log(n) + 2lM(\theta) + l^2M(\theta)^2/\{\alpha' \log(n)\}}
{2lv(\theta)+2K^2\{\alpha'\log(n)+lM(\theta)\}/3}
\right] \right\}.
\end{eqnarray*}
Now there exists a $D>0$ such that $v(\theta)< D M(\theta)$ for all $\theta$. If we further write $\psi=l M(\theta)/\{\alpha' \log(n)\}$, and note that $\psi\geq 0$, we get
\begin{eqnarray*}
\Pr\left\{S_l(\theta) \leq -\alpha' \log(n)\right\} &\leq& \exp\left\{-\alpha' \log(n) \left( 
\frac{1 + 2\psi + \psi^2}
{2K^2/3+(2D+2K^2/3)\psi}
\right) \right\} % \\
%&\leq& \exp\left\{-\alpha' \log(n)\left(
% \frac{3}{3D+2K^2}
%\right)\right\}.
\end{eqnarray*}
For $\psi\geq 0$, we can lower bound the bracketed term in the exponent by some strictly positive constant.
Thus for sufficiently large $\alpha'$ we will have that this probability is less than $n^{-3-\delta}$.

Now the above argument is for $S_l(\theta)$ at a specific value of $\theta$, but we are interested in $\min_{|\theta|\leq 2K} S_l(\theta)$. To deal with this minimisation we use the Lipschitz property of $\gamma(y)$, which gives that
\[
|S_l(\theta)-S_l(\theta')| \leq 2Kl |\theta-\theta'|.
\]
Thus if we choose $\epsilon>0$ we can partition the interval $[-2K,2K]$ into $\lceil 4K^2n/\epsilon \rceil$ intervals of width at most $\epsilon/(Kn)$. For one such interval, as $l\leq n$,
\[
\Pr\left\{ \min_{\theta:|\theta-\theta'|<\epsilon/(2Kn)} S_l(\theta) \leq -\alpha\log(n)\right\} \leq
\Pr\left\{S_l(\theta') \leq - \alpha\log(n) + \epsilon\right\}.
\]
Thus if we choose $\alpha>\alpha'$ such that, for sufficiently large $n$, $-\alpha\log(n)+\epsilon< -\alpha' \log(n)$, we have that, for sufficiently large $n$, this probability is less than $n^{-3-\delta}$. We require the bound to hold for all $O(n)$ intervals, and thus we get that 
\[
\Pr\left( \min_{\theta: |\theta|\leq 2K} S_l(\theta) \leq -\alpha \log(n) \right) \leq C_2 \frac{1}{n^{2+\delta}},
\]
for some constant $C_2$, as required.

We now consider the case where $|\theta|>2K$. We will use the following bound
\[
X_i(\theta)=\gamma(Z_i-\theta)-\gamma(Z_i) 
\geq \left\{ \begin{array}{cl} K^2-Z_i^2 & \mbox{if $|Z_i|<K$}, \\
-K^2 & \mbox{otherwise},
\end{array}\right.
\]
for $|\theta|>2K$. 
Let $\tilde{X}_i$ be the random variable defined by the right-hand side of this equation, and $\tilde{S}_l=\sum_{i=1}^l \tilde{X}_i$. Then we have
\[
\Pr\left\{\min_{\theta: |\theta|>2K} S_l(\theta) < -\alpha \log(n)\right\} \leq \Pr\left\{ \tilde{S}_l < -\alpha \log(n)\right\}.
\]

Using the notation in assumption (\ref{ass:2}), where $p=\Pr(|Z_i|>K)$ and, if $\tilde{Z}$ is a random variable whose distribution is that of $Z_i$ conditional on $|Z_i|\leq K$, $\sigma^2=\mbox{E}(\tilde{Z}^2)$, we have
\[
\mbox{E}(\tilde{X}_i)\geq K^2(1-2p)-(1-p)\sigma^2,
\]
which we will denote by $\tilde{M}$. By assumption (\ref{ass:2}), $\tilde{M}>0$. As $|\tilde{X}_i|\leq K^2$ we have $\mbox{E}(\tilde{X}_i^2)< K^4$. 

Thus using Bennett's inequality, and a similar argument to above,
\begin{eqnarray*}
\Pr\left\{ \tilde{S}_l < -\alpha \log(n) \right\}& \leq& \exp\left\{
\frac{-\{\alpha\log(n)+l\tilde{M}\}^2}
{ 2[lK^4+K^2\{\alpha\log(n)+l\tilde{M}\}/3] }
\right\} \\
&=& \exp\left\{ -\alpha \log(n) \left[
\frac{(1+\psi)^2}
{ 2(D\psi+K^2(1+\psi)/3) } \right]
\right\},
\end{eqnarray*}
where $\psi=l\tilde{M}/\{\alpha \log(n)\}$, and $D=K^4/\tilde{M}$. We can bound from below the bracketed term in the exponent for all $\psi>0$. Thus we can choose $\alpha$ large enough so that the right-hand side is less than $n^{-2-\delta}$ for all $l$ as required. \hfill $\Box$

 { \bf Proof of Lemma \ref{lem:2}.}

Fix $\Delta>0$. Let $Z_1',Z_2',\ldots$ be IID copies of $Z_1$. Let $X'_i(\theta)=\gamma(Z_i'-\theta)-\gamma(Z_i')$. Then
\[
S_l(\theta)+S_l'(\theta-\Delta)=\sum_{i=1}^l \left\{\gamma(Z_i-\theta)-\gamma(Z_i)+\gamma(Z_i'-\theta+\Delta)-\gamma(Z_i') \right\}=\sum_{i=1}^l \left\{X_i(\theta)+X_i'(\theta-\Delta)\right\}.
\]
As we have fixed $\Delta$, we will write $\bar{X}_i(\theta)=X_i(\theta)+X_i'(\theta-\Delta)$, and $\bar{S}_l(\theta)=\sum_{i=1}^l \bar{X}_i(\theta)$. We are thus interested in bounding
\[
\Pr \left( \min_{\theta} \bar{S}_l < \alpha l \right).
\]

To get the required bound, our argument will closely follow that of Lemma \ref{lem:1}. In particular we will show that we get the required exponential bound on this probability first for the case where we minimise $\theta$ over $|\theta|\leq \Delta+2K$ and second for the case where we minimise $\theta$ over $|\theta|>\Delta+2K$.

By condition (\ref{ass:1}) we have that, for any $\theta$,
\[
\mbox{E}\{\bar{X}_i(\theta)\}=M(\theta)+M(\theta-\Delta)\geq\min\left\{c_1\frac{\Delta^2}{2},c_2\right\},
\]
and we denote the right-hand side, which is a constant as we have fixed $\Delta$, by $\bar{M}$. As $|\bar{X}_i(\theta)|\leq 2K$ we have $\mbox{E}\{\bar{X}_i(\theta)^2\}\leq 4K^2$.

Thus, for any chosen $\theta$, we can use Bennett's inequality to bound the lower tail for $\bar{S}_l(\theta)$. We get
\begin{eqnarray*}
\Pr\left\{\bar{S}_l(\theta) \leq l \bar{M}/2\right\}  &=& \Pr\left\{\bar{S}_l(\theta)-l\bar{M} \leq -l\bar{M}/2\right\} \\
&\leq&\exp\left\{ -\frac{l^2\bar{M}^2}{8(4lK^4+Kl\bar{M}/3)} \right\} \\
&\leq&\exp\left\{ -l C_2'  \right\},
\end{eqnarray*}
where $C_2'=\bar{M}^2/(32K^4+8K\bar{M}/3)$.

The above inequality is for a single, fixed, value of $\theta$. To deal with the minimisation over $\theta$ such
that $|\theta|\leq \Delta +2K$ we partition this interval into intervals of length $\bar{M}/(8K)$, or smaller. As 
$\bar{X}_i(\theta)$  is Lipschitz in $\theta$, with constant $4K$, we have that for $\theta$ in an interval of length $\bar{M}/(8K)$ the value of $\bar{S}_l(\theta)$  can vary by at most $l\bar{M}/4$ from the value in the centre of the interval. Thus if
$\bar{S}_l(\theta) \leq l \bar{M}/2$ then we must have $\bar{S}_l(\theta') \leq l \bar{M}/4$ for all $\theta'\in[\theta-\bar{M}/(8K),\theta+\bar{M}/(8K)]$.

Our partition of the region $|\theta|\leq \Delta +2K$ requires a fixed number of intervals of length at most $\bar{M}/(8K)$. Thus we have
\[
\Pr\left\{\min_{\theta:|\theta|<\Delta+2K} \bar{S}_l(\theta) \leq l \bar{M}/4\right\} \leq C_1'\exp\left\{ -l
C_2' \right\},
\]
where $C_1$ is the number of intervals. %, $\lceil (\Delta+2K)(8K)/\bar{M} \rceil$.

We now consider the case where we minimise $\bar{S}_l(\theta)$ over $|\theta|>\Delta+2K$. As in the proof to Lemma \ref{lem:1}, for such $\theta$ we can  bound from below both $X_i(\theta)$ and $X_i(\theta-\Delta)$ by a random variable that takes the value $K^2-Z_i^2$ if $|Z_i|<K$ and that takes the value $-K^2$ otherwise. Let $\tilde{X}_i$ denote this random variable, and $\tilde{X}_i'$ the corresponding random variable defined from $Z'_i$. We then have that for any $\alpha$
\[
\Pr\left\{\min_{\theta:|\theta|>\Delta+2K} \bar{S}_l(\theta) \leq \alpha l\right\} \leq
 \Pr\left\{\sum_{i=1}^l (\tilde{X}_i+\tilde{X}_i') \leq \alpha l \right\}.
\]
Defining $\tilde{M}$ to be the lower bound on $\mbox{E}(\tilde{X}_i)$ that was given in the proof of Lemma \ref{lem:1} we
have, using Bennett's inequality again,
\begin{eqnarray*}
\Pr\left\{\sum_{i=1}^l (\tilde{X}_i+\tilde{X}_i') \leq l \tilde{M}  \right\} & = &
\Pr\left\{\sum_{i=1}^l (\tilde{X}_i+\tilde{X}_i' - 2 \tilde{M}) \leq -l\tilde{M} \right\} \\
&\leq& \exp\left\{ - \frac{l^2 \tilde{M}}{2(4K^4l+ 2K\tilde{M}l/3) }
\right\}.
\end{eqnarray*}
The final expression can be written as $\exp\{-lC_2''\}$ for some suitable $C_2''$. 

Putting this together with the result when we minimise over $|\theta|\leq \Delta+2K$  we get the required inequality with 
$\alpha=\min\{\bar{M}/4,\tilde{M}\}$,
$C_1=C_1'+1$ and $C_2=\min\{C_2',C_2''\}$. \hfill $\Box$

\section{Proofs from Section \ref{sec:Model}}

{\bf Proof of Theorem \ref{thm:1}.}
Consider any segmentation of the data that does not include changepoints at both $t-1$ and $t$. We will show that for sufficiently large $y_t$, that adding changepoints at both $t-1$ and $t$ (or at just one of these if the segmentation has a change at the other time-point) will reduce the penalised cost. Thus the optimal segmentation must have changes at both $t-1$ and $t$.

Let the segment in the original segmentation that contains $y_t$ contain $y_{s:u}$ for $s<t$ and $u>t$. The change in cost between this segmentation and one with changepoints added at $t-1$ and $t$ will be
\begin{equation} \label{eq:2}
\min_\theta \sum_{i=s}^{t-1} \gamma(y_i;\theta)
+\min_\theta \sum_{i=t+1}^{u} \gamma(y_i;\theta)+2\beta-
\min_\theta \sum_{i=s}^{u} \gamma(y_i;\theta).
\end{equation}
Here we have used the fact that the only change in cost will be for fitting $y_{u:t}$ as the other segmentations are unchanged. The new segmentation has segments which include $y_{s:t-1}$, $y_t$ and $y_{t+1:u}$ and introduces two extra changepoints. The cost of the segmentation which just includes $y_t$ is 0. We need to show that for large enough $y_t$ this will always be negative. To see this we use the fact that 
\[
\min_\theta \sum_{i=s}^{u} \gamma(y_i;\theta) \geq \min \{\gamma(y_t;(y_t+y_{t+1})/2),\gamma(y_{t+1};(y_t+y_{t+1})/2)\},
\]
and this tends to infinity as $y_t\rightarrow \infty$. The other terms in (\ref{eq:2}) do not depend on $y_t$, and hence (\ref{eq:2}) will be negative for sufficiently large $y_t$.

The proof for $s\leq t$ or $u\geq t$ follows similarly. $\hfill \blacksquare$

{\bf Proof of Theorem \ref{thm:2}.}
We will prove this by showing that for any segmentation with a segment which is shorter than $\beta/K$, we can reduce the penalised cost by removing the changepoint at either the start or end of the segment.

Consider a segmentation of the data with neighbouring segments $y_{s:t}$ and $y_{t+1:u}$. If either $t-s<\beta/K$ or $u-t-1<\beta/K$ then removing the changepoint at $t$ will reduce the penalised cost. Without loss of generality assume $t-s<\beta/K$. The change in cost of removing the changepoint at $t$ is
\begin{eqnarray*}
\lefteqn{
\min_\theta \sum_{i=s}^{u} \gamma(y_i;\theta) 
-\min_\theta \sum_{i=s}^{t} \gamma(y_i;\theta)
-\min_\theta \sum_{i=t+1}^{u} \gamma(y_i;\theta)-\beta}\\
&< & 
(t-s)K+ \min_\theta \sum_{i=t+1}^{u} \gamma(y_i;\theta) 
-\min_\theta \sum_{i=s}^{t} \gamma(y_i;\theta) 
-\min_\theta \sum_{i=t+1}^{u} \gamma(y_i;\theta)-\beta \\ 
&=& (t-s)K-\beta
\end{eqnarray*}
The first  inequality uses the fact that the cost for segmenting $y_{s:u}$ is less than the cost if we fix $\theta$ to the optimal value for segmenting $y_{t+1:u}$. We then bound the contribution of the cost for each of $y_{s:t}$ by $K$ .The second inequality the fact that the costs are positive. We have that this change in cost is negative, as required, because $(t-s)<\beta/K$.
$\hfill \blacksquare$

\section{Proofs from Section \ref{sec:CompCost}}

{\bf Proof of Theorem \ref{thm:3}.}
Given that $\gamma$ is convex then, following the proof of the worst case complexity of the pDPA (\cite{rigaill2015pruned} appendix A),
we get that the function $\FC_t(\theta)$ can be described in at most $2t-1$ intervals such that for each interval
there is a single value for the best time of the most recent changepoint. That is we can define intervals
$I_k$ for $k=1,\ldots,K$, with $K<2t$, such that for a given $k$ there exists $s$ such that 
$$\forall \theta \ \in  I_k = [s_k, e_k] \qquad Q_s + \sum_{i=s+1}^n \gamma(y_i;\theta)%\Fc_{s+1:n}(\theta)
+ \beta = \FC_t(\theta).$$
%That is on each interval there is a single value for the best time of the most recent changepoint.

All $\Fc_{t+1:n}(\theta)$ are themselves 
defined in pieces, as sums of $\gamma(y_i, \theta)$.
For each $\gamma(y_i, \theta)$ we need to consider $L$ intervals 
and thus $L-1$ points between intervals. 
We call these points $T_{i,j}$ for $j$ in $1$ to $L-1$. %Considering all $i$ 
%from $1$ to $t$ we have $t (l-1)$ such points. 
For a given $I_k=[s_k, e_k]$ 
define $N_k$ to be the number of the $T_{i,j}$ points that are 
in the open interval $(s_k, e_k)$. 
As all $I_k$ are disjoint then each $T_{i,j}$ can only appear in a single interval. So
$\sum_{k=1}^{K} N_k\leq t(L-1)$. 
%(we do not need to consider ${y_i+T_j}$ that are exactly  equal to $s_k$ or $e_k$ as they will not define additional intervals).

On $I_k$ R-FPOP will thus define $\FC_t(\theta)$ using $N_k+1$ intervals. 
Thus R-FPOP uses 
\[\sum_{k=1}^{K} (N_k+1) \leq t(L-1)+2t-1\]
intervals to define $\FC_t(\theta)$
$\blacksquare$.

{\bf Proof of Corollary \ref{corollary:lgamma}.} 
The space complexity is obtained using the fact that the number 
of intervals is bounded by $2n-1 + n(L-1)$ and the fact that on 
each interval we need to store a quadratic.

As for the time complexity, at step $t$ R-FPOP needs to consider 
at most $2t-1 + t(L-1)$ ordered intervals. The key to bounding the time-complexity is that
we can split up the operations at step $t$ of R-FPOP into a series of operations on each interval.
The cost for each operations on an interval is $\Oc(1)$, and as there are $\Oc(t)$ intervals the 
overall cost of the $t$ iteration is $\Oc(t)$. The details are as follows.

On each of these intervals we will compute the roots of $\FC_t(\theta)$ minus a constant function. 
According to the number of roots the interval will 
be split in at most three (because on each interval $\FC_t(\theta)$ is convex). 
Calculating these roots can  be done in $\Oc(1)$ for each interval. 
Thus we get a new ordered list of intervals in $\Oc(t)$ time. 
Iterating on this list, successive intervals having the same analytical 
decomposition can be fused in $\Oc(t)$ time.
Once this is done, it is possible to add $\gamma(y_t, \mu)$ to $\FC_t(\mu)$ 
on all intervals in $\Oc(t)$ time.
Finally the minimum of $\FC_{t+1}(\theta)$ on each interval is recovered
in $\Oc(t)$ time. 
Overall, step $t$ is performed in $\Oc(t)$. Summing over all $t$
we get a quadratic complexity $\blacksquare$

{\bf Proof of Theorem \ref{thm:4}.}
From observations up to and including time $t$, the biweight loss function will define (at most) $2t+1$ intervals which
are separated by points of non-diffentiability of the loss function, $y_i-K$ and $y_i+K$ for $i=1,\ldots,t$.
Denote these intervals as $I_k$, $k=1,\ldots,2t+1$. 
Following the proof of the worst case complexity of the pDPA (\cite{rigaill2015pruned} appendix A)
we see that R-FPOP will need at most $2t-1$ intervals to describe $\FC_{t}(\theta)$ on any given $I_k$.
Summing over $k$  we recover a quadratic complexity. $\blacksquare$

{\bf Proof of Corollary \ref{cor:2}}
%First we note that property \ref{prop:worst} is true for the outlier loss.
The proof followa the proof of corollary \ref{corollary:lgamma} replacing the $\Oc(n)$ bound on the number of intervals
by the $\Oc(n^2)$ bound on the number of intervals. $\blacksquare$

\section{Pseudo Code for R-FPOP} \label{App:R-FPOP}

Here we provide some pseudo-code of the algorithm. In FPOP the algorithm is working on candidate change-points $\tau$.
Each of those change can be associated to one or more intervals on which it is optimal. In R-FPOP we directly work
on these intervals. R-FPOP essentially relies on three sub-routines or sub-algorithms
that manipulate the functions $Q^*_{t}(\theta)$ and $Q_{t}(\theta)$:
\begin{itemize}
\item Sub-routine \ref{algo_sum} to compute the function $Q_t(\theta) = Q^*_t(\theta) + \gamma(y_t, \theta) $ ;
\item Sub-routine \ref{algo_min} to recover the minimum and best change of the function $Q_t(\theta)$ ;
\item Sub-routine \ref{algo_compare} to compare the function $Q_t(\theta)$ to a constant and recover the function $Q^*_{t+1}(\theta)$.
\end{itemize}
Using these sub-routines the pseudo-code of R-FPOP is fairly simple and provided below in Algorithm \ref{algo_RFpop}.
We then provide the pseudo-code of each sub-routine in Algorithms \ref{algo_sum}, \ref{algo_min} and \ref{algo_compare}.

%%%%%%%%%%%%%%%%%%%%%%%%%%%% R-FPop

\IncMargin{1em}
\begin{algorithm}[!h]
\SetKwData{Left}{left}\SetKwData{This}{this}\SetKwData{Up}{up}
\SetKwFunction{Union}{Union}\SetKwFunction{FindCompress}{FindCompress}
\SetKwInOut{Input}{Input}\SetKwInOut{Output}{Output}
\Input{Set of data of the form $\mathbf{y}_{1:n}=(y_1,\hdots,y_n)$,\\
A measure of fit $\gamma(\cdot, \cdot)$ dependent on the data and the mean,\\
A penalty $\beta$ which does not depend on the number or location of the changepoints.}
\BlankLine
Set $Q^*_{1}(\theta) = 0$ on the interval $(\min_i \{ y_i\}, \max_i \{ y_i \}]$ \;

\For{$t=1,\hdots,n$}{

Compute $Q_{t}(\theta)= Q^*_{t}(\theta)+ \gamma(y_{t},\theta)$ using sub-routine \ref{algo_sum} \;
Compute $ Q_t$ and $\tau_t$ the $\min$ and $\arg \min$ of $ Q_{t}(\theta) $ using sub-routine \ref{algo_min} \;
Set $cp(t) = (Q_t, \tau_t)$ \;
Compute $Q^*_{t+1}(\theta)$ by comparing the function $Q_{t}(\theta)$ to $Q_t+\beta$ using sub-routine \ref{algo_compare} \;
}
\Output{The changepoints recorded in $cp(n)$.}
\BlankLine
\caption{Robust FPOP algorithm}\label{algo_RFpop}
\end{algorithm}\DecMargin{1em}

%%%%%%%%%%%%%%%%%%%%% ADD

\IncMargin{1em}
\begin{algorithm}[!h]
\SetKwData{Left}{left}\SetKwData{This}{this}\SetKwData{Up}{up}
\SetKwFunction{Union}{Union}\SetKwFunction{FindCompress}{FindCompress}
\SetKwInOut{Input}{Input}\SetKwInOut{Output}{Output}
\Input{$Q^*_t(\theta)$ a function defined on $N^*_t$ intervals: $(a_i^{(t)}, b_i^{(t)}]$ \\
For each interval we also have a change: $\tau_i^{*(t)}$ \\
$\gamma(y_t, \theta)$ a function defined on l intervals: $(c_j^{(t)}, d_j^{(t)}]$ \\
We assume $a_1^{(t)} = c_1^{(t)} $ and $b_{N^*_t}^{(t)} = d_l^{(t)} $ 
}

\BlankLine
Set current number of intervals for $Q_t(\theta)$ to $N_t=0$ \;
Set current $Q^*_t$ interval to $i=1$ \;
Set current $\gamma(y_t, \theta)$ interval to $j=1$ \;

\While{  $i \leq N^*_t$ and $j \leq l$ }{
$N_t = N_t +1$ \;
Create the new interval $(A_{N_t}^{(t)}, B_{N_t}^{(t)}] =  \left( \max \{ a_i^{(t)}, c_j^{(t)} \}, \min \{b_i^{(t)}, d_j^{(t)} \} \right]$ \;
For $\theta$ in interval $(A_{N_t}^{(t)}, B_{N_t}^{(t)}]$ set $Q_t(\theta) = Q^*_t(\theta) + \gamma(y_t, \theta) $ \;
Set $\tau_{N_t}^{(t)} = \tau_{i}^{*(t)}$ \;
\If{  $B_{N_t}^{(t)} = b_i^{(t)}$ }{
$i=i+1$\;}
\If{  $B_{N_t}^{(t)} = d_j^{(t)}$ }{
$j=j+1$\;}

}
\Output{The function $Q_t(\theta) = Q^*_t(\theta) + \gamma(y_t, \theta)$ defined on $N_t$ intervals: $(A_i^{(t)}, B_i^{(t)}]$}
\BlankLine
\caption{Sub-routine to compute $Q_t(\theta) =  Q^*_t(\theta) + \gamma(y_t, \theta)$}\label{algo_sum}
\end{algorithm}\DecMargin{1em}

%%%%%%%%%%%%%%%%%%%%%%%%%%%% MIN

\IncMargin{1em}
\begin{algorithm}[!h]
\SetKwData{Left}{left}\SetKwData{This}{this}\SetKwData{Up}{up}
\SetKwFunction{Union}{Union}\SetKwFunction{FindCompress}{FindCompress}
\SetKwInOut{Input}{Input}\SetKwInOut{Output}{Output}
\Input{$Q_t(\theta)$ a function defined on $N_t$ intervals: $(A_i^{(t)}, B_i^{(t)}]$ \\
For each $(A_i^{(t)}, B_i^{(t)}]$ we also have an associated change: $\tau_i^{(t)}$ \\
}
\BlankLine
Set $Q_t = \infty$ \;
Set $\tau_t = 0$ \;

\While{ $i \leq N_t$ }{
On the intervals $(A_i^{(t)}, B_i^{(t)}]$ recover $m = \min \{ Q_t(\theta) \}$ \;
\If{ $m < Q_t$}{
Set $Q_t = m$ \;
Set $\tau_t =\tau_i^{(t)}$
}
$i = i+1$ \;

}
\Output{$Q_t$ and $\tau_t$}
\BlankLine
\caption{Sub-routine to recover the minimum $Q_t$ and best change $\tau_t$ of $Q_t(\theta)$ }\label{algo_min}
\end{algorithm}\DecMargin{1em}

%%%%%%%%%%%%%%%%%%%%%%%%%%%% COMPARE
\IncMargin{1em}
\begin{algorithm}[!h]
\SetKwData{Left}{left}\SetKwData{This}{this}\SetKwData{Up}{up}
\SetKwFunction{Union}{Union}\SetKwFunction{FindCompress}{FindCompress}
\SetKwInOut{Input}{Input}\SetKwInOut{Output}{Output}
\Input{$Q_t(\theta)$ a function defined on $N_t$ intervals: $(A_i^{(t)}, B_i^{(t)}]$ \\
For each $(A_i^{(t)}, B_i^{(t)}]$ we have an associated change: $\tau_i^{(t)}$ \\
$C$ a constant function ($= Q_t + \beta $)\\
}
\BlankLine
Set current number of intervals for $Q^*_{t+1}(\theta)$ to $N^*_{t+1}=0$ \;

\While{  $i \leq N_t$ }{
Find the $n_t$ roots of $(Q_t(\theta) - C)$ in the intervals $(A_i^{(t)}, B_i^{(t)})$ \;
Sort the $n_t$ roots and store them in a vector $R_{tmp}$ \;
Create the vector $R = (A_i^{(t)}, R_{tmp}, B_i^{(t)})$ \;
\For{ $j = 1$ to $j=n_t+1$ }{
$N^*_{t+1}= N^*_{t+1} +1$ \;
Create a new interval $\left(a_{N^*_{t+1}}^{(t+1)}, b_{N^*_{t+1}}^{(t+1)}\right] =  (R_j, R_{j+1}]$ \;
\If{ $Q_t(\theta) \geq C$ on $(R_j, R_{j+1}]$ }{
For $\theta$ in $(R_j, R_{j+1}]$ set $Q^*_t(\theta) = C$ \;
Set $\tau_{t+1}^{*(N^*_{t+1})} = t$ \;
}
\Else{
For $\theta$ in $(R_j, R_{j+1}]$ set $Q^*_t(\theta) = Q_t(\theta)$ \;
Set $\tau_{t+1}^{*(N^*_{t+1})} = \tau_{t}^{(i)}$ \;
}
}
$i = i+1$ \;
}
\Output{The function $Q^*_{t+1}(\theta)$ defined on $N^*_{t+1}$ intervals: $(a_i^{(t+1)}, b_i^{(t+1)}]$}
\BlankLine
\caption{Sub-routine to compare $Q_t(\theta)$ to a constant function and recover $Q^*_{t+1}(\theta)$}\label{algo_compare}
\end{algorithm}\DecMargin{1em}

\pagebreak

\section{ROC curve for all tumor fractions} \label{App:all results}

\begin{figure}[h]
\center{
\begin{tabular}{ccc}
Tumor Fraction = 0.34 & Tumor Fraction = 0.5  & \\
\includegraphics[width=6.1cm, trim = 0cm 0cm 3.5cm 0cm, clip]{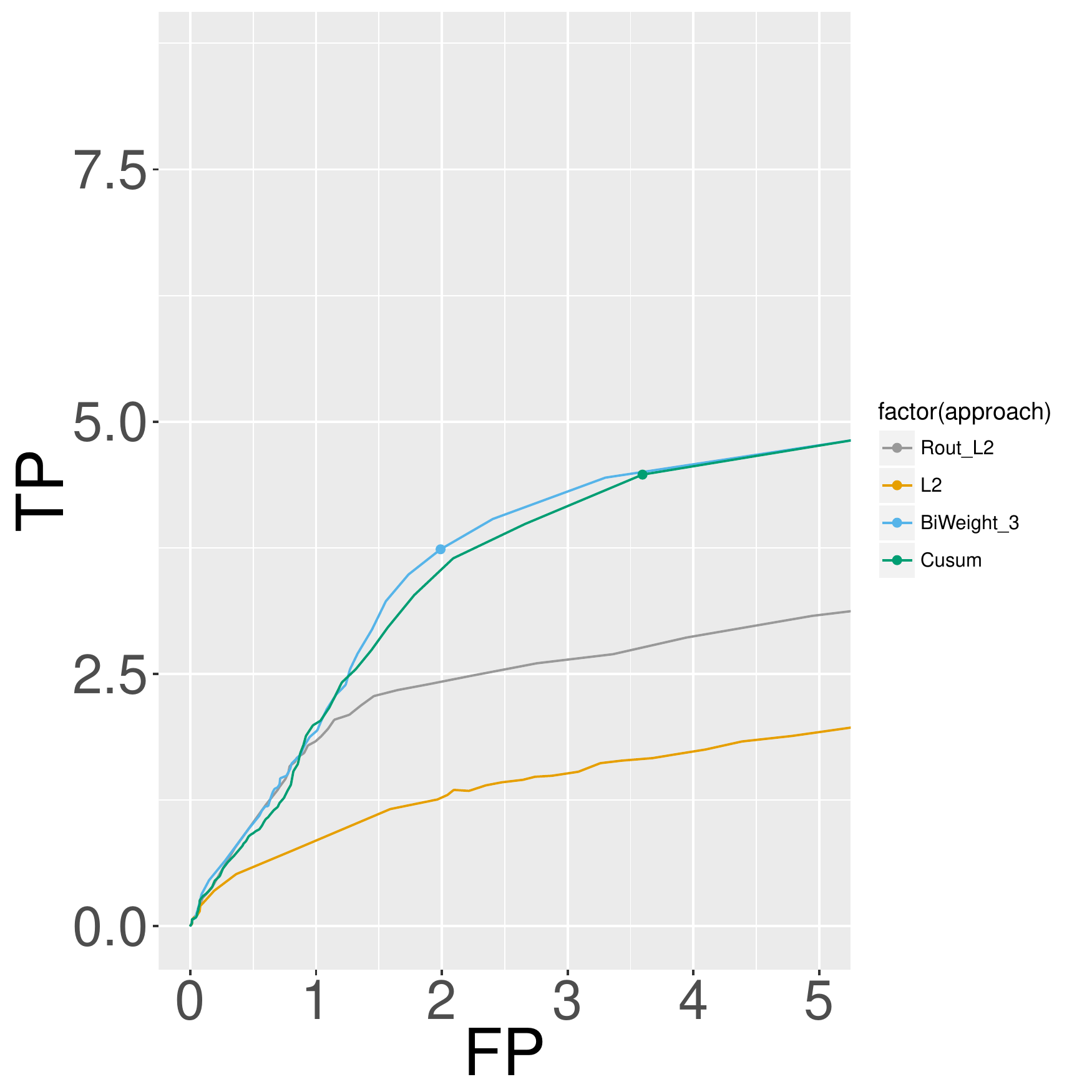} &
\includegraphics[width=6.1cm, trim = 0cm 0cm 3.5cm 0cm, clip]{1_GSE11976ROC_0_5.pdf} &
\includegraphics[width=3cm, trim = 14cm 6cm 0cm 7cm, clip]{1_GSE11976ROC_0_5.pdf} \\
Tumor Fraction = 0.79 & Tumor Fraction = 1  & \\
\includegraphics[width=6.1cm, trim = 0cm 0cm 3.5cm 0cm, clip]{1_GSE11976ROC_0_79.pdf} &
\includegraphics[width=6.1cm, trim = 0cm 0cm 3.5cm 0cm, clip]{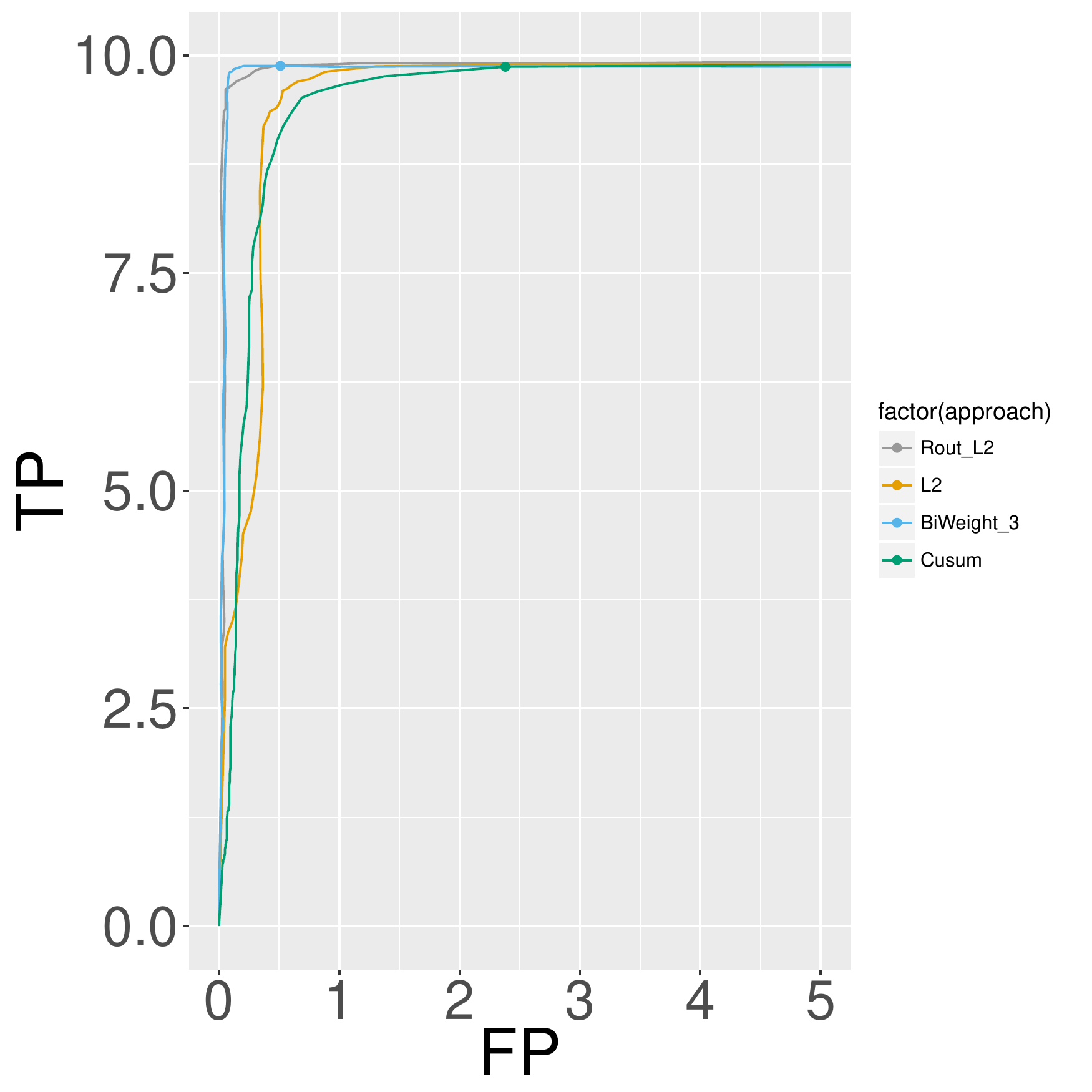} & \\
\end{tabular}}
\caption{Average ROC on the GSE11976 datasets for the Cusum, L2, L2 with outlier removal (Rout L2) and our robust biweight loss (Biweight 3) for four tumor fraction.}\label{fig:DNA_copy2}
\end{figure}

\begin{figure}
\center{
\begin{tabular}{ccc}
Tumor Fraction = 0.3 & Tumor Fraction = 0.5  & \\
\includegraphics[width=6.1cm, trim = 0cm 0cm 3.5cm 0cm, clip]{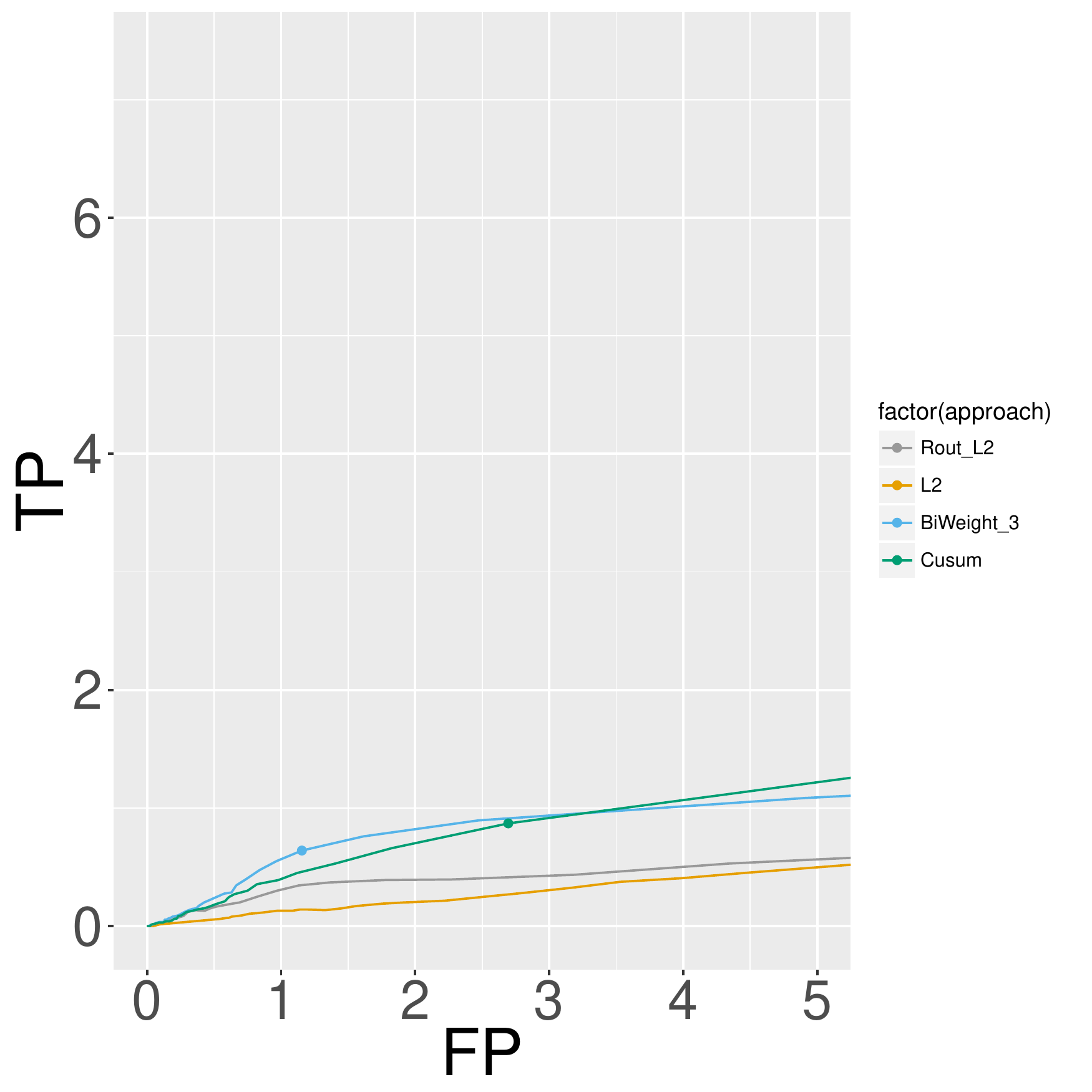} &
\includegraphics[width=6.1cm, trim = 0cm 0cm 3.5cm 0cm, clip]{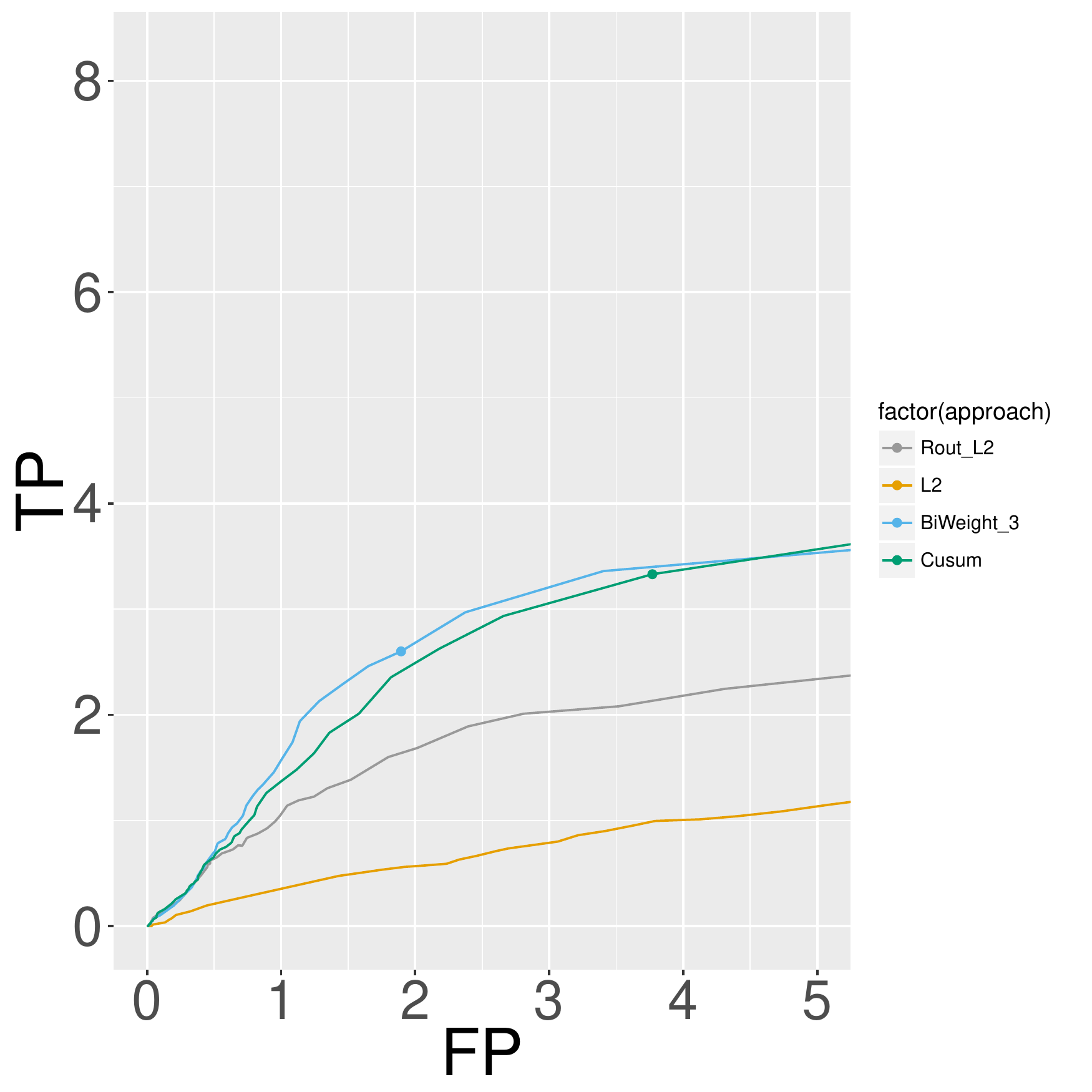} &
\includegraphics[width=3cm, trim = 14cm 6cm 0cm 7cm, clip]{2_GSE29172ROC_0_5.pdf} \\
\\
Tumor Fraction = 0.7 & Tumor Fraction = 1  \\
\includegraphics[width=6.1cm, trim = 0cm 0cm 3.5cm 0cm, clip]{2_GSE29172ROC_0_7.pdf} &
\includegraphics[width=6.1cm, trim = 0cm 0cm 3.5cm 0cm, clip]{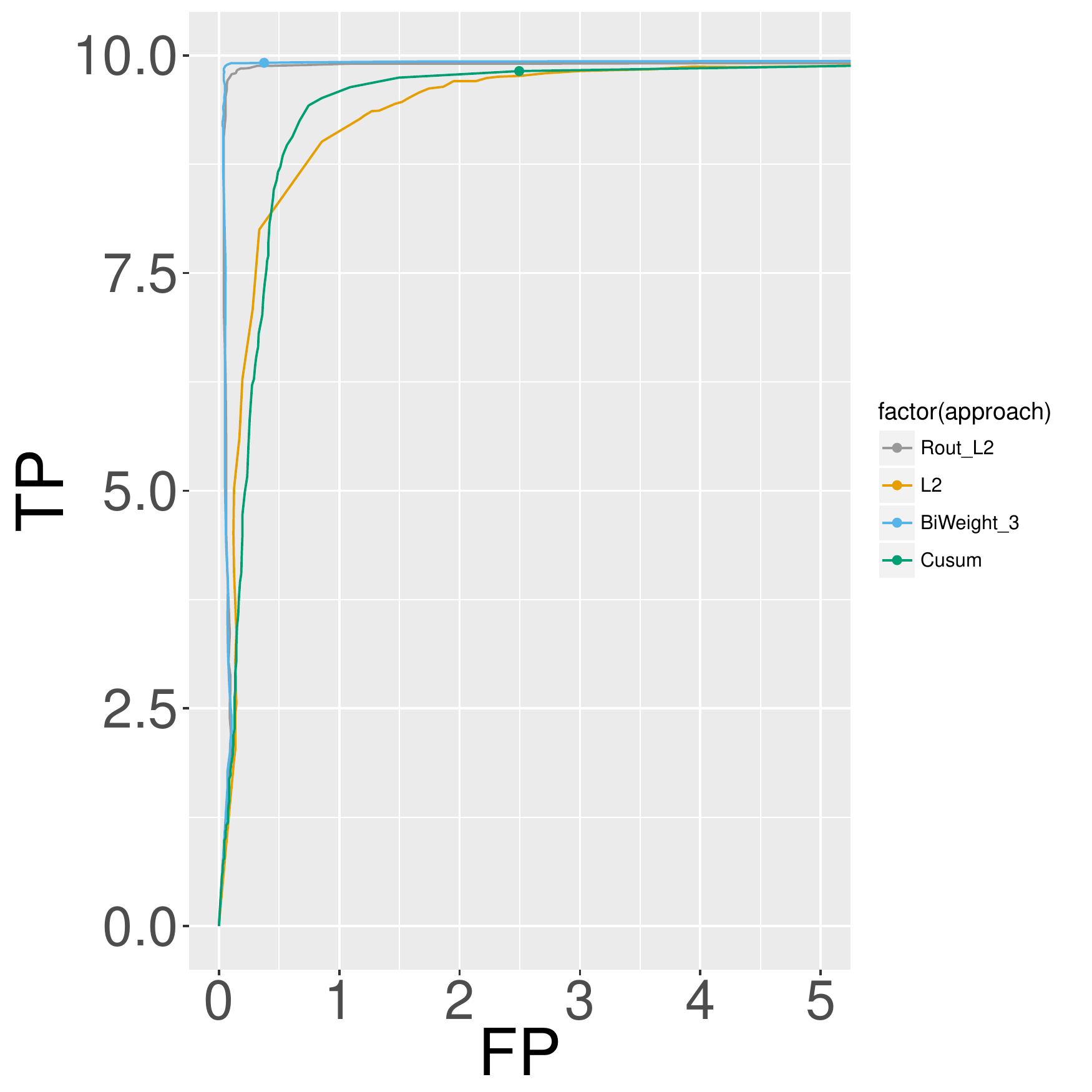} & \\
\end{tabular}}
\caption{Average ROC on the GSE29172 datasets for the Cusum, L2, L2 with outlier removal (Rout L2) and our robust biweight loss (Biweight 3) for four tumor fraction.}\label{fig:DNA_copy1}
\end{figure}

%\pagebreak
\putbib[refs]
\end{bibunit}

\end{appendices}
\end{document}

%% file: header.tex
\usepackage{graphicx}
\usepackage{natbib}
\usepackage{amsmath}
\usepackage{amsfonts}
\usepackage{url}

\newcounter{ctr}

\newcounter{ctr1}

\newcounter{ctr2}

\newcounter{ctr3}

\newtheorem{definition}{Definition}[section]    %for book style
\newtheorem{theorem}[definition]{Theorem}
\newenvironment{theorem*}[1]{{\bf Theorem #1} \begin{itshape}}{\end{itshape}}
\newtheorem{lemma}[definition]{Lemma}
\newtheorem{corollary}[definition]{Corollary}
\newenvironment{corollary*}[1]{{\bf Corollary #1} \begin{itshape}}{\end{itshape}}

\newenvironment{proposition*}[1]{{\bf Proposition #1} \begin{itshape}}{\end{itshape}}

\newcommand{\ud}{\, {\rm d} \kern-.015em }

\newcommand{\modulus}[1]{\left| \kern.05em #1 \kern.05em \right|}
\newcommand{\norm}[1]{\left\| \kern.05em #1 \kern.05em \right\|}
\newcommand{\inner}[1]{\left\langle \kern.05em #1 \kern.05em \right\rangle }

\newcommand{\bm}[1]{\mbox{\protect\boldmath $ #1 $}}

\newcommand{\pick}[2]{\renewcommand{\arraystretch}{0.6}
\left( \kern-.4em \begin{array}{c} #1 \\ #2 \end{array} \kern-.4em \right) }

%% file: Robust_Cpt_Rev2.bbl
\begin{thebibliography}{}

\bibitem[Adams and MacKay, 2007]{Adams:2007}
Adams, R.~P. and MacKay, D.~J. (2007).
\newblock Bayesian online changepoint detection.
\newblock {\em arXiv:0710.3742}.

\bibitem[Bagci, 2016]{Bagci:2016}
Bagci, I.~E. (2016).
\newblock {\em Novel security mechanisms for wireless sensor networks}.
\newblock PhD thesis, Lancaster University, Lancaster, UK.

\bibitem[Bagci et~al., 2015]{Bagci:2015}
Bagci, I.~E., Roedig, U., Martinovic, I., Schulz, M., and Hollick, M. (2015).
\newblock Using channel state information for tamper detection in the internet
  of things.
\newblock In {\em Proceedings of the 31st Annual Computer Security Applications
  Conference}, pages 131--140. ACM.

\bibitem[Bai, 1997]{Bai:1997}
Bai, J. (1997).
\newblock Estimating multiple breaks one at a time.
\newblock {\em Econometric theory}, 13(3):315--352.

\bibitem[Baranowski et~al., 2016]{baranowski2016narrowest}
Baranowski, R., Chen, Y., and Fryzlewicz, P. (2016).
\newblock Narrowest-over-threshold detection of multiple change-points and
  change-point-like features.
\newblock {\em arXiv:1609.00293}.

\bibitem[Bengtsson et~al., 2016]{Bengtsson:2016}
Bengtsson, H., Neuvial, P., Seshan, V.~E., Olshen, A.~B., Spellman, P.~T., and
  Olshen, R.~A. (2016).
\newblock Package ‘pscbs’.

\bibitem[Cao and Wu, 2015]{cao2015changepoint}
Cao, H. and Wu, W.~B. (2015).
\newblock Changepoint estimation: another look at multiple testing problems.
\newblock {\em Biometrika}, 102(4):974--980.

\bibitem[Fearnhead, 2006]{Fearnhead:2006SC}
Fearnhead, P. (2006).
\newblock Exact and efficient inference for multiple changepoint problems.
\newblock {\em Statistics and Computing}, 16:203--213.

\bibitem[Frick et~al., 2014]{Frick:2014}
Frick, K., Munk, A., and Sieling, H. (2014).
\newblock {Multiscale change-point inference}.
\newblock {\em Journal of the Royal Statistical Society: Series B},
  76(3):495--580.

\bibitem[Fryzlewicz, 2014]{Fryzlewicz:2014}
Fryzlewicz, P. (2014).
\newblock Wild binary segmentation for multiple change-point detection.
\newblock {\em Annals of Statistics}, 42:2243--2281.

\bibitem[Futschik et~al., 2014]{Futschik:2014}
Futschik, A., Hotz, T., Munk, A., and Sieling, H. (2014).
\newblock {Multiscale DNA partitioning: statistical evidence for segments}.
\newblock {\em Bioinformatics}, 30:2255--2262.

\bibitem[Haynes et~al., 2017a]{Haynes:2015}
Haynes, K., Eckley, I.~A., and Fearnhead, P. (2017a).
\newblock Computationally efficient changepoint detection for a range of
  penalties.
\newblock {\em Journal of Computational and Graphical Statistics}, 26:134--143.

\bibitem[Haynes et~al., 2017b]{Haynes:2016}
Haynes, K., P, F., and Eckley, I. (2017b).
\newblock A computationally efficient nonparametric approach for changepoint
  detection.
\newblock {\em Statistics and Computing}, 27:1293--1305.

\bibitem[Hinkley, 1971]{Hinkley:1971}
Hinkley, D.~V. (1971).
\newblock Inference about the change-point from cumulative sum tests.
\newblock {\em Biometrika}, 58(3):509--523.

\bibitem[Hotz et~al., 2013]{Hotz:2013}
Hotz, T., Sch{\"u}tte, O.~M., Sieling, H., Polupanow, T., Diederichsen, U.,
  Steinem, C., and Munk, A. (2013).
\newblock Idealizing ion channel recordings by a jump segmentation
  multiresolution filter.
\newblock {\em IEEE Transactions on Nanobioscience}, 12(4):376--386.

\bibitem[Huber, 2011]{Huber:2011}
Huber, P.~J. (2011).
\newblock {\em Robust statistics}.
\newblock Springer.

\bibitem[Hu{\v{s}}kov{\'a}, 1991]{Huvskova:1991}
Hu{\v{s}}kov{\'a}, M. (1991).
\newblock {Recursive M-tests for the change-point problem}.
\newblock In {\em Economic Structural Change}, pages 13--33. Springer.

\bibitem[Hu{\v{s}}kov{\'a}, 2013]{Huvskova:2013}
Hu{\v{s}}kov{\'a}, M. (2013).
\newblock Robust change point analysis.
\newblock In {\em Robustness and Complex Data Structures}, pages 171--190.
  Springer.

\bibitem[Hu{\v{s}}kov{\'a} and Maru{\v{s}}iakov{\'a}, 2012]{Huvskova:2012}
Hu{\v{s}}kov{\'a}, M. and Maru{\v{s}}iakov{\'a}, M. (2012).
\newblock M-procedures for detection of changes for dependent observations.
\newblock {\em Communications in Statistics-Simulation and Computation},
  41(7):1032--1050.

\bibitem[Hu{\v{s}}kov{\'a} and Picek, 2005]{Huvskova:2005}
Hu{\v{s}}kov{\'a}, M. and Picek, J. (2005).
\newblock Bootstrap in detection of changes in linear regression.
\newblock {\em Sankhy{\=a}: The Indian Journal of Statistics}, pages 200--226.

\bibitem[Hu{\v{s}}kov{\'a} and Sen, 1989]{Huvskova:1989}
Hu{\v{s}}kov{\'a}, M. and Sen, P.~K. (1989).
\newblock Nonparametric tests for shift and change in regression at an unknown
  time point.
\newblock In {\em Statistical Analysis and Forecasting of Economic Structural
  Change}, pages 71--85. Springer.

\bibitem[Johnson, 2013]{Johnson:2013}
Johnson, N.~A. (2013).
\newblock A dynamic programming algorithm for the fused lasso and
  $l_0$-segmentation.
\newblock {\em Journal of Computational and Graphical Statistics},
  22(2):246--260.

\bibitem[Killick et~al., 2010]{Killick:2010}
Killick, R., Eckley, I.~A., Ewans, K., and Jonathan, P. (2010).
\newblock {Detection of changes in variance of oceanographic time-series using
  changepoint analysis}.
\newblock {\em Ocean Engineering}, 37(13):1120--1126.

\bibitem[Killick et~al., 2012]{Killick:2012}
Killick, R., Fearnhead, P., and Eckley, I.~A. (2012).
\newblock {Optimal detection of changepoints with a linear computational cost}.
\newblock {\em Journal of the American Statistical Association},
  107(500):1590--1598.

\bibitem[Kim et~al., 2005]{Kim:2012}
Kim, C.-J., Morley, J.~C., and Nelson, C.~R. (2005).
\newblock The structural break in the equity premium.
\newblock {\em Journal of Business \& Economic Statistics}, 23:181--191.

\bibitem[Lavielle and Moulines, 2000]{Lavielle:2000}
Lavielle, M. and Moulines, E. (2000).
\newblock Least-squares estimation of an unknown number of shifts in a time
  series.
\newblock {\em Journal of time series analysis}, 21(1):33--59.

\bibitem[Ma and Yau, 2016]{ma2016pairwise}
Ma, T.~F. and Yau, C.~Y. (2016).
\newblock A pairwise likelihood-based approach for changepoint detection in
  multivariate time series models.
\newblock {\em Biometrika}, 103(2):409--421.

\bibitem[Maidstone et~al., 2017]{Maidstone:2016}
Maidstone, R., Hocking, T., Rigaill, G., and Fearnhead, P. (2017).
\newblock On optimal multiple changepoint algorithms for large data.
\newblock {\em Statistics and Computing}, 27:519--533.

\bibitem[{National Research Council}, 2013]{Frontiers2013}
{National Research Council} (2013).
\newblock Frontiers in massive data analysis.

\bibitem[Olshen et~al., 2004]{Olshen:2004}
Olshen, A.~B., Venkatraman, E.~S., Lucito, R., and Wigler, M. (2004).
\newblock {Circular binary segmentation for the analysis of array-based DNA
  copy number data.}
\newblock {\em Biostatistics}, 5(4):557--72.

\bibitem[{\'{O} Ruanaidh} and Fitzgerald, 1996]{Fitzgerald:1996}
{\'{O} Ruanaidh}, J. J.~K. and Fitzgerald, W.~J. (1996).
\newblock {\em {Numerical Bayesion Methods Applied to Signal Processing}}.
\newblock New York: Springer.

\bibitem[Page, 1954]{Page:1954}
Page, E. (1954).
\newblock Continuous inspection schemes.
\newblock {\em Biometrika}, 41(1/2):100--115.

\bibitem[Pierre-Jean et~al., 2015]{pierre2014performance}
Pierre-Jean, M., Rigaill, G., and Neuvial, P. (2015).
\newblock {Performance evaluation of DNA copy number segmentation methods}.
\newblock {\em Briefings in Bioinformatics}, 16:600--615.

\bibitem[Reeves et~al., 2007]{Reeves:2007}
Reeves, J., Chen, J., Wang, X.~L., Lund, R., and Lu, Q.~Q. (2007).
\newblock A review and comparison of changepoint detection techniques for
  climate data.
\newblock {\em Journal of Applied Meteorology and Climatology}, 46(6):900--915.

\bibitem[Rigaill, 2015]{rigaill2015pruned}
Rigaill, G. (2015).
\newblock A pruned dynamic programming algorithm to recover the best
  segmentations with 1 to {$K_{\mbox{max}}$} change-points.
\newblock {\em Journal de la Soci{\'e}t{\'e} Fran{\c{c}}aise de Statistique},
  156(4):180--205.

\bibitem[Rigaill et~al., 2013]{Rigaill:2013}
Rigaill, G., Hocking, T.~D., Bach, F., and Vert, J.-P. (2013).
\newblock Learning sparse penalties for change-point detection using max margin
  interval regression.
\newblock In {\em Proceedings of the 30th International Conference on Machine
  Learning, JMLR W\&CP}, volume~28, pages 172--180.

\bibitem[Ruggieri and Antonellis, 2016]{Ruggieri:2016}
Ruggieri, E. and Antonellis, M. (2016).
\newblock {An exact approach to Bayesian sequential change point detection}.
\newblock {\em Computational Statistics \& Data Analysis}, 97:71--86.

\bibitem[Vostrikova, 1981]{Vostrikova:1981}
Vostrikova, L. (1981).
\newblock Detection of the disorder in multidimensional random-processes.
\newblock {\em Doklady Akademii Nauk SSSR}, 259(2):270--274.

\bibitem[Worsley, 1979]{Worsley:1979}
Worsley, K. (1979).
\newblock On the likelihood ratio test for a shift in location of normal
  populations.
\newblock {\em Journal of the American Statistical Association},
  74(366a):365--367.

\bibitem[Wyse et~al., 2011]{Wyse:2011}
Wyse, J., Friel, N., et~al. (2011).
\newblock {Approximate simulation-free Bayesian inference for multiple
  changepoint models with dependence within segments}.
\newblock {\em Bayesian Analysis}, 6(4):501--528.

\bibitem[Yao, 1984]{Yao:1984}
Yao, Y.-C. (1984).
\newblock {Estimation of a noisy discrete-time step function: Bayes and
  empirical Bayes approaches}.
\newblock {\em The Annals of Statistics}, pages 1434--1447.

\bibitem[Yao and Au, 1989]{Yao:1989}
Yao, Y.-C. and Au, S. (1989).
\newblock Least-squares estimation of a step function.
\newblock {\em Sankhy{\=a}: The Indian Journal of Statistics, Series A}, pages
  370--381.

\end{thebibliography}


\begin{thebibliography}{}

\bibitem[Boucheron et~al., 2013]{boucheron2013concentration}
Boucheron, S., Lugosi, G., and Massart, P. (2013).
\newblock {\em Concentration inequalities: A nonasymptotic theory of
  independence}.
\newblock Oxford University Press.

\bibitem[Rigaill, 2015]{rigaill2015pruned}
Rigaill, G. (2015).
\newblock A pruned dynamic programming algorithm to recover the best
  segmentations with 1 to {$K_{\mbox{max}}$} change-points.
\newblock {\em Journal de la Soci{\'e}t{\'e} Fran{\c{c}}aise de Statistique},
  156(4):180--205.

\end{thebibliography}
